\pdfoutput=1

\documentclass[11pt,twoside,a4paper,cmspaper,final,collab]{cms-tdr}
\def\svnVersion{457141P}\def\cmsCernNoTag{CERN-EP-2017-193}\def\cmsCernDate{\today}\def\cmsMessage{Published in Physical Review C as \href{http://dx.doi.org/10.1103/PhysRevC.97.044912}{\doi{10.1103/PhysRevC.97.044912.}}}

\begin{document}\cmsNoteHeader{HIN-17-001}

\hyphenation{had-ron-i-za-tion}
\hyphenation{cal-or-i-me-ter}
\hyphenation{de-vices}
\RCS$HeadURL: svn+ssh://svn.cern.ch/reps/tdr2/papers/HIN-17-001/trunk/HIN-17-001.tex $
\RCS$Id: HIN-17-001.tex 455706 2018-04-16 16:25:47Z ztu $
\newlength\cmsFigWidth
\ifthenelse{\boolean{cms@external}}{\setlength\cmsFigWidth{0.98\columnwidth}}{\setlength\cmsFigWidth{0.7\textwidth}}
\ifthenelse{\boolean{cms@external}}{\providecommand{\cmsLeft}{upper\xspace}}{\providecommand{\cmsLeft}{left\xspace}}
\ifthenelse{\boolean{cms@external}}{\providecommand{\cmsRight}{lower\xspace}}{\providecommand{\cmsRight}{right\xspace}}
\ifthenelse{\boolean{cms@external}}{\providecommand{\CL}{C.L.\xspace}}{\providecommand{\CL}{CL\xspace}}
\ifthenelse{\boolean{cms@external}}{\providecommand{\CLp}{C.L\xspace}}{\providecommand{\CLp}{CL\xspace}}
\newcommand {\rootsNN}  {\ensuremath{\sqrt{\smash[b]{s_{_{\mathrm{NN}}}}}}}
\newcommand {\deta}     {\ensuremath{\Delta\eta}}
\newcommand {\ptbar}       {\ensuremath{\overline{p}_\mathrm{T}}\xspace}
\newcommand {\ptalpha}       {\ensuremath{p_{\mathrm T,\alpha}}}
\newcommand {\ptbeta}       {\ensuremath{p_{\mathrm T,\beta}}}
\newcommand {\PbPb}  {\ensuremath{\text{PbPb}}\xspace}
\newcommand {\pPb}  {\ensuremath{\text{pPb}}\xspace}
\newcommand {\AonA}  {\ensuremath{\text{AA}}\xspace}
\newcommand{\noff}    {\ensuremath{N_\text{trk}^\text{offline}}\xspace}
\newcommand{\Pb}{\ensuremath{\mathrm{Pb}}\xspace}
\renewcommand{\Pp}{\ensuremath{\mathrm{p}}\xspace}

\cmsNoteHeader{HIN-17-001}
\title{Constraints on the chiral magnetic effect using charge-dependent azimuthal correlations in \texorpdfstring{\pPb}{pPb} and \texorpdfstring{\PbPb}{PbPb} collisions at the LHC}

\date{\today}

\abstract{
Charge-dependent azimuthal correlations of same- and opposite-sign pairs with respect to the second- and third-order event planes
have been measured in \pPb collisions at $\rootsNN = 8.16\TeV$ and \PbPb
collisions at 5.02\TeV with the CMS experiment at the LHC. The measurement is motivated
by the search for the charge separation phenomenon predicted by the chiral magnetic effect (CME) in heavy ion collisions.
Three- and two-particle azimuthal correlators are extracted
as functions of the pseudorapidity difference, the transverse momentum (\pt) difference, and the \pt average of
same- and opposite-charge pairs in various event multiplicity ranges.
The data suggest that the charge-dependent three-particle correlators with respect to the second- and third-order event planes share a common origin, predominantly arising from charge-dependent two-particle azimuthal correlations coupled with an anisotropic flow. The CME is expected to lead to a
$v_{2}$-independent three-particle correlation when the magnetic field is fixed. Using an event shape engineering technique, upper limits on the $v_{2}$-independent fraction of the three-particle correlator are estimated to be 13\% for \pPb and 7\% for \PbPb collisions at 95\% confidence level. The results of this analysis,
both the dominance of two-particle correlations as a source of the three-particle
results and the similarities seen between \PbPb and \pPb, provide stringent constraints on the
origin of charge-dependent three-particle azimuthal correlations and challenge their interpretation
as arising from a chiral magnetic effect in heavy ion collisions.
}

\hypersetup{%
pdfauthor={CMS Collaboration},%
pdftitle={Constraints on the chiral magnetic effect using charge-dependent azimuthal correlations in pPb and PbPb collisions at the LHC},%
pdfsubject={CMS},%
pdfkeywords={CMS, heavy ion physics, CME, small systems}}

\maketitle

\section{Introduction}
\label{sec:intro}

It has been suggested that in high-energy nucleus-nucleus (\AonA) collisions, metastable
domains of gluon fields with nontrivial topological configurations may
form~\cite{Lee:1973iz,Lee:1974ma,Morley1985,Kharzeev:1998kz}.
These domains can carry an imbalance between left- and right-handed quarks
arising from interactions of chiral quarks with topological gluon fields,
leading to a local parity ($P$) violation~\cite{Morley1985,Kharzeev:1998kz}.
This chirality imbalance, in the presence of the extremely strong magnetic field, which can be produced in a
noncentral \AonA collision, is expected to lead to
an electric current perpendicular to the reaction plane, resulting in a final-state charge
separation phenomenon known as the chiral magnetic effect (CME)~\cite{Kharzeev:2004ey,Kharzeev:2007jp,Fukushima:2008xe}.
Such macroscopic phenomena arising from quantum anomalies are a subject of interest for a wide range
of physics communities. The chiral-anomaly-induced phenomena have been observed in magnetized relativistic matter in three-dimensional
Dirac and Weyl materials~\cite{Lv:2015pya,Huang:2015eia,Li:2014bha}.
The search for the charge separation from the CME in \AonA collisions
was first carried out at RHIC at BNL~\cite{Abelev:2009ac,Abelev:2009ad,Adamczyk:2013kcb,Adamczyk:2014mzf,Adamczyk:2013hsi}
and later at the CERN LHC~\cite{Abelev:2012pa} at various center-of-mass energies.
In these measurements, a charge-dependent azimuthal correlation with respect to
the reaction plane was observed, which is qualitatively consistent with the expectation
of charge separation from the CME. No strong collision energy dependence of the signal
is observed going from RHIC to LHC energies, although some theoretical predictions
suggested that the possible CME signal could be much
smaller at the LHC than at RHIC because of a shorter lifetime of the magnetic field~\cite{Kharzeev:2015znc}. Nevertheless, theoretical estimates of the time evolution of the magnetic field have large uncertainties~\cite{Kharzeev:2015znc}.

The experimental evidence for the CME in heavy ion collisions remains
inconclusive because of several identified sources of background
correlations that can account for part or all of the observed charge-dependent
azimuthal correlations~\cite{Wang:2009kd,Bzdak:2010fd,Schlichting:2010qia}.
Moreover, the charge-dependent azimuthal correlation
in high-multiplicity \pPb collisions has been recently found to have a nearly
identical value to that observed in \PbPb collisions~\cite{Khachatryan:2016got}. This is a strong indication that
the observed effect in heavy ion collisions might predominantly result from background contributions. The CME-induced charge separation effect is predicted to be negligible in \pPb collisions, as the angle between
the magnetic field direction and the event plane
is expected to be randomly distributed~\cite{Khachatryan:2016got,Belmont:2016oqp}.

The charge separation can be characterized by the first $P$-odd sine term ($a_{1}$)
in a Fourier decomposition of the charged-particle azimuthal distribution~\cite{Voloshin:2004vk}:
\begin{linenomath}
\begin{equation}
\label{azimuthal}
\frac{\rd N}{\rd\phi} \propto  1 + 2\sum_{n} \bigl\{v_{n}\cos[n(\phi-\Psi_\mathrm{RP})] + a_{n}\sin[n(\phi-\Psi_\mathrm{RP})]\bigr\},
\end{equation}
\end{linenomath}
where $\phi - \Psi_\mathrm{RP}$ represents the particle azimuthal angle
with respect to the reaction plane angle $\Psi_\mathrm{RP}$ in heavy ion collisions (determined
by the impact parameter and beam axis), and $v_{n}$ and $a_{n}$ denote the
coefficients of $P$-even and $P$-odd Fourier terms, respectively. Although
the reaction plane is not an experimental observable, it can be approximated in heavy ion collisions by the second-order event
plane, $\Psi_{2}$, determined by the direction of the beam and the maximal particle
density in the elliptic azimuthal anisotropy. The $P$-odd terms will vanish after averaging over events, because the sign of the chirality imbalance changes event by event. Therefore, the observation of such an effect is only possible through the measurement of particle azimuthal correlations. An azimuthal three-particle correlator, $\gamma_{112}$,
proposed to explore the first coefficient, $a_1$, of the $P$-odd Fourier terms
characterizing the charge separation~\cite{Voloshin:2004vk} is:
\begin{linenomath}
\ifthenelse{\boolean{cms@external}}{
\begin{multline}
\label{2pcorrelatorEP}
\gamma_{112} \equiv \left\langle \cos(\phi_{\alpha} + \phi_{\beta} - 2\Psi_{2}) \right\rangle  = \\
\langle \cos(\phi_{\alpha}-\Psi_{2})
\cos(\phi_{\beta}-\Psi_{2}) \rangle\\
- \langle \sin(\phi_{\alpha}-\Psi_{2})\sin(\phi_{\beta}-\Psi_{2}) \rangle.
\end{multline}
}{
\begin{equation}
\label{2pcorrelatorEP}
\gamma_{112} \equiv \left\langle \cos(\phi_{\alpha} + \phi_{\beta} - 2\Psi_{2}) \right\rangle  = \left\langle \cos(\phi_{\alpha}-\Psi_{2})\cos(\phi_{\beta}-\Psi_{2}) \right\rangle
- \left\langle \sin(\phi_{\alpha}-\Psi_{2})\sin(\phi_{\beta}-\Psi_{2}) \right\rangle.
\end{equation}
}
\end{linenomath}
Here, $\alpha$ and $\beta$ denote particles with the same or opposite electric
charge sign and the angle brackets reflect an averaging over particles and events.
Assuming particles $\alpha$ and $\beta$ are uncorrelated, except for their individual correlations
with respect to the event plane, the first term on the right-hand side of Eq.~(\ref{2pcorrelatorEP})
becomes $\left\langle v_{1,\alpha}v_{1,\beta} \right\rangle$, which is generally small
and independent of the charge~\cite{Abelev:2009ad}, while the second term is sensitive to the
charge separation and can be expressed as $\left\langle a_{1,\alpha}a_{1,\beta} \right\rangle$.

While the similarity of the \pPb and \PbPb data at 5.02\TeV analyzed by the CMS experiment pose a considerable challenge to the CME interpretation of
the charge-dependent azimuthal correlations observed in \AonA collisions~\cite{Khachatryan:2016got},
important questions still remain to be addressed: is the correlation
signal observed in pPb collisions entirely a consequence of background
correlations? What is the underlying mechanism for those
background correlations that are almost identical in \pPb and \PbPb collisions? Can the background contribution
be quantitatively constrained with data and, if so, is there still
evidence for a statistically significant CME signal?

In particular, among the proposed mechanisms for background
correlations, one source is related to the charge-dependent two-particle correlation
from local charge conservation in decays of resonances or clusters (\eg, jets)~\cite{Schlichting:2010qia}.
By coupling with the anisotropic particle emission, an effect
resembling charge separation with respect to the reaction plane can be
generated. The observed characteristic range of the two-particle correlation in
data is around one unit of rapidity, consistent with short-range cluster decays.
In this mechanism of local charge conservation coupled with
the elliptic flow, a background contribution to the three-particle
correlator, $\gamma_{112}$, is expected to be~\cite{Bzdak:2012ia}:
\begin{linenomath}
\begin{equation}
\label{eq:lcc}
\gamma^\text{bkg}_{112} = \kappa_{2} \left\langle \cos(\phi_{\alpha}-\phi_{\beta}) \right\rangle \left\langle \cos 2(\phi_{\beta}-\Psi_\mathrm{RP}) \right\rangle \\
= \kappa_{2} \, \delta \, v_{2}.
\end{equation}
\end{linenomath}
Here, $\delta \equiv \left\langle \cos(\phi_{\alpha}-\phi_{\beta}) \right\rangle$
represents the charge-dependent two-particle azimuthal correlator and $\kappa_{2}$ is a constant parameter, independent of $v_2$, but mainly determined
by the kinematics and acceptance of particle detection~\cite{Bzdak:2012ia}.
As both the charge conservation effect and anisotropic flow are known to be
present in heavy ion collisions, the primary goal of this paper is to conduct
a systematic investigation of how much of the observed charge-dependent correlations
in the data can be accounted for by this mechanism.

Although the background contribution from local charge conservation
is well defined in Eq.~(\ref{eq:lcc}) and has been long
recognized~\cite{Schlichting:2010qia,Bzdak:2012ia,Kharzeev:2015znc}, it is still not known to what extent background contributions account for the observed $\gamma_{112}$ correlator.
The main difficulty lies in determining the unknown value of $\kappa_2$ in a model-independent way. The other difficulty is to demonstrate directly
the linear dependence on $v_2$ of $\gamma^\text{bkg}_{112}$, which is nontrivial
as one has to ensure the magnetic field, and thus the CME, does not change when selecting events with different $v_2$ values. Therefore, selecting events with a quantity that directly relates to the magnitude of $v_2$ is essential.

This paper aims to overcome the difficulties mentioned above and achieve a better understanding as to the contribution of the local charge conservation background to the
charge-dependent azimuthal correlation data. The results should serve as a new baseline for the search
for the CME in heavy ion collisions. Two approaches are employed as outlined below.

\begin{enumerate}
\item {Higher-order harmonic three-particle correlator:}
in heavy ion collisions,
the charge separation effect from the CME is only expected along the direction
of the induced magnetic field normal to the reaction plane, approximated by the second-order event plane, $\Psi_{2}$.
As the symmetry plane of the third-order Fourier term (``triangular flow"
~\cite{Alver:2010gr}), $\Psi_{3}$, is expected to have a weak
correlation with $\Psi_{2}$~\cite{Aad:2014fla}, the charge separation effect with respect to
$\Psi_{3}$ is expected to be negligible. By constructing a charge-dependent
correlator with respect to the third-order event plane,
\begin{linenomath}
\begin{equation}
\label{eq:gamma123}
\gamma_{123} \equiv \left\langle \cos(\phi_{\alpha}+2\phi_{\beta}-3\Psi_{3}) \right\rangle,
\end{equation}
\end{linenomath}
charge-dependent background effects unrelated to the CME can be explored.
In particular, in the context of the local charge conservation mechanism, the
$\gamma_{123}$ correlator is also expected to have a background contribution, with
\ifthenelse{\boolean{cms@external}}{
\begin{linenomath}
\begin{equation}\begin{aligned}
\label{eq:lcc123}
\gamma^\text{bkg}_{123} &= \kappa_{3} \left\langle \cos(\phi_{\alpha}-\phi_{\beta}) \right\rangle \left\langle \cos 3(\phi_{\beta}-\Psi_{3}) \right\rangle \\
&= \kappa_{3} \,  \delta \,  v_{3},
\end{aligned}\end{equation}
\end{linenomath}
}{
\begin{linenomath}
\begin{equation}
\label{eq:lcc123}
\gamma^\text{bkg}_{123} = \kappa_{3} \left\langle \cos(\phi_{\alpha}-\phi_{\beta}) \right\rangle \left\langle \cos 3(\phi_{\beta}-\Psi_{3}) \right\rangle \\
= \kappa_{3} \,  \delta \,  v_{3},
\end{equation}
\end{linenomath}
}
similar to that for the $\gamma_{112}$ correlator as given in Eq.~(\ref{eq:lcc}).
As the $\kappa_2$ and $\kappa_3$ parameters mainly depend on particle kinematics and detector acceptance effects,
they are expected to be similar, largely independent of harmonic event plane orders.
The relation in Eq.~(\ref{eq:lcc123}) can be generalized for all ``higher-order harmonic''
three-particle correlators, $\gamma_{1,n-1;n} = \kappa_n \,  \delta \,  v_{n}$.
Derivation of Eq.~(\ref{eq:lcc123}) as well as generalization to all higher-order harmonics
can be found in Appendix~\ref{app:a}, which follows similar steps as for that of Eq.~(\ref{eq:lcc})
given in Ref.~\cite{Bzdak:2012ia}. One caveat here is that when averaging
over a wide $\eta$ and $\pt$ range, the $\kappa_{n}$ value
may also depend on the $\eta$ and $\pt$ dependence of the $v_n$ harmonic,
which is similar, but not exactly identical between the $v_2$ and $v_3$ coefficients~\cite{Chatrchyan:2013kba,Chatrchyan:2013nka}.

By taking the difference of correlators between same- and opposite-sign pairs
(denoted as $\Delta\gamma_{112}$ and $\Delta\gamma_{123}$ among three particles, and $\Delta\delta$ between two particles) to
eliminate all charge-independent background sources, the following relation is expected to hold
if the charge dependence of three-particle correlators is dominated by the effect of
local charge conservation coupled with the anisotropic flow:
\begin{linenomath}
\begin{equation}
\label{eq:kappa}
\frac{\Delta\gamma_{112}}{\Delta\delta \,  v_{2}} \approx \frac{\Delta\gamma_{123}}{\Delta\delta \,  v_{3}}.
\end{equation}
\end{linenomath}
Therefore, an examination of Eq.~(\ref{eq:kappa}) will quantify to what
extent the proposed background from charge conservation contributes to the $\gamma_{112}$ correlator,
and will be a critical test of the CME interpretation in heavy ion collisions.

\item {Event shape engineering (ESE):}
to establish directly a linear relationship between
the $\gamma$ correlators and $v_n$ coefficients, the ESE technique~\cite{Schukraft:2012ah}
is employed. In a narrow centrality or multiplicity range (so that the magnetic field does not change
significantly), events are further classified based on the magnitude of the event-by-event
Fourier harmonic related to the anisotropy measured in the forward rapidity region.
Within each event class, the $\gamma$ correlators and $v_n$ values are measured and compared
to test the linear relationship. A nonzero intercept value of the $\gamma$ correlators with a linear fit would reflect the strength of the CME.
\end{enumerate}

With a higher luminosity \pPb run at $\rootsNN = 8.16$\TeV and using the high-multiplicity trigger in CMS,
the \pPb data sample gives access to multiplicities comparable to those in peripheral
\PbPb collisions, allowing for a detailed comparison and
study of the two systems with very different expected CME contributions in the collisions~\cite{Khachatryan:2016got}.
Measurements of three-particle correlators, $\gamma_{112}$ and $\gamma_{123}$,
and the two-particle correlator, $\delta$, are presented in different charge combinations
as functions of the pseudorapidity ($\eta$) difference ($\abs{\deta}$), the transverse momentum (\pt) difference ($|\Delta\pt|$),
and the average \pt of correlated particles (\ptbar). Integrated over $\eta$ and \pt, the event multiplicity
dependence of three- and two-particle correlations is also presented in \pPb and \PbPb collisions.
In \pPb collisions, the particle correlations are explored separately with respect to the event planes that are obtained
using particles with $4.4<\abs{\eta}<5.0$ from the \Pp- and \Pb-going beam directions.
The ESE analysis is performed for $\gamma_{112}$ as a function of $v_2$ in both \pPb and \PbPb collisions.

This paper is organized as follows. After a brief description of the detector and data samples in Section~\ref{sec:data}, the event and track selections are discussed in Section~\ref{sec:track}, followed by the discussion of the analysis technique in Section~\ref{sec:analysis}. The results are presented in Section~\ref{sec:results}, and the paper is summarized in Section~\ref{sec:conclusion}.

\section{Detector and data samples}
\label{sec:data}

The central feature of the CMS apparatus is a superconducting solenoid of 6\unit{m}
internal diameter, providing a magnetic field of 3.8\unit{T}. Within the solenoid volume,
there are four primary subdetectors, including a silicon
pixel and strip tracker detector, a lead tungstate crystal electromagnetic calorimeter
(ECAL), and a brass and scintillator hadron calorimeter (HCAL), each composed of a
barrel and two endcap sections. The silicon tracker
measures charged particles within the range $\abs{\eta}< 2.5$. Iron and quartz-fiber Cherenkov hadron forward
(HF) calorimeters cover the range $2.9 < \abs{\eta} < 5.2$. The HF calorimeters are constituted of towers, each of which is a two-dimensional cell with a granularity of 0.5 units in $\eta$ and 0.349 radians in $\phi$. For charged
particles with $1 < \pt < 10\GeV$ and $\abs{\eta} < 1.4$, the track resolutions are
typically 1.5\% in \pt and 25--90 (45--150)\mum in the transverse (longitudinal)
impact parameter~\cite{TRK-11-001}. A detailed description of the CMS detector,
together with a definition of the coordinate system used and the relevant kinematic variables, can be found in
Ref.~\cite{Chatrchyan:2008aa}.

The \pPb data at $\rootsNN = 8.16\TeV$ used in this analysis were
collected in 2016, and correspond to an integrated luminosity
of $186\unit{nb}^{-1}$. The beam energies are 6.5\TeV for the protons and
2.56\TeV per nucleon for the lead nuclei. The data were collected in two different
run periods: one with the protons circulating in the clockwise direction
in the LHC ring, and one with them circulating in the counterclockwise direction.
By convention, the proton beam rapidity is taken to be positive when combining the
data from the two run periods.
A subset of \PbPb data at $\rootsNN = 5.02\TeV$ collected
in 2015 (30--80\% centrality, where centrality is defined as the fraction of the total
inelastic cross section, with 0\% denoting the most central collisions) is used.
The \PbPb  data were reprocessed using the same reconstruction algorithm
as the \pPb data, in order to compare directly the two colliding systems at similar final-state multiplicities.
The three-particle correlator, $\gamma_{112}$, data for \pPb collisions at \rootsNN = 8.16\TeV
are compared to those previously published at \rootsNN = 5.02\TeV~\cite{Khachatryan:2016got}
to examine any possible collision energy dependence. Because of statistical limitations, new analyses of
higher-order harmonic three-particle correlator and event shape engineering introduced in this paper
cannot be performed with the 5.02\TeV \pPb data.

\section{Selection of events and tracks}
\label{sec:track}

The event reconstruction, event selections, and the triggers, including the dedicated triggers
to collect a large sample of high-multiplicity \pPb events at $\rootsNN = 8.16$\TeV, are similar
to those used in previous CMS particle correlation measurements at lower
energies~\cite{Chatrchyan:2013nka,Khachatryan:2010gv,Khachatryan:2012dih,Khachatryan:2016txc}, as discussed below. For \PbPb events, they are identical to those in Ref.~\cite{Khachatryan:2016got}.

Minimum bias \pPb events at 8.16\TeV were selected by requiring energy deposits in at least one of the two
HF calorimeters above a threshold of approximately 1\GeV and
the presence of at least one track with $\pt > 0.4$\GeV in the pixel tracker.
In order to collect a large sample of high-multiplicity \pPb collisions, a dedicated
trigger was implemented using the CMS level-1 (L1) and high-level
trigger (HLT) systems. At L1, the total number of towers of ECAL+HCAL above
a threshold of 0.5\GeV in transverse energy (\ET) was required to be greater than
a given threshold (120 and 150 towers), where a tower is defined by $\Delta\eta{\times}\Delta\phi = 0.087{\times}0.087$\unit{radians}. Online track reconstruction for the HLT was based on
the same offline iterative tracking algorithm to maximize the trigger efficiency.
For each event, the vertex reconstructed with the greatest number of tracks was selected.
The number of tracks with $\abs{\eta}<2.4$, $\pt > 0.4\GeV$, and a distance of closest
approach less than 0.12\unit{cm} to this vertex, was determined for each event and required
to exceed a certain threshold (120, 150, 185, 250).

In the offline analysis of \pPb (\PbPb) collisions, hadronic events
are selected by requiring the presence of at least one (three) energy deposit(s) greater
than 3\GeV in each of the two HF calorimeters. Events are also required to
contain a primary vertex within 15\unit{cm} of the nominal interaction point
along the beam axis and 0.15\unit{cm} in the transverse direction.
In the \pPb data sample, the average pileup (number of interactions per bunch crossing) varied between 0.1 to 0.25 \pPb
interactions per bunch crossing. A procedure similar to that described in Ref.~\cite{Chatrchyan:2013nka}
is used for identifying and rejecting pileup events. It is based on the
number of tracks associated with each reconstructed vertex and the distance between multiple
vertices. The pileup in \PbPb data is negligible.

For track selections, the impact parameter significance of the track with
respect to the primary vertex in the direction along the beam axis and in the transverse plane, $d_z/\sigma(d_z)$ and $d_\mathrm{T}/\sigma(d_\mathrm{T})$,
are required to be less than 3. The relative uncertainty in \pt, $\sigma(\pt)/\pt$, must be less than 10\%.
Primary tracks, i.e., tracks that originate at the primary
vertex and satisfy the high-purity criteria of Ref.~\cite{TRK-11-001},
are used to define the event charged-particle multiplicity (\noff). To perform correlation measurements, each track is also required to leave
at least one hit in one of the three layers of the pixel tracker.
Only tracks with $\abs{\eta}<2.4$ and $\pt > 0.3\GeV$ are used in this analysis to ensure high tracking efficiency.

The \pPb and \PbPb data are compared in classes of \noff,
where primary tracks with $\abs{\eta}<2.4$ and $\pt >0.4$\GeV are counted. To compare
with results from other experiments, the \PbPb data are also analyzed based on centrality
classes for the 30--80\% centrality range.

\section{Analysis technique}
\label{sec:analysis}

The analysis technique of three-particle correlations
employed in this paper is based on that established in Ref.~\cite{Khachatryan:2016got},
with the extension of charge-dependent two-particle correlations,
higher-order harmonic three-particle correlations, and correlation studies
in different event shape classes (i.e., ESE analysis). The details
are outlined below.

\subsection{Calculations of two- and three-particle correlators}
\label{subsec:analysis_mixed}

Without directly reconstructing the event plane, the expression given in
Eq.~(\ref{2pcorrelatorEP}) can be alternatively evaluated using a three-particle
correlator with respect to a third particle~\cite{Abelev:2009ac,Abelev:2009ad},
$\left\langle \cos(\phi_{\alpha} + \phi_{\beta} - 2\phi_{c}) \right\rangle / v_{2,c}$,
where $v_{2,c}$ is the elliptic flow anisotropy of particle $c$ with inclusive charge sign.
The three-particle correlator is
measured via the scalar-product method of $Q$ vectors.
A complex $Q$ vector for each event is defined as $Q_n$ $\equiv$ $\sum_{i=1}^M w_{i} \re^{in\phi_i} / W$,
where $\phi_{i}$ is the azimuthal angle of particle $i$, $n$ is the Fourier harmonic order, $M$ is the number of particles in the $Q_n$ calculation in each event, $w_i$ is a weight assigned to each particle for efficiency correction, which is derived from a simulation using the \textsc{hijing} event generator~\cite{Gyulassy:1994ew}. The $W=\sum_{i=1}^M w_{i}$ represents the
weight of the $Q$ vector. In this way,
the three-particle correlator can be expressed in terms of the product
of $Q$ vectors, i.e., $Q_{1,\alpha}$ and $Q_{1,\beta}$, when particles $\alpha$ and $\beta$ are chosen from different detector phase-space regions or carry different charge signs,
\ifthenelse{\boolean{cms@external}}{
\begin{linenomath}
\begin{multline}
\gamma_{112} = \frac{\left\langle \cos(\phi_{\alpha} + \phi_{\beta} - 2\phi_{c})\right\rangle}{v_{2,c}} \\
= \frac{\left\langle Q_{1,\alpha}Q_{1,\beta}Q^*_{2,{\mathrm{HF}\pm}} \right\rangle}{\sqrt{\frac{\left\langle Q_{2,{\mathrm{HF}\pm}} Q^*_{2,{\mathrm{HF}\mp}}\right\rangle \left\langle Q_{2,{\mathrm{HF}\pm}} Q^*_{2,\text{trk}}\right\rangle}{\left\langle Q_{2,{\mathrm{HF}\mp}} Q^*_{2,\text{trk}}\right\rangle} }},
\label{3pcorrelatorQVector}
\end{multline}
\end{linenomath}
}{
\begin{linenomath}
\begin{equation}
\gamma_{112} = \frac{\left\langle \cos(\phi_{\alpha} + \phi_{\beta} - 2\phi_{c})\right\rangle}{v_{2,c}} \\
= \frac{\left\langle Q_{1,\alpha}Q_{1,\beta}Q^*_{2,{\mathrm{HF}\pm}} \right\rangle}{\sqrt{\frac{\left\langle Q_{2,{\mathrm{HF}\pm}} Q^*_{2,{\mathrm{HF}\mp}}\right\rangle \left\langle Q_{2,{\mathrm{HF}\pm}} Q^*_{2,\text{trk}}\right\rangle}{\left\langle Q_{2,{\mathrm{HF}\mp}} Q^*_{2,\text{trk}}\right\rangle} }},
\label{3pcorrelatorQVector}
\end{equation}
\end{linenomath}
}
where the angle brackets on the right-hand side
denote an event average of
the $Q$-vector products, weighted by the product of their respective total weights $W$. Here $Q_{2,\text{trk}}$ is the charge inclusive $Q_2$ vector of all particles in the tracker region, and $Q_{2,\text{HF}\pm}$ denotes the $Q_2$-vector for particles $c$ detected in the HF towers.
When particles $\alpha$ and $\beta$ are of the same sign and share the same phase space region (denoted as $\alpha = \beta$), an extra term is needed to remove the contribution of a particle pairing with itself, so evaluation of the three-particle correlator is modified as
\ifthenelse{\boolean{cms@external}}{
\begin{linenomath}
\begin{multline}
\gamma_{112} = \frac{\left\langle \cos(\phi_{\alpha} + \phi_{\beta} - 2\phi_{c})\right\rangle}{v_{2,c}} \\
= \frac{\left\langle Q_{112}Q^*_{2,{\mathrm{HF}\pm}} \right\rangle}{\sqrt{\frac{\left\langle Q_{2,{\mathrm{HF}\pm}} Q^*_{2,{\mathrm{HF}\mp}}\right\rangle \left\langle Q_{2,{\mathrm{HF}\pm}} Q^*_{2,\text{trk}}\right\rangle}{\left\langle Q_{2,{\mathrm{HF}\mp}} Q^*_{2,\text{trk}}\right\rangle} }},
\label{3pcorrelatorQVector_overlap}
\end{multline}
\end{linenomath}
}{
\begin{linenomath}
\begin{equation}
\gamma_{112} = \frac{\left\langle \cos(\phi_{\alpha} + \phi_{\beta} - 2\phi_{c})\right\rangle}{v_{2,c}} \\
= \frac{\left\langle Q_{112}Q^*_{2,{\mathrm{HF}\pm}} \right\rangle}{\sqrt{\frac{\left\langle Q_{2,{\mathrm{HF}\pm}} Q^*_{2,{\mathrm{HF}\mp}}\right\rangle \left\langle Q_{2,{\mathrm{HF}\pm}} Q^*_{2,\text{trk}}\right\rangle}{\left\langle Q_{2,{\mathrm{HF}\mp}} Q^*_{2,\text{trk}}\right\rangle} }},
\label{3pcorrelatorQVector_overlap}
\end{equation}
\end{linenomath}
}
where the $Q_{112}$ is defined as,
\ifthenelse{\boolean{cms@external}}{
\begin{linenomath}
\begin{multline}
Q_{112} \equiv \frac{\left(\sum \limits_{i=1} w_{i} \re^{i\phi_i}\right)^{2}-\sum \limits_{i=1} w^{2}_{i} \re^{i2\phi_i}}{\left(\sum \limits_{i=1} w_{i}\right)^{2}-\sum \limits_{i=1} w^{2}_{i}},
\label{3pcorrelatorQVector_overlap_1}
\end{multline}
\end{linenomath}
}{
\begin{linenomath}
\begin{equation}
Q_{112} \equiv \frac{\left(\sum \limits_{i=1} w_{i} \re^{i\phi_i}\right)^{2}-\sum \limits_{i=1} w^{2}_{i} \re^{i2\phi_i}}{\left(\sum \limits_{i=1} w_{i}\right)^{2}-\sum \limits_{i=1} w^{2}_{i}},
\label{3pcorrelatorQVector_overlap_1}
\end{equation}
\end{linenomath}
}
and the denominator of Eq.~(\ref{3pcorrelatorQVector_overlap_1}) is the respective event weight associated with $Q_{112}$.

In the numerators of Eqs.~(\ref{3pcorrelatorQVector}) and (\ref{3pcorrelatorQVector_overlap}),
the particles $\alpha$ and $\beta$ are identified in the tracker, with
$\abs{\eta} < 2.4$ and $0.3<\pt<3$\GeV, and are assigned a weight factor $w_{i}$
to correct for tracking inefficiency.
The particle $c$ is selected by using the tower energies and positions in the HF calorimeters
with $4.4 < \abs{\eta} < 5.0$. This choice of $\eta$ range for the HF towers
imposes an $\eta$ gap of at least 2 units with respect to particles $\alpha$
and $\beta$ from the tracker, to minimize possible short-range correlations.
To account for any occupancy effect of the HF detectors resulting from the large
granularities in $\eta$ and $\phi$, each tower is assigned a weight factor $w_{i}$ corresponding to
its \ET value when calculating the $Q$ vector. The denominator of the right-hand side of Eqs.~(\ref{3pcorrelatorQVector}) and (\ref{3pcorrelatorQVector_overlap})
corresponds to the $v_{2,c}$ using the scalar-product method~\cite{Abelev:2009ac,Abelev:2009ad},
with $Q_{2,\text{trk}}$ and $Q_{2,{\mathrm{HF}\pm}}$ denoting $Q_2$
vectors obtained from the tracker and the two HF detectors (positive and negative $\eta$ side) with the same kinematic requirements
as for the numerator.
The three-particle correlator is evaluated for particles $\alpha$ and $\beta$
carrying the same sign (SS) and opposite sign (OS). The SS combinations, ($+,+$) and ($-,-$), give consistent results and are therefore combined.
For \pPb collisions, the three-particle correlator is also measured with particle $c$
from HF$+$ and HF$-$, corresponding to the \Pp- and \Pb-going direction, respectively.
For symmetric \PbPb collisions, the results from HF$+$ and HF$-$ are consistent with each other and thus combined.

The higher-order harmonic three-particle correlator, $\gamma_{123}$,
defined in Eq.~(\ref{eq:gamma123}), is evaluated in exactly the same way as the
$\gamma_{112}$ correlator as follows when particles $\alpha$ and $\beta$ do not overlap,
\ifthenelse{\boolean{cms@external}}{
\begin{linenomath}
\begin{multline}
\gamma_{123} = \frac{\left\langle \cos(\phi_{\alpha} + 2\phi_{\beta} - 3\phi_{c})\right\rangle}{v_{3,c}} \\
= \frac{\left\langle Q_{1,\alpha}Q_{2,\beta}Q^*_{3,{\mathrm{HF}\pm}} \right\rangle}{\sqrt{\frac{\left\langle Q_{3,{\mathrm{HF}\pm}} Q^*_{3,{\mathrm{HF}\mp}}\right\rangle \left\langle Q_{3,{\mathrm{HF}\pm}} Q^*_{3,\text{trk}}\right\rangle}{\left\langle Q_{3,{\mathrm{HF}\mp}} Q^*_{3,\text{trk}}\right\rangle} }},
\label{3pcorrelatorQVector_123}
\end{multline}
\end{linenomath}
}{
\begin{linenomath}
\begin{equation}
\gamma_{123} = \frac{\left\langle \cos(\phi_{\alpha} + 2\phi_{\beta} - 3\phi_{c})\right\rangle}{v_{3,c}} \\
= \frac{\left\langle Q_{1,\alpha}Q_{2,\beta}Q^*_{3,{\mathrm{HF}\pm}} \right\rangle}{\sqrt{\frac{\left\langle Q_{3,{\mathrm{HF}\pm}} Q^*_{3,{\mathrm{HF}\mp}}\right\rangle \left\langle Q_{3,{\mathrm{HF}\pm}} Q^*_{3,\text{trk}}\right\rangle}{\left\langle Q_{3,{\mathrm{HF}\mp}} Q^*_{3,\text{trk}}\right\rangle} }},
\label{3pcorrelatorQVector_123}
\end{equation}
\end{linenomath}
}
with higher-order $Q$ vectors for particles $\alpha$ and $\beta$ of SS and OS. Similarly to Eq.~(\ref{3pcorrelatorQVector_overlap})
when particles $\alpha$ and $\beta$ can overlap, the $\gamma_{123}$ can be evaluated via
\ifthenelse{\boolean{cms@external}}{
\begin{linenomath}
\begin{multline}
\gamma_{123} = \frac{\left\langle \cos(\phi_{\alpha} + 2\phi_{\beta} - 3\phi_{c})\right\rangle}{v_{3,c}} \\
= \frac{\left\langle Q_{123}Q^*_{3,{\mathrm{HF}\pm}} \right\rangle}{\sqrt{\frac{\left\langle Q_{3,{\mathrm{HF}\pm}} Q^*_{3,{\mathrm{HF}\mp}}\right\rangle \left\langle Q_{3,{\mathrm{HF}\pm}} Q^*_{3,\text{trk}}\right\rangle}{\left\langle Q_{3,{\mathrm{HF}\mp}} Q^*_{3,\text{trk}}\right\rangle} }},
\label{3pcorrelatorQVector_overlap_123}
\end{multline}
\end{linenomath}
}{
\begin{linenomath}
\begin{equation}
\gamma_{123} = \frac{\left\langle \cos(\phi_{\alpha} + 2\phi_{\beta} - 3\phi_{c})\right\rangle}{v_{3,c}} \\
= \frac{\left\langle Q_{123}Q^*_{3,{\mathrm{HF}\pm}} \right\rangle}{\sqrt{\frac{\left\langle Q_{3,{\mathrm{HF}\pm}} Q^*_{3,{\mathrm{HF}\mp}}\right\rangle \left\langle Q_{3,{\mathrm{HF}\pm}} Q^*_{3,\text{trk}}\right\rangle}{\left\langle Q_{3,{\mathrm{HF}\mp}} Q^*_{3,\text{trk}}\right\rangle} }},
\label{3pcorrelatorQVector_overlap_123}
\end{equation}
\end{linenomath}
}
where $Q_{123}$ is defined as
\begin{linenomath}
\begin{equation}
Q_{123} \equiv \frac{\left(\sum \limits_{i=1} w_{i} \re^{i\phi_i}\sum \limits_{i=1} w_{i} \re^{i2\phi_i} \right)-\sum \limits_{i=1} w^{2}_{i} \re^{i3\phi_i}}{\left(\sum \limits_{i=1} w_{i}\right)^{2}-\sum \limits_{i=1} w^{2}_{i}},
\label{3pcorrelatorQVector_overlap_123_1}
\end{equation}
\end{linenomath}
and the respective event weight associated with $Q_{123}$ is the denominator of Eq.~(\ref{3pcorrelatorQVector_overlap_123_1}).

Similarly, the charge-dependent two-particle
correlator, $\delta \equiv \left\langle \cos(\phi_{\alpha}-\phi_{\beta}) \right\rangle$,
is also evaluated with $Q$ vectors as $\delta = \left\langle Q_{1,\alpha}Q^*_{1,\beta} \right\rangle$
when particles $\alpha$ and $\beta$ are chosen from different detector phase-space regions
or have opposite signs, or otherwise,
\begin{linenomath}
\begin{equation}
\delta = \left\langle \frac{\left(\sum \limits_{i=1} w_{i} \re^{i\phi_i}\sum \limits_{i=1} w_{i} \re^{-i\phi_i} \right)-\sum \limits_{i=1} w^{2}_{i}}{\left(\sum \limits_{i=1} w_{i}\right)^{2}-\sum \limits_{i=1} w^{2}_{i}} \right\rangle,
\label{2pcorrelatorQVector_overlap_1}
\end{equation}
\end{linenomath}
and the respective event weight is the denominator of Eq.~(\ref{2pcorrelatorQVector_overlap_1}).

The effect of the nonuniform detector acceptance is corrected by evaluating
the cumulants of $Q$-vector products~\cite{Selyuzhenkov:2007zi}. While the correction
is found to be negligible for the $\gamma_{112}$ and $\delta$ correlators,
there is a sizable effect of 5--10\% correction to the $\gamma_{123}$ correlator.

\subsection{Event shape engineering}
\label{subsec:analysis_ese}

In the ESE analysis, within each
multiplicity range of \pPb or centrality range of \PbPb data, events are
divided into different $q_{2}$ classes, where $q_2$ is defined
as the magnitude of the $Q_{2}$ vector. In this analysis, the
$q_2$ value is calculated from one side of the HF region
within the range $3 < \eta < 5$ for both \pPb and \PbPb collisions (weighted by the tower \ET), where in \pPb collisions only the \Pb-going side of HF is used because of the poor resolution from a relatively low charged-particle multiplicity on the proton-going side. In each
$q_2$ class, the $v_2$ harmonic is measured with the scalar product method using a common resolution term ($v_{2,c}$) as in the $\gamma_{112}$ correlator. Therefore, the $v_{2}$ from the tracker region can be expressed in terms of the Q-vectors as
\begin{linenomath}
\begin{equation}
v_{2} = \frac{\left\langle Q_{2,\alpha}Q^{*}_{2,{\mathrm{HF}\pm}} \right\rangle}{\sqrt{\frac{\left\langle Q_{2,{\mathrm{HF}\pm}} Q^*_{2,{\mathrm{HF}\mp}}\right\rangle \left\langle Q_{2,{\mathrm{HF}\pm}} Q^*_{2,\text{trk}}\right\rangle}{\left\langle Q_{2,{\mathrm{HF}\mp}} Q^*_{2,\text{trk}}\right\rangle} }},
\label{v2_SP}
\end{equation}
\end{linenomath}
where particles from the HF are selected from the same region as particle $c$ in the $\gamma_{112}$ correlator.

In \PbPb collisions, the particle $c$ in the $\gamma_{112}$ correlator is taken
from the HF detector that is at the opposite $\eta$ side to the one used to
calculate $q_2$. However, the results are in good agreement with those
where the particle $c$ for $\gamma_{112}$ and $q_2$ is measured from the same side of the HF detector, which can be found in Appendix~\ref{app:b}. In \pPb collisions, the particle $c$ in the
$\gamma_{112}$ correlator with respect to the \Pb- and \Pp-going sides is
studied, when $q_2$ is measured only in the \Pb-going side. The results are
found to be independent of the side in which the particle $c$ is detected.

\begin{figure}[thb]
\centering
\includegraphics[width=0.48\textwidth]{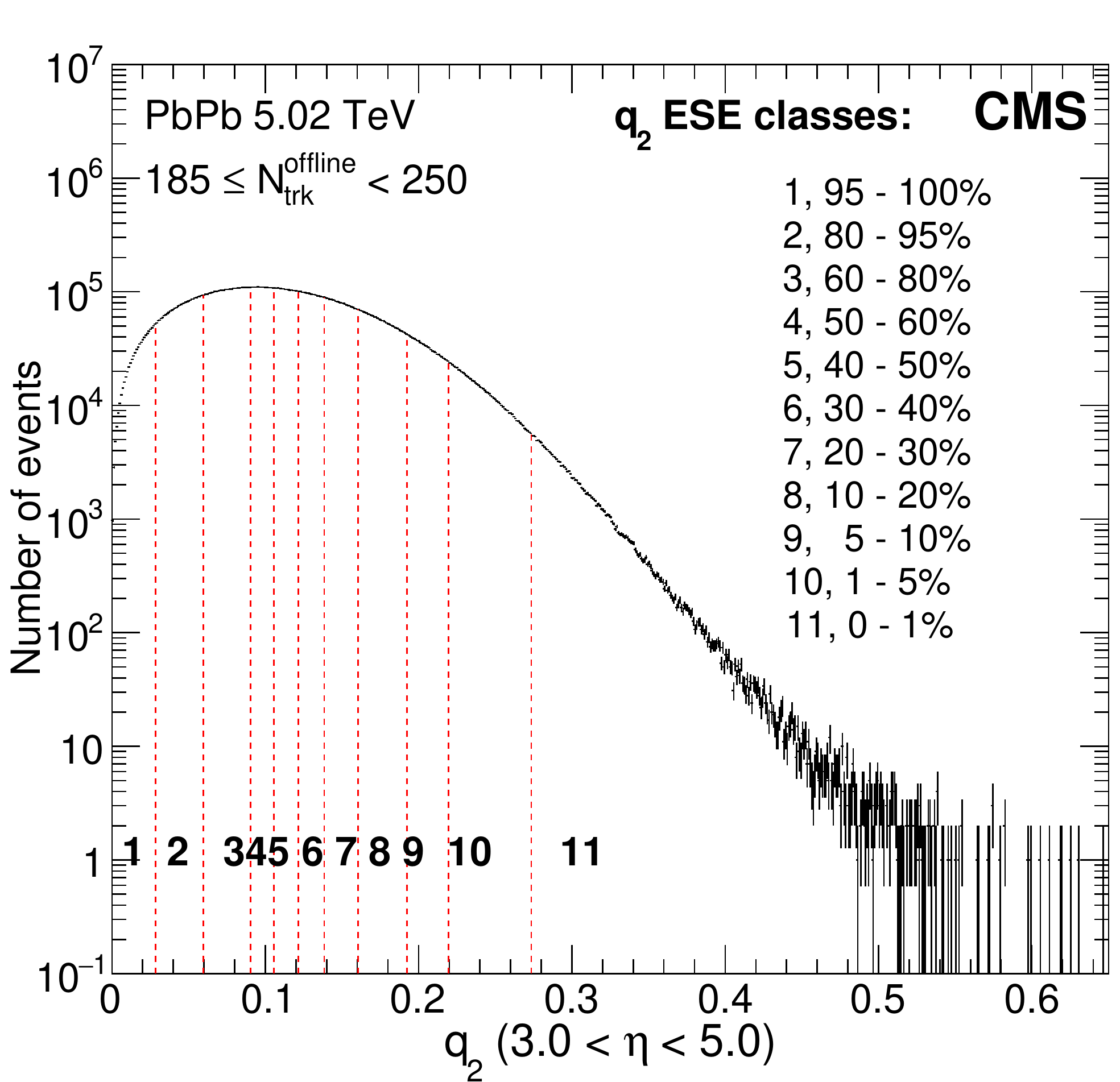}
\includegraphics[width=0.48\textwidth]{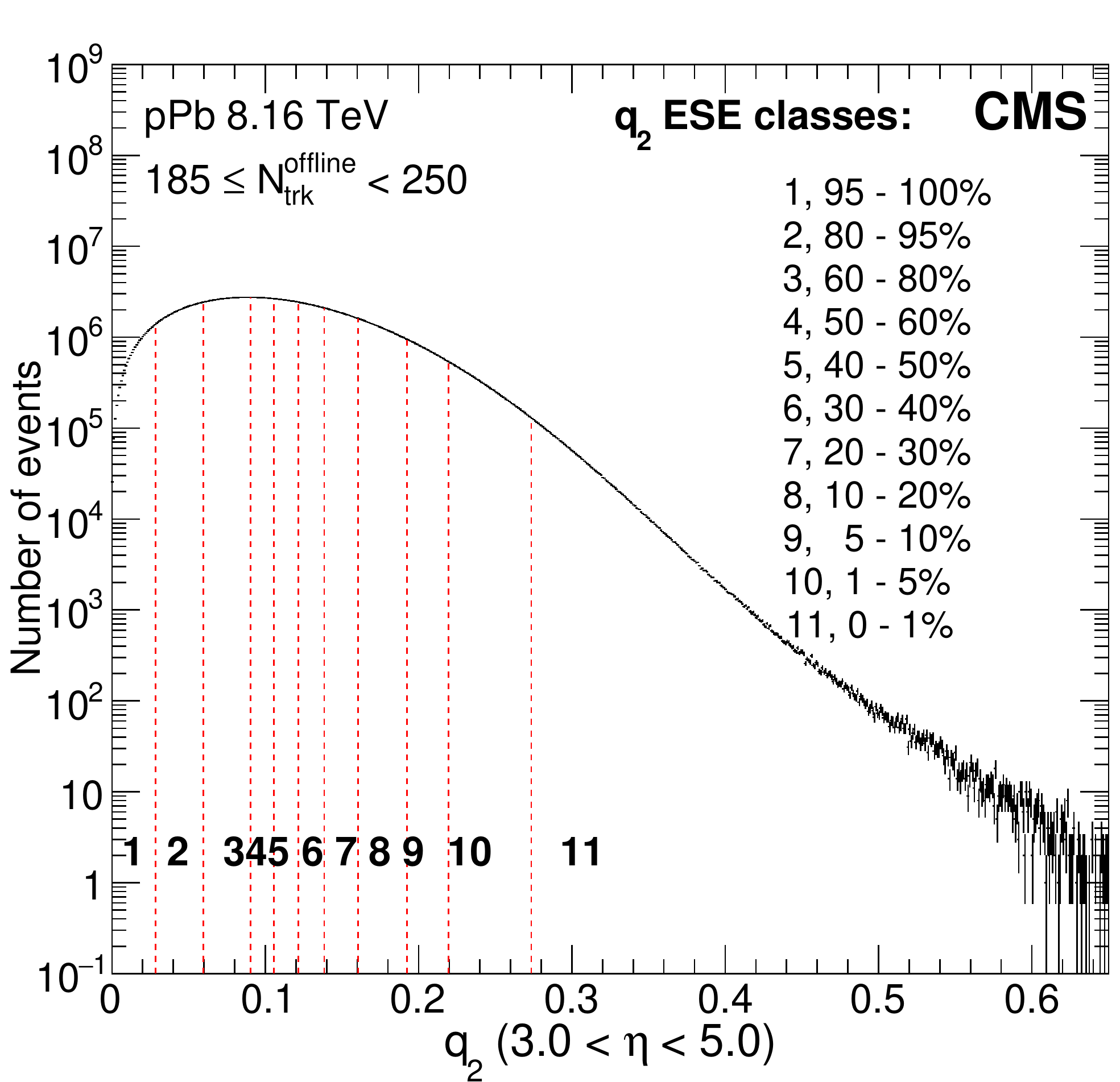}
\caption{The $q_{2}$ classes are shown in different fractions with respect to the total number of events in multiplicity range $185 \leq \noff < 250$ in \PbPb (\cmsLeft) and \pPb (\cmsRight) collisions at $\rootsNN = 5.02$ and 8.16\TeV, respectively.}
\label{fig:Q2_bins}
\end{figure}

In Fig.~\ref{fig:Q2_bins}, the HF $q_2$ distributions are shown
for \PbPb and \pPb collisions in the multiplicity range $185 \leq \noff < 250$, where most of the high-multiplicity \pPb events were recorded by the high-multiplicity trigger in this range.
As indicated by the vertical dashed lines, the distribution is divided into several intervals with each corresponding to a fraction of the full distribution,
where 0--1\% represents the highest $q_{2}$ class. For each $q_{2}$ class,
the three-particle $\gamma_{112}$ is calculated with the default kinematic
regions for particles $\alpha,\beta$, and $c$, and the $v_{2}$
harmonics from the tracker ($\abs{\eta} < 2.4$) are also obtained by the
scalar-product method~\cite{Bilandzic:2013kga}. The \pPb and \PbPb results are presented in Section~\ref{sec:results} for
both SS and OS pairs, as well as
the differences found for the two charge combinations.

\begin{figure}[thb]
\centering
\includegraphics[width=\cmsFigWidth]{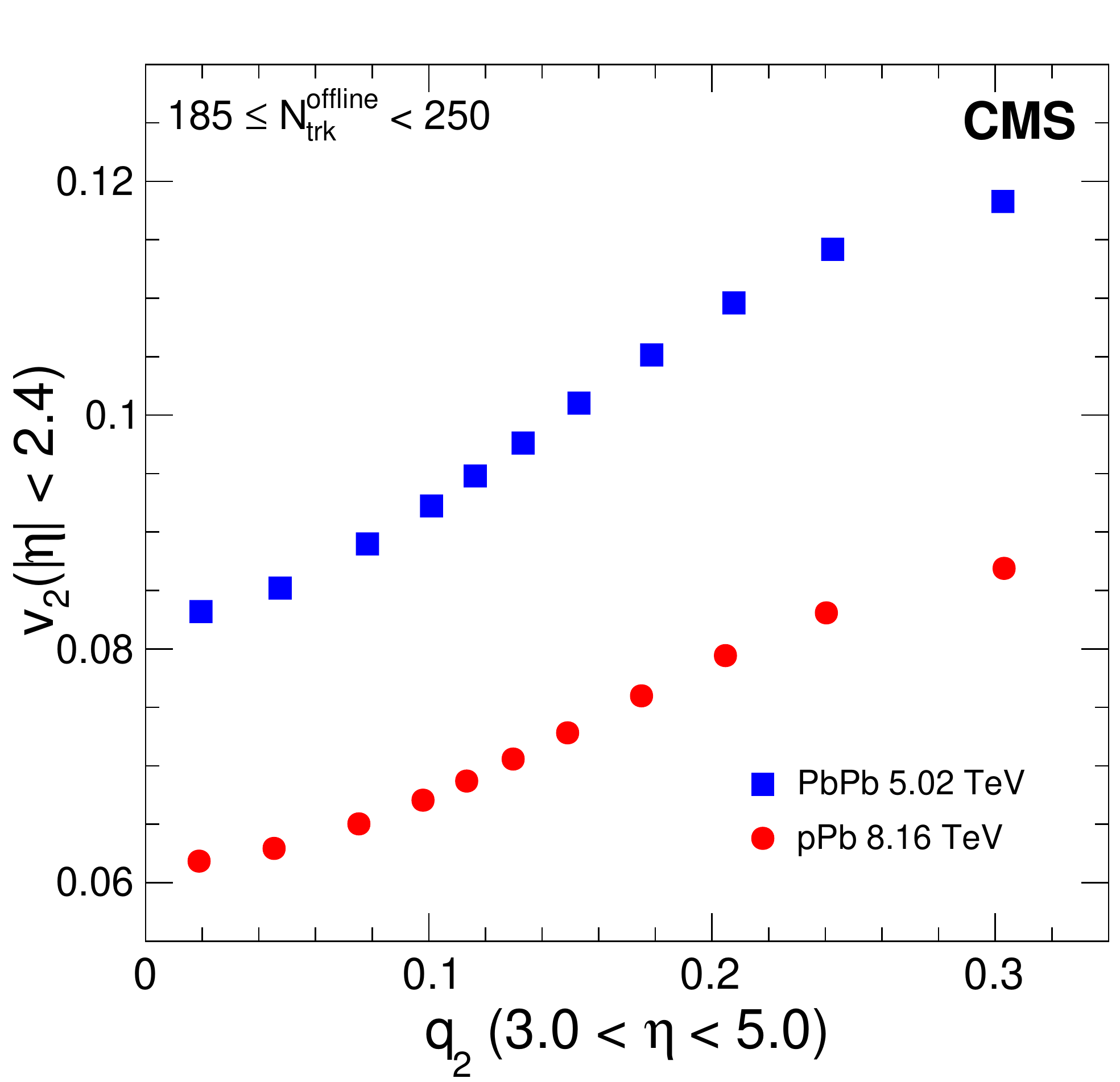}
\caption{The correlation between the tracker $v_{2}$ and the HF $q_{2}$
is shown for \pPb and \PbPb collisions at collisions at $\rootsNN = 8.16$ and 5.02\TeV, respectively.}
\label{fig:Q2vsV2}
\end{figure}

In Fig.~\ref{fig:Q2vsV2}, the $v_2$ values for tracker particles
as a function of the average $q_2$ in each HF $q_2$ class are shown.
A proportionality close to linear is seen, indicating the two quantities
are strongly correlated because of the initial-state geometry~\cite{Adam:2015eta}.

\subsection{Systematic uncertainties}
\label{subsec:systematics}
The absolute systematic uncertainties of the two-particle correlator $\delta$, and three-particle correlators $\gamma_{112}$ and $\gamma_{123}$, have been studied. Varying
the $d_z/\sigma(d_z)$ and $d_\mathrm{T}/\sigma(d_\mathrm{T})$ from less than 3
(default) to less than 2 and 5, and the $\sigma(\pt)/\pt < 10\%$ (default) to
$\sigma(\pt)/\pt < 5\%$, together yield the systematic uncertainties of ${\pm}1.0\times 10^{-5}$ for the $\gamma_{112}$, ${\pm}4.0\times 10^{-5}$ for the $\gamma_{123}$, and ${\pm}1.0\times 10^{-4}$ for the $\delta$ correlator.
The longitudinal primary vertex position ($V_z$) has been varied, using ranges
$\abs{V_{z}} < 3$\unit{cm} and $3 < \abs{V_z} < 15$\unit{cm}, where the differences with respect to the default range $\abs{V_z} < 15$\unit{cm} are ${\pm}1.0\times 10^{-5}$ for the $\gamma_{112}$, ${\pm}3.0\times 10^{-5}$ for the $\gamma_{123}$, and ${\pm}1.0\times 10^{-4}$ for the $\delta$ correlator, taken as the systematic uncertainty.
In the \pPb collisions only, using the lower-threshold of the high-multiplicity trigger with respect to the default trigger, yields a systematic uncertainty of $\pm 3.0\times 10^{-5}$ for all three correlators, which accounts for the possible trigger bias from the inefficiency of the default trigger around the threshold. In the \pPb data sample, the average pileup can be as high as 0.25 and therefore the systematic effects from pileup have been evaluated. The full sample has been split into 4 different sets of events with different average pileup, according to their instantaneous luminosity during each run. The systematic effects for $\gamma_{112}$ and $\delta$ have been found to be ${\pm}1.0\times 10^{-5}$, and for $\gamma_{123}$ is to be ${\pm}3.0\times 10^{-5}$.

A final test of the analysis procedures is
done by comparing ``known" charge-dependent signals based on the \textsc{epos}
event generator~\cite{Pierog:2013ria} to those found after events are passed through a {\sc \GEANTfour}~\cite{GEANT4,Allison:2006ve}
simulation of the CMS detector response. Based on this test, a systematic uncertainty of
${\pm}2.5\times 10^{-5}$ is assigned for the $\gamma_{112}$, ${\pm}4.0\times 10^{-5}$ for the $\gamma_{123}$, and ${\pm}5.0\times 10^{-4}$ for the $\delta$ correlators, by taking the difference in the correlators between the reconstructed and the generated level. Note that this uncertainty for the $\delta$ correlator is based on differential variables, where the uncertainty covers the maximum deviation from the closure test. For results that averaged over $\abs{\Delta\eta} < 1.6$, the systematic uncertainty is found to be ${\pm}2.0\times 10^{-4}$ when directly evaluating the average. The tracking efficiency and acceptance of positively and negatively charged particles have been evaluated separately, and the difference has been found to be negligible.
All sources of systematic uncertainty are uncorrelated and added in quadrature to obtain
the total absolute systematic uncertainty. No dependence of the systematic uncertainties on the
sign combination, multiplicity, $\deta$, $\Delta\pt$, or average-\pt is found. The systematic uncertainties in our results are point-to-point correlated. In \pPb collisions, the systematic uncertainty is also observed to be independent of particle $c$ pointing to the \Pb- or \Pp-going direction, and thus it is quoted to be the same for these two situations. The systematic uncertainties are summarized in Table~\ref{tab:syst-table}.

\begin{table*}[ht]
\topcaption{\label{tab:syst-table} Summary of systematic uncertainties in SS and OS three-particle correlators $\gamma_{112}$ and $\gamma_{123}$ , and two-particle correlator $\delta$ in \pPb collisions at $\rootsNN = 8.16\TeV$ and \PbPb collisions at 5.02\TeV.}
\centering
\begin{scotch}{lccc}
 Source	&	   $\gamma_{112}$ ($\times10^{-5}$) & $\gamma_{123}$ ($\times10^{-5}$) & $\delta$ ($\times10^{-4}$)  \\
\hline
 Track selections	& 1.0 &  4.0 & 1.0 \\
 Vertex Z position	& 1.0 &	3.0 & 1.0 \\
 Pileup (\pPb only) & 1.0 &  3.0 & 0.1 \\
 High multiplicity trigger bias (\pPb only) & 3.0	& 3.0  & 0.3 \\
 MC closure	&	2.5 	& 4.0 & 5.0\\
\hline
 Total in \pPb &	4.3 &  7.7 & 5.2 \\
 Total in \PbPb &  2.9 &  6.4 & 5.2 \\
\end{scotch}
\end{table*}

\section{Results}
\label{sec:results}

\subsection{Charge-dependent two- and three-particle correlators}
\label{subsec:results_mixed}

\begin{figure}[thb]
\centering
  \includegraphics[width=\cmsFigWidth]{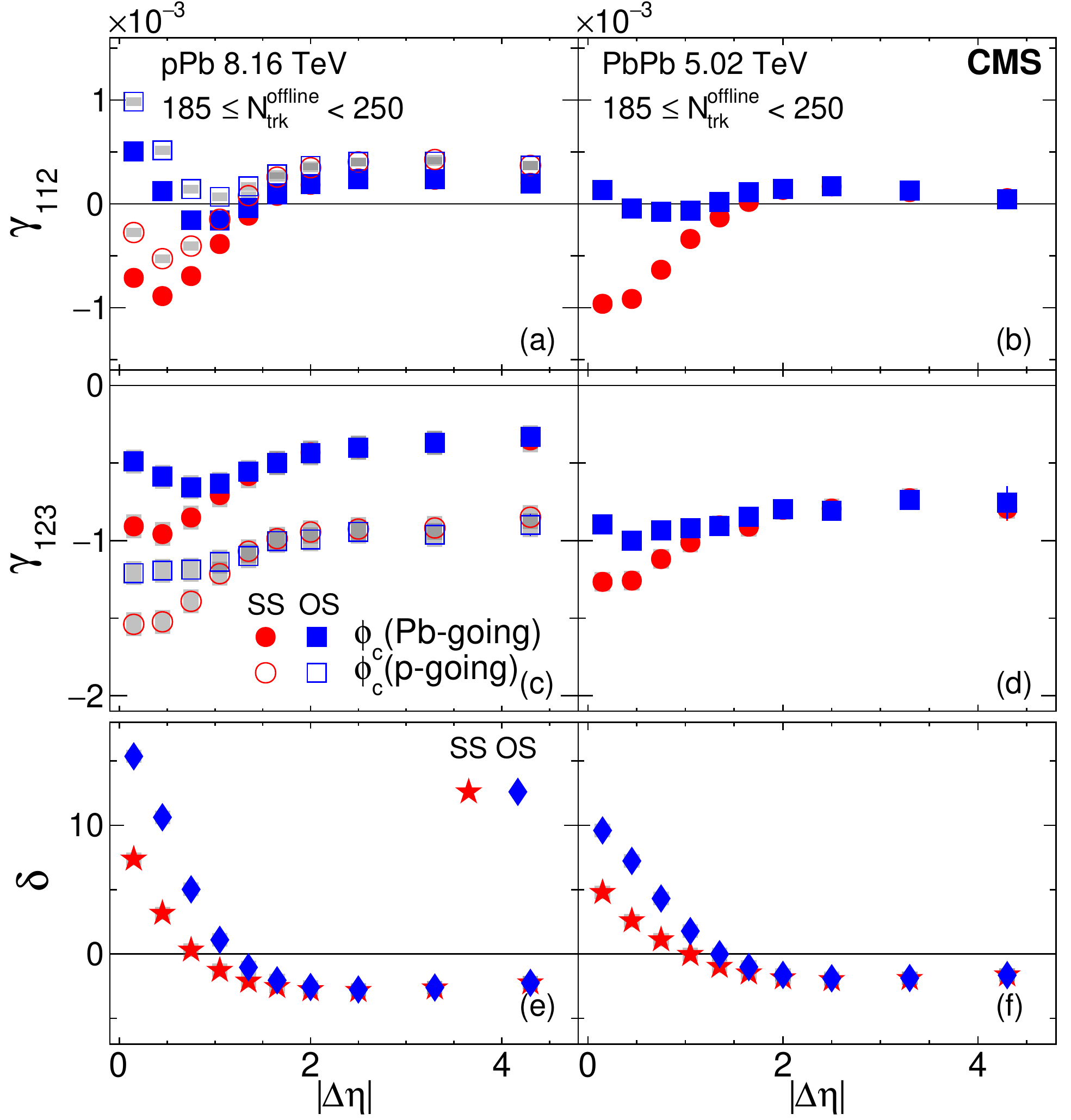}
  \caption{ \label{fig1_a}
The SS and OS three-particle correlators,
$\gamma_{112}$ (upper)  and $\gamma_{123}$ (middle), and two-particle
correlator, $\delta$ (lower), as a function of $\abs{\deta}$
for $185 \leq \noff < 250$ in \pPb collisions at $\rootsNN = 8.16$\TeV~(left) and
\PbPb collisions at 5.02\TeV~(right). The \pPb results obtained with particle $c$
in \Pb-going (solid markers) and \Pp-going (open markers) sides are
shown separately. The SS and OS two-particle correlators are denoted by different markers for both \pPb and \PbPb collisions.
Statistical and systematic uncertainties
are indicated by the error bars and shaded regions, respectively.
   }
\end{figure}

Measurements of the charge-dependent three-particle ($\gamma_{112}$, $\gamma_{123}$)
and two-particle ($\delta$) correlators are shown in Fig.~\ref{fig1_a}
as functions of the pseudorapidity difference ($\abs{\deta} \equiv \abs{\eta_{\alpha}-\eta_{\beta}}$)
between SS and OS particles $\alpha$ and $\beta$, in the multiplicity range $185 \leq \noff < 250$
for \pPb collisions at $\rootsNN = 8.16\TeV$ and \PbPb collisions at 5.02\TeV. The SS and OS of $\delta$ correlators are shown with different markers to differentiate the two-particle correlation from the three-particle correlation with a particle $c$ in the forward rapidity.
The \pPb data are obtained with particle $c$ in the \Pb- and \Pp-going
sides separately. The multiplicity range $185 \leq \noff < 250$ for \PbPb data roughly
corresponds to the centrality range 60--65\%.

\begin{figure}[thb]
\centering
  \includegraphics[width=\cmsFigWidth]{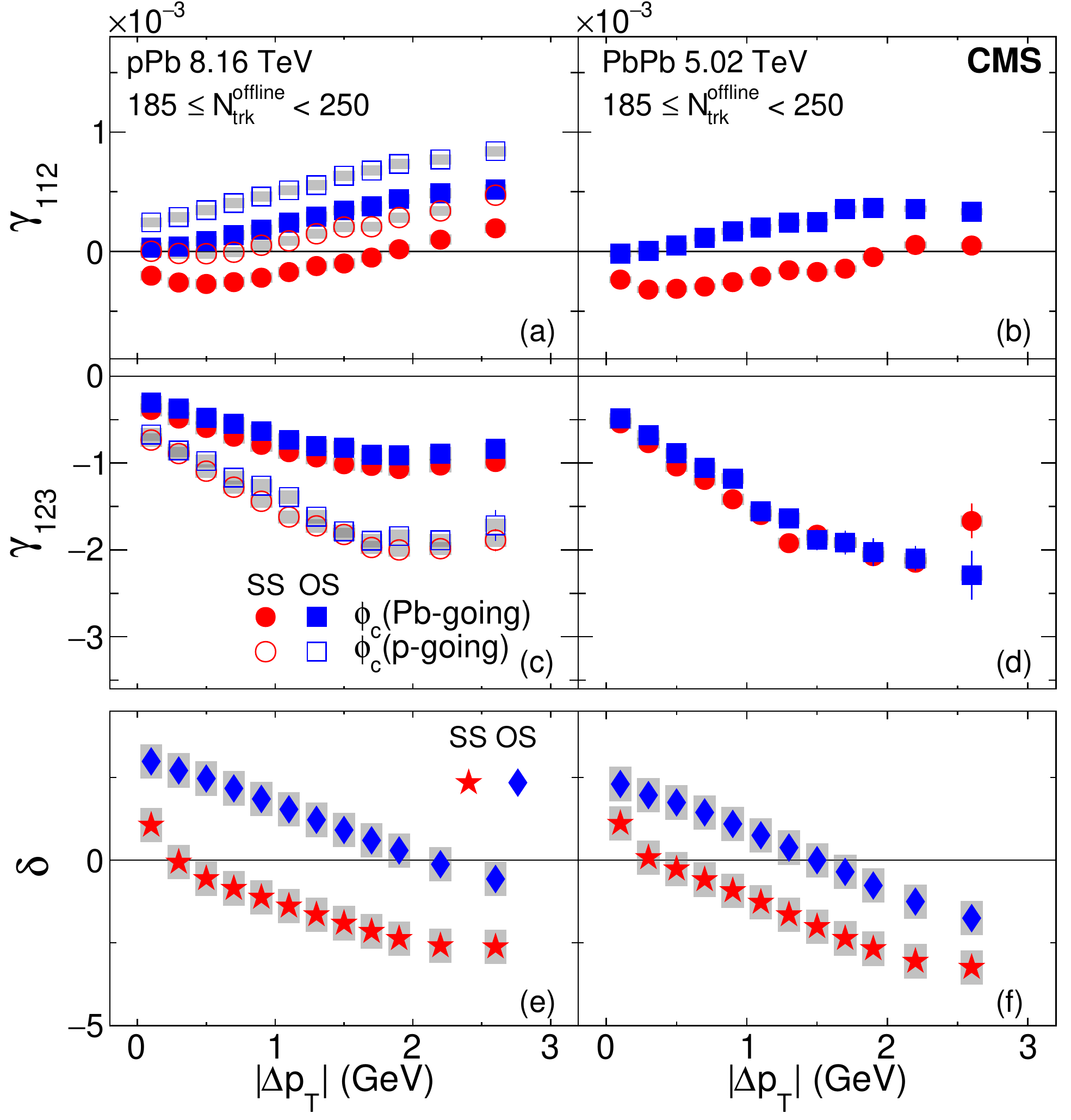}
  \caption{ \label{fig1_b}
The SS and OS three-particle correlators,
$\gamma_{112}$ (upper)  and $\gamma_{123}$ (middle), and two-particle
correlator, $\delta$ (lower), as a function of $\abs{\Delta\pt}$
for $185 \leq \noff < 250$ in \pPb collisions at $\rootsNN = 8.16$\TeV~(left)
and \PbPb collisions at 5.02\TeV~(right) collisions. The \pPb results
obtained with particle $c$ in \Pb-going (solid markers) and \Pp-going
(open markers) sides are shown separately. The SS and OS two-particle correlators are denoted by different markers for both \pPb and \PbPb collisions. Statistical and
systematic uncertainties are indicated by the error bars and
shaded regions, respectively.
   }
\end{figure}

Similar to the observation reported in Ref.~\cite{Khachatryan:2016got},
the three-particle $\gamma_{112}$ (Figs.~\ref{fig1_a}a and \ref{fig1_a}b) and $\gamma_{123}$
(Figs.~\ref{fig1_a}c and \ref{fig1_a}d) correlators show a charge dependence for $\abs{\deta}$
up to about 1.6, in both \pPb (5.02~\cite{Khachatryan:2016got} and 8.16\TeV) and \PbPb (5.02\TeV) systems.
Little collision energy dependence of the $\gamma_{112}$ data for \pPb collisions
is found from $\rootsNN = 5.02\TeV$ to 8.16\TeV within uncertainties
(as will be shown later in Figs.~\ref{fig3} and \ref{fig4} as a function of event multiplicity).
For $\abs{\deta} > 1.6$, the SS and OS correlators converge to a common value,
which is weakly dependent on $\abs{\deta}$ out to about 4.8 units.
In \pPb collisions, the $\gamma_{112}$ correlator obtained with
particle $c$ from the \Pp-going side is shifted toward more positive values than
that from the \Pb-going side by approximately the same amount for both the SS
and OS pairs. This trend is reversed for the higher-order harmonic $\gamma_{123}$
correlator, where the \Pb-going side data are more positive than the \Pp-going side data. The
\Pb-going side results for the $\gamma_{112}$ correlator for the \pPb collisions
are of similar magnitude as the results for \PbPb collisions, although a more pronounced
peak structure at small $\abs{\deta}$ is observed in \pPb collisions. The common shift of SS and OS correlators
between the \Pp- and \Pb-going side reference ($c$) particle may be related to
sources of correlation that are charge independent, such as directed flow (the first-order azimuthal anisotropy in Eq.~(\ref{azimuthal})) and
the momentum conservation effect, the latter being sensitive to the difference
in multiplicity between \Pp- and \Pb-going directions. The two-particle $\delta$
correlators (Figs.~\ref{fig1_a}e and \ref{fig1_a}f) for both SS and OS pairs
also show a decreasing trend as $\abs{\deta}$ increases and converge to the same
values at $\abs{\deta} \approx 1.6$, similar to that for the three-particle
correlators. The values of both OS and SS $\delta$ correlators are found to
be larger in \pPb than in \PbPb collisions at similar multiplicities.
As the $\delta$ correlator is sensitive to short-range jet-like correlations, reflected by the low-$\abs{\deta}$ region,
this effect may be related to the higher-\pt jets or clusters in \pPb compared to \PbPb\
collisions at similar multiplicities, as suggested in Ref.~\cite{Chatrchyan:2013nka},
because of short-range two-particle $\Delta\eta$--$\Delta\phi$ correlations.

\begin{figure}[thb]
\centering
  \includegraphics[width=\cmsFigWidth]{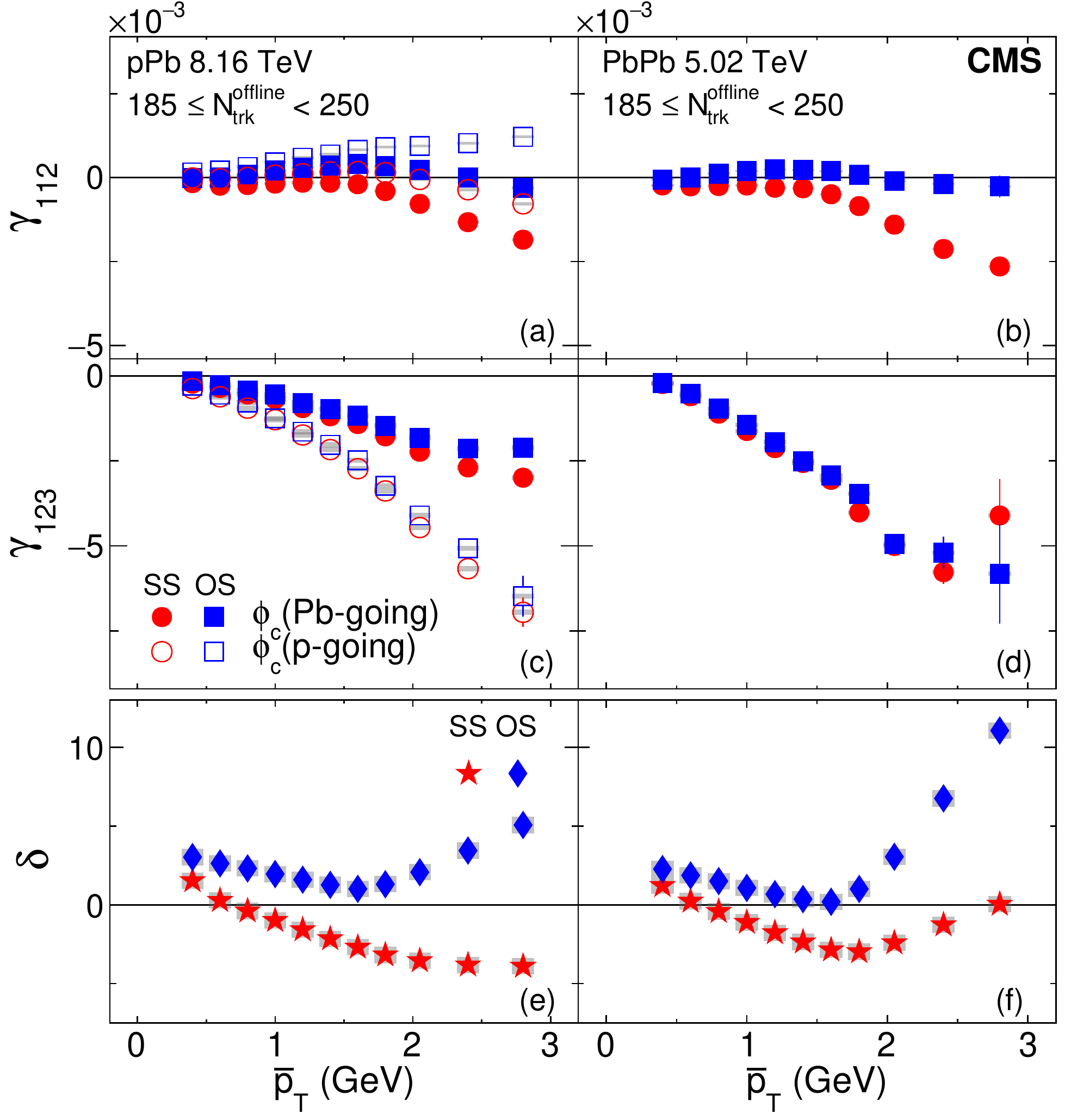}
  \caption{ \label{fig1_c}
The SS and OS three-particle correlators,
$\gamma_{112}$ (upper)  and $\gamma_{123}$ (middle), and two-particle
correlator, $\delta$ (lower), as a function of $\ptbar$
for $185 \leq \noff < 250$ in \pPb collisions at $\rootsNN = 8.16$\TeV~(left) and
\PbPb collisions at 5.02\TeV~(right). The \pPb results
obtained with particle $c$ in \Pb-going (solid markers) and \Pp-going (open markers)
sides are shown separately. The SS and OS two-particle correlators are denoted by different markers for both \pPb and \PbPb collisions. Statistical and systematic uncertainties
are indicated by the error bars and shaded regions, respectively.
   }
\end{figure}

To provide more detailed information on the particle \pt dependence of
the correlations, the $\gamma_{112}$, $\gamma_{123}$, and $\delta$ correlators
are measured as functions of the \pt difference ($\abs{\Delta\pt} \equiv \abs{\ptalpha-\ptbeta}$)
and average ($\ptbar \equiv (\ptalpha+\ptbeta)/2$) of the SS and OS pairs in \pPb and \PbPb collisions, and shown in Figs.~\ref{fig1_b} and \ref{fig1_c}.
The $|\Delta\pt|$- and \ptbar-dependent results are averaged over the full $|\eta|<2.4$ range.
In particular, the charge-dependent correlations from the CME are expected to be
strongest in the low-\pt region~\cite{Kharzeev:2007jp}.

For all correlators, similar behaviors between \pPb and \PbPb data are again observed.
The trends in $|\Delta\pt|$ for $\gamma_{112}$ and $\gamma_{123}$ correlators seem
to be opposite. The $\gamma_{112}$ correlator increases as a function of $|\Delta\pt|$,
while a decreasing trend is seen for the $\gamma_{123}$ correlator up to $|\Delta\pt|
\approx 2$\GeV, where $\gamma_{123}$ becomes constant in $|\Delta\pt|$.
The opposite behavior observed between the $\gamma_{112}$ and $\gamma_{123}$ correlators
is related to back-to-back jet-like correlations, which give a positive (negative) contribution
to even- (odd-) order Fourier harmonics~\cite{Chatrchyan:2012wg}.
The $\delta$ correlators decrease monotonically as functions of $|\Delta\pt|$ for both
SS and OS pairs in \pPb and \PbPb collisions. This trend of decreasing for $\delta$ is consistent with the expectation
from either transverse momentum conservation or back-to-back jet correlations~\cite{Bzdak:2010fd}.

In terms of the $\ptbar$ dependence in Fig.~\ref{fig1_c}, all three correlators
for both SS and OS pairs show very similar behaviors in the
low-$\ptbar$ region, which is likely a consequence of the same physical origin.
However, an opposite trend starts emerging at $\ptbar \approx 1.6$\GeV,
most evidently for $\gamma_{112}$ and $\delta$. Within the $0.3<\pt<3$\GeV range,
as $\ptbar$ increases toward 3\GeV, both particles of a pair tend to be
selected with a high-\pt value, while for low-$\ptbar$ or any $|\Delta\pt|$ values,
the pair usually consists of at least one low-\pt particle. This
may be the reason for a different trend seen at high $\ptbar$.
The qualitative behavior of the data is captured by the A Multi-Phase Transport model~\cite{Lin:2004en,Ma:2011uma}. In Appendix~\ref{app:c}, all three correlators as functions of $\abs{\deta}$, $\Delta\pt$, and $\ptbar$ in different multiplicity and centrality ranges in \pPb and \PbPb collisions, can be found.

\begin{figure}[thbp]
\centering
  \includegraphics[width=\cmsFigWidth]{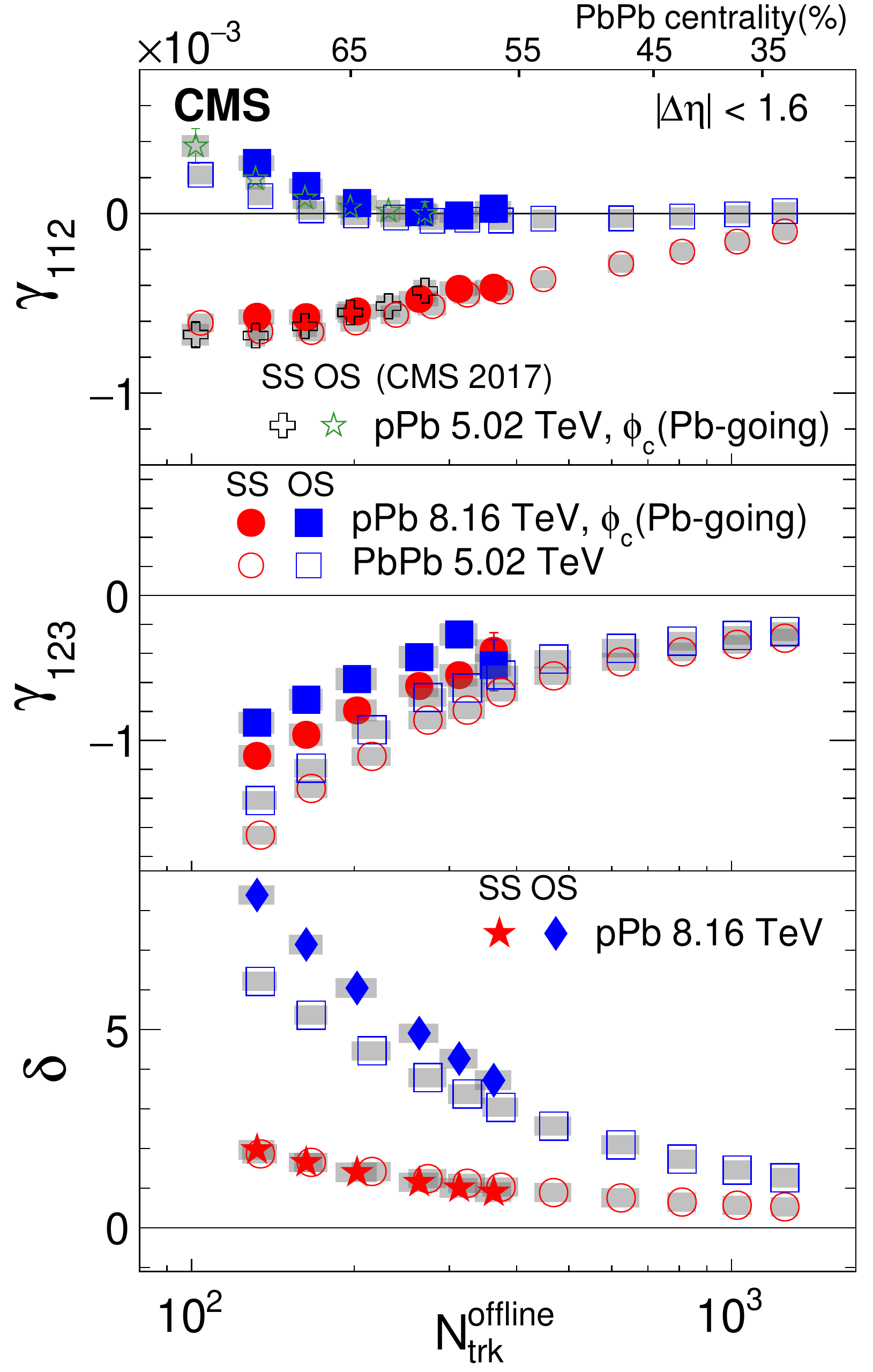}
  \caption{ \label{fig3}
The SS and OS three-particle correlators,
$\gamma_{112}$ (upper)  and $\gamma_{123}$ (middle), and
two-particle correlator, $\delta$ (lower), averaged over
$\abs{\deta}<1.6$ as a function of \noff in \pPb collisions at
$\rootsNN = 8.16$\TeV and \PbPb collisions at 5.02\TeV. The SS and OS two-particle correlators are denoted by different markers for \pPb collisions. The results of $\gamma_{112}$
for \pPb collisions at 5.02\TeV from CMS Collaboration (CMS 2017:~\cite{Khachatryan:2016got}), are also shown for comparison.
Statistical and systematic uncertainties are indicated by the error
bars and shaded regions, respectively.
   }
\end{figure}

\begin{figure*}[thb]
\centering
  \includegraphics[width=0.8\textwidth]{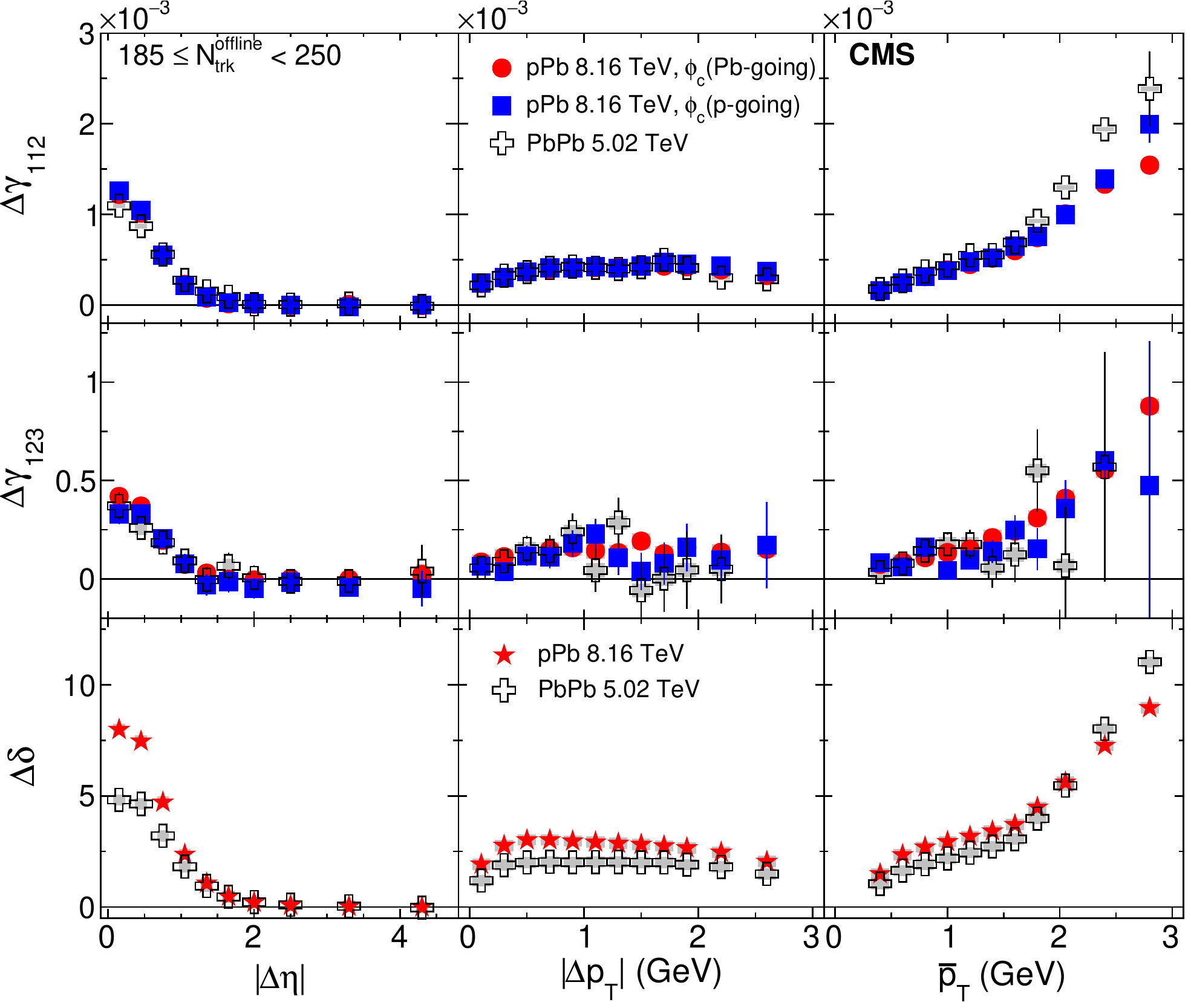}
  \caption{ \label{fig2}
The difference of the OS and SS three-particle correlators,
$\gamma_{112}$ (upper)  and $\gamma_{123}$ (middle), and two-particle correlator, $\delta$ (lower)
as functions of $\Delta\eta$~(left), $\Delta\pt$~(middle), and \ptbar~(right) for
$185 \leq \noff < 250$ in \pPb collisions at $\rootsNN = 8.16$\TeV
and \PbPb collisions at 5.02\TeV. The $\Delta\delta$ correlator is denoted by a different marker for \pPb collisions.
The \pPb results are obtained with particle $c$ from \Pb- and \Pp-going
sides separately. Statistical and systematic uncertainties
are indicated by the error bars and shaded regions, respectively.
   }
\end{figure*}

To explore the multiplicity or centrality dependence of the three-
and two-particle correlators, an average of the data is taken
over $\abs{\deta}<1.6$, corresponding to the region in Fig.~\ref{fig1_a} which exhibits charge dependence. The average over $\abs{\deta}<1.6$ is
weighted by the density of particle pairs in $\abs{\deta}$, and all further plots averaged over $\abs{\deta}<1.6$ are weighted similarly. The resulting
$\abs{\deta}$-averaged data of $\gamma_{112}$, $\gamma_{123}$ and $\delta$
are shown in Fig.~\ref{fig3} for both OS and SS pairs, as functions of
\noff for \pPb collisions at $\rootsNN = 8.16$\TeV (particle $c$ from the \Pb-going
side) and \PbPb collisions at 5.02\TeV. Previously published \pPb\
data at 5.02\TeV are also shown for comparison~\cite{Khachatryan:2016got}.
The centrality scale on the top of Fig.~\ref{fig3} relates to the \PbPb
experimental results. Up to $\noff = 400$, the \pPb and \PbPb results are measured
in the same \noff ranges. The new \pPb data at 8.16\TeV extend the
multiplicity reach further than the previously published \pPb data at 5.02\TeV (which stopped at $\noff \approx 300$).

Within the uncertainties, the SS and OS $\gamma_{112}$ correlators in \pPb and \PbPb
collisions exhibit the same magnitude and trend as functions of event multiplicity.
The \pPb data are independent of collision energy from 5.02 to 8.16\TeV at similar multiplicities.
This justifies the comparison of new pPb data and PbPb data at somewhat different energies.
For both \pPb and \PbPb collisions, the OS correlator reaches a value close to zero for $\noff > 200$, while the
SS correlator remains negative, but the magnitude gradually decreases as
\noff increases. Part of the observed multiplicity (or centrality) dependence
is understood as a dilution effect that falls with the inverse of event multiplicity~\cite{Abelev:2009ad}.
The notably similar magnitude and multiplicity dependence
of the three-particle correlator, $\gamma_{112}$, observed in \pPb collisions relative to that in \PbPb
collisions again indicates that the dominant contribution of the signal is not related to the CME.
The results of SS and OS three-particle correlators as functions of centrality in \PbPb collisions
at $\rootsNN = 5.02$\TeV are also found to be consistent with the results from
lower energy \AonA collisions~\cite{Abelev:2009ad, Abelev:2012pa}. However,
values of $\gamma_{123}$ correlators between \pPb and \PbPb are observed to be different,
unlike those for $\gamma_{112}$ correlators. As the CME contribution to $\gamma_{123}$ is not expected,
the data suggest different properties of backgrounds in \pPb and \PbPb systems.
If the $\gamma_{112}$ correlator in \pPb data is expected to be background dominated,
as argued earlier, the similarity found to the \PbPb data in $\gamma_{112}$ requires further understanding.
The two-particle $\delta$ correlators show a similar trend in multiplicity between \pPb
and \PbPb systems, but a larger splitting between OS and SS pairs is observed in \pPb than in \PbPb data.

\begin{figure}[thbp]
\centering
  \includegraphics[width=\cmsFigWidth]{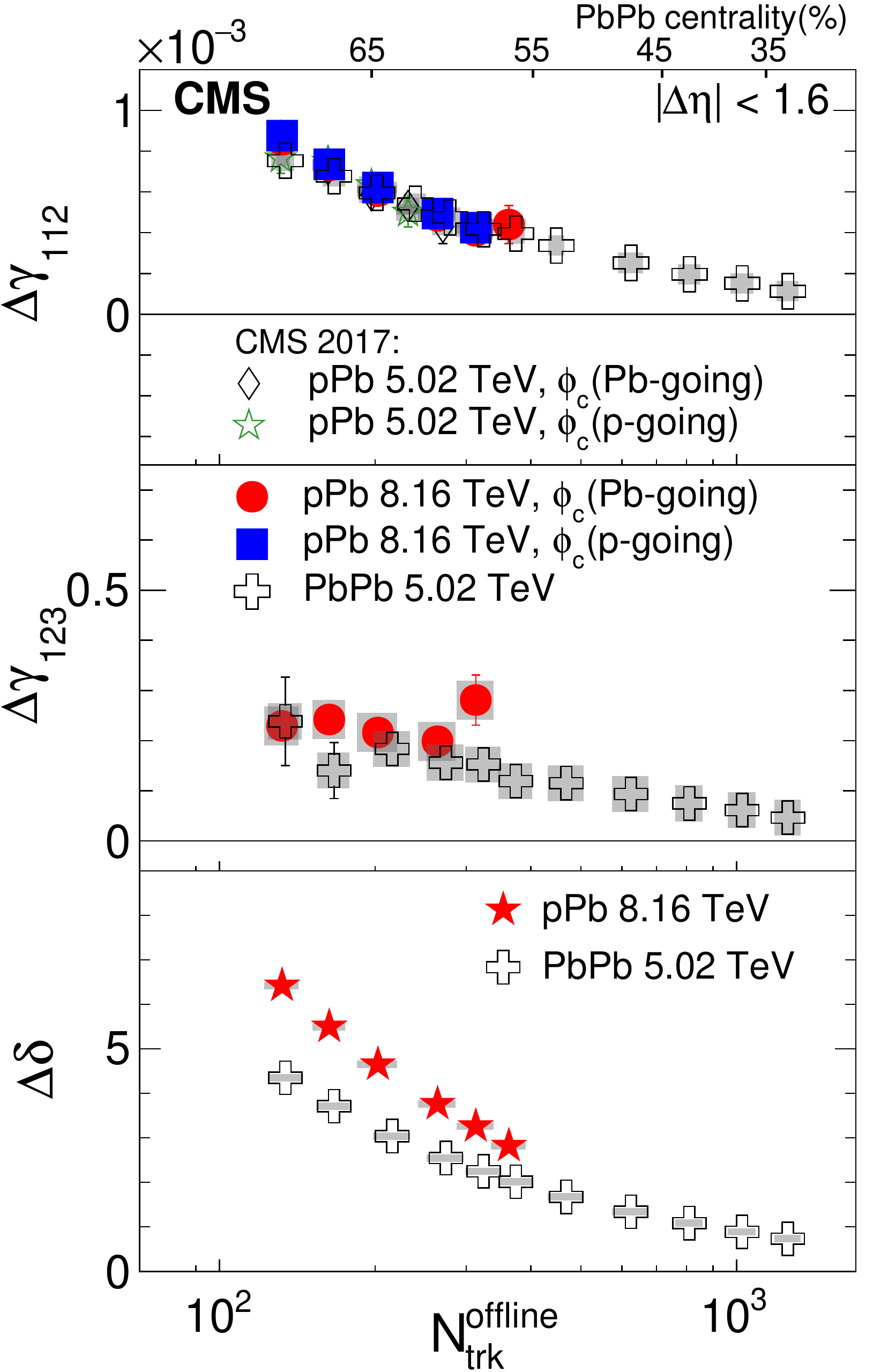}
  \caption{ \label{fig4}
The difference of the OS and SS three-particle correlators,
$\gamma_{112}$ (upper)  and $\gamma_{123}$ (middle), and
two-particle correlator, $\delta$ (lower), averaged over
$\abs{\deta}<1.6$ as a function of \noff in \pPb collisions at
$\rootsNN = 8.16$\TeV and \PbPb collisions at 5.02\TeV.
The \pPb results are obtained with particle $c$ from \Pb- and \Pp-going
sides separately. The $\Delta\delta$ correlator is denoted by a different marker for \pPb collisions. The results of $\gamma_{112}$ for \pPb collisions at 5.02\TeV from CMS Collaboration (CMS 2017:~\cite{Khachatryan:2016got}),
are also shown for comparison. Statistical and systematic uncertainties
are indicated by the error bars and shaded regions, respectively.
   }
\end{figure}

To eliminate sources of correlations that are charge independent
(\eg, directed flow, $v_1$) and to explore a possible charge separation effect
generated by the CME or charge-dependent background correlations,
the differences of three-particle correlators, $\Delta\gamma_{112}$
and $\Delta\gamma_{123}$, and two-particle correlator, $\Delta\delta$, between
OS and SS are shown in Fig.~\ref{fig2} as functions of $\abs{\deta}$, $|\Delta\pt|$, and \ptbar\
in the multiplicity range $185 \leq \noff < 250$ for \pPb collisions
at $\rootsNN = 8.16$\TeV and \PbPb collisions at 5.02\TeV.

After taking the difference, the three-particle correlators, $\Delta\gamma_{112}$
and $\Delta\gamma_{123}$, in \pPb collisions with particle $c$ from either the \Pp- or \Pb-going side, and in \PbPb collisions, show nearly identical values, except in the high \ptbar region.
Note that for OS and SS correlators separately, this similarity between
\pPb and \PbPb is only observed for the $\gamma_{112}$ correlator.
As a function of $\abs{\deta}$, the charge-dependent difference is largest
at $\abs{\deta} \approx 0$ and drops to zero for $\abs{\deta} > 1.6$ for both systems.
The striking similarity in the observed charge-dependent azimuthal correlations
between \pPb and \PbPb as functions of $\abs{\deta}$, $|\Delta\pt|$ and \ptbar\
strongly suggests a common physical origin. As argued in Ref.~\cite{Khachatryan:2016got},
a strong charge separation signal from the CME is not expected in a very
high-multiplicity \pPb collisions, and not with respect to $\Psi_{3}$ (for the $\gamma_{123}$
correlator) in either the \pPb or \PbPb system. The similarity
seen between high-multiplicity \pPb and peripheral \PbPb collisions for both
$\Delta\gamma_{112}$ and $\Delta\gamma_{123}$ further challenges the attribution of the
observed charge-dependent correlations to the CME. The two-particle correlator,
$\Delta\delta$, on the other hand, is found to show a larger value in \pPb than in \PbPb collisions.

\begin{figure}[thb]
\centering
  \includegraphics[width=\cmsFigWidth]{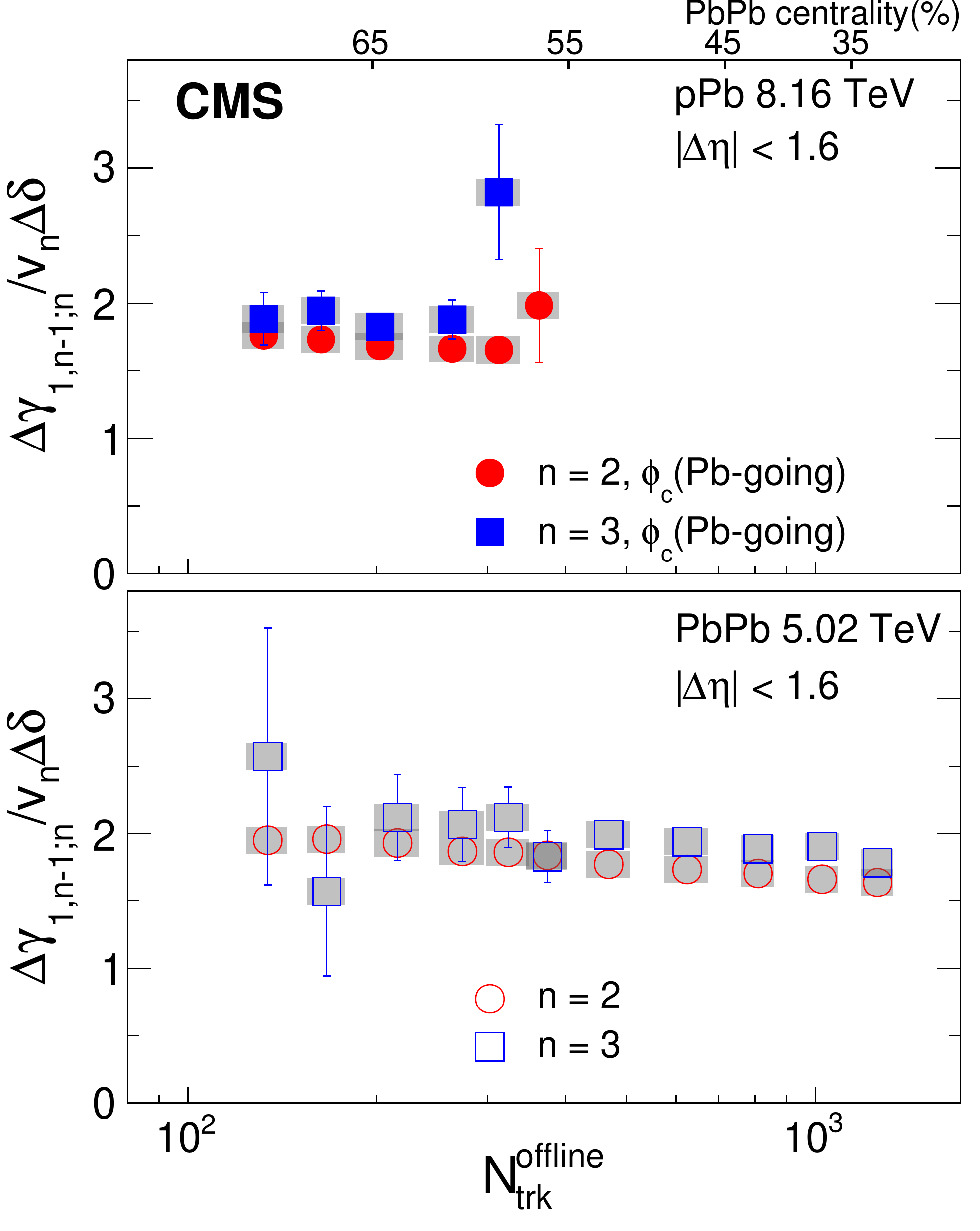}
  \caption{ \label{fig5}
The ratio of $\Delta\gamma_{112}$ and $\Delta\gamma_{123}$ to the product of $v_{n}$  and $\delta$,
averaged over $\abs{\deta}<1.6$, in \pPb collisions for the \Pb-going direction at
$\rootsNN = 8.16$\TeV (upper)  and \PbPb collisions at 5.02\TeV (lower).
Statistical and systematic uncertainties are indicated by the
error bars and shaded regions, respectively.
   }
\end{figure}

\begin{figure*}[thb]
\centering
  \includegraphics[width=0.8\textwidth]{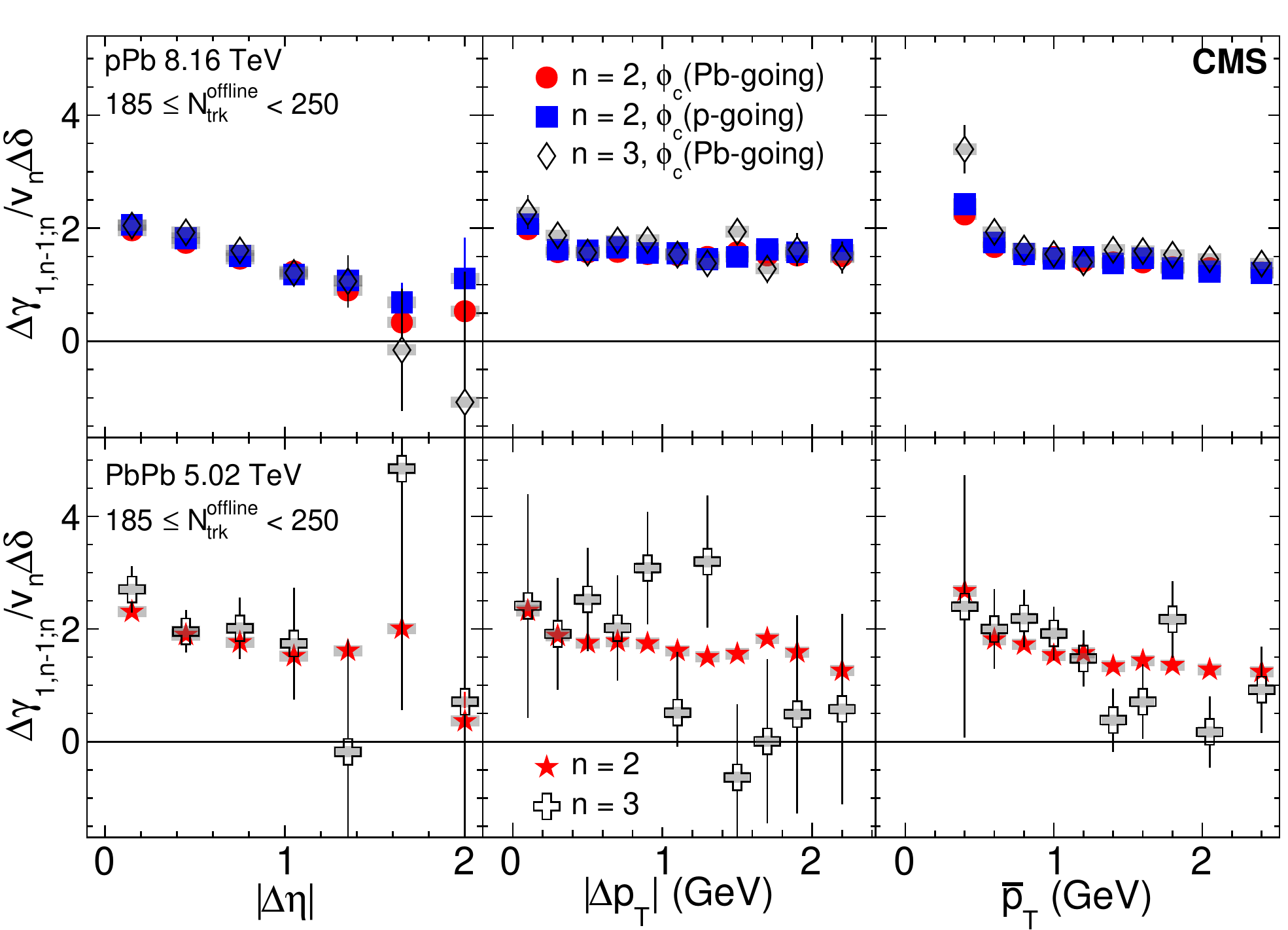}
  \caption{ \label{fig6}
The ratio of $\Delta\gamma_{112}$ and $\Delta\gamma_{123}$ to the product of $v_{n}$ and $\delta$,
as functions of $\Delta\eta$ (left), $\Delta\pt$ (middle), and $\ptbar$ (right) for $185 \leq \noff < 250$ in \pPb collisions at
$\rootsNN = 8.16$\TeV (upper)  and \PbPb collisions at 5.02\TeV (lower).
Statistical and systematic uncertainties are indicated by the
error bars and shaded regions, respectively.
   }
\end{figure*}

The differences of three-particle correlators, $\Delta\gamma_{112}$
and $\Delta\gamma_{123}$, and two-particle correlator, $\Delta\delta$, between
OS and SS are shown in Fig.~\ref{fig4} as functions of \noff averaged over $\abs{\deta}< 1.6$
for \pPb collisions at $\rootsNN = 8.16$\TeV and \PbPb collisions at 5.02\TeV. For comparison, previously published \pPb\
data at 5.02\TeV are also shown~\cite{Khachatryan:2016got}.
Similar to those shown in Fig.~\ref{fig2}, the observed difference between OS and SS pairs in
$\Delta\gamma_{112}$ and $\Delta\gamma_{123}$ is strikingly similar in \pPb and \PbPb collisions
over the entire overlapping multiplicity range (and also independent of collision energy
for $\Delta\gamma_{112}$ in \pPb), while higher values of an OS-SS difference in
$\Delta\delta$ are found for the \pPb system.

To check if the mechanism of local charge conservation coupled with anisotropic flow
can explain the observed charge dependence of the $\Delta\gamma_{112}$
and $\Delta\gamma_{123}$ correlators, the relation in Eq.~(\ref{eq:kappa}) is used. The ratios
of $\Delta\gamma_{112}$ and $\Delta\gamma_{123}$
to the product of $\Delta\delta$ and $v_n$ are shown in
Fig.~\ref{fig5}, averaged over $\abs{\deta}<1.6$,
as functions of event multiplicity in \pPb and \PbPb collisions.
The $v_2$ and $v_3$ values for particles $\alpha$ or $\beta$ are calculated with
the scalar-product method with respect to the particle $c$. In \pPb collisions, only results with the \Pb-going direction are shown because the \Pp-going direction data lack statistical precision, except for the multiplicity range $185 \leq \noff < 250$.

The ratios shown in Fig.~\ref{fig5} for both systems are found to be similar
between $n$=2 and $n$=3, on average with values slightly less than 2. This observation indicates that the measured charge dependence of
three-particle correlators is consistent with mostly being dominated by
charge-dependent two-particle correlations (e.g., from local charge conservation)
coupled with the anisotropic flow $v_n$. For a given $n$ value, the ratios are also similar between \pPb and \PbPb collisions
(and may reflect similar particle kinematics and acceptances), and approximately constant as
functions of event multiplicity. Notably, the $\Delta\delta$ in Fig.~\ref{fig4} are
different between the \pPb and \PbPb systems. However, the anisotropic flow harmonics $v_n$
are larger for \PbPb collisions than for \pPb collisions~\cite{Chatrchyan:2013nka}. As a result, the product of
$\Delta\delta$ and $v_n$ leads to similar values of $\Delta\gamma_{112}$ and
$\Delta\gamma_{123}$ correlators between the \pPb and \PbPb systems, implying the $\kappa_{2}$ is similar to $\kappa_{3}$.

The ratios of $\Delta\gamma_{112}$ and
$\Delta\gamma_{123}$ to the product of $\Delta\delta$
and $v_n$ can also be studied as functions of $\abs{\deta}$, $\Delta\pt$, and $\ptbar$ in \pPb and
\PbPb collisions, as shown in Fig.~\ref{fig6} for the multiplicity range of $185 \leq \noff < 250$.
Here, the $v_n$ are calculated as the average $v_n$ of particles $\alpha$ and $\beta$,
$v_n = (v_{n,\alpha} + v_{n,\beta})/2$ (based on the relation derived in Eq.~(\ref{eq:lcc123_5})
in Appendix~\ref{app:a}), and are weighted by the number of pairs of particles $\alpha$
and $\beta$ in the given kinematic ranges when averaged over $\eta$ or \pt.
The ratios involving $\Delta\gamma_{112}$
and $\Delta\gamma_{123}$ are again found to be similar differentially for all three variables
in both \pPb and \PbPb collisions. This observation further supports a common origin of
$\Delta\gamma_{112}$ and $\Delta\gamma_{123}$ from charge-dependent two-particle
correlations coupled with the anisotropic flow.

\subsection{Event shape engineering}
\label{subsec:results_ese}

To explore directly the background scenario in Eq.~(\ref{eq:lcc}) in terms of a linear
dependence on $v_2$ for the $\gamma_{112}$ correlator, results based on the ESE analysis are presented in this section.

\begin{figure}[thb]
\centering
  \includegraphics[width=\cmsFigWidth]{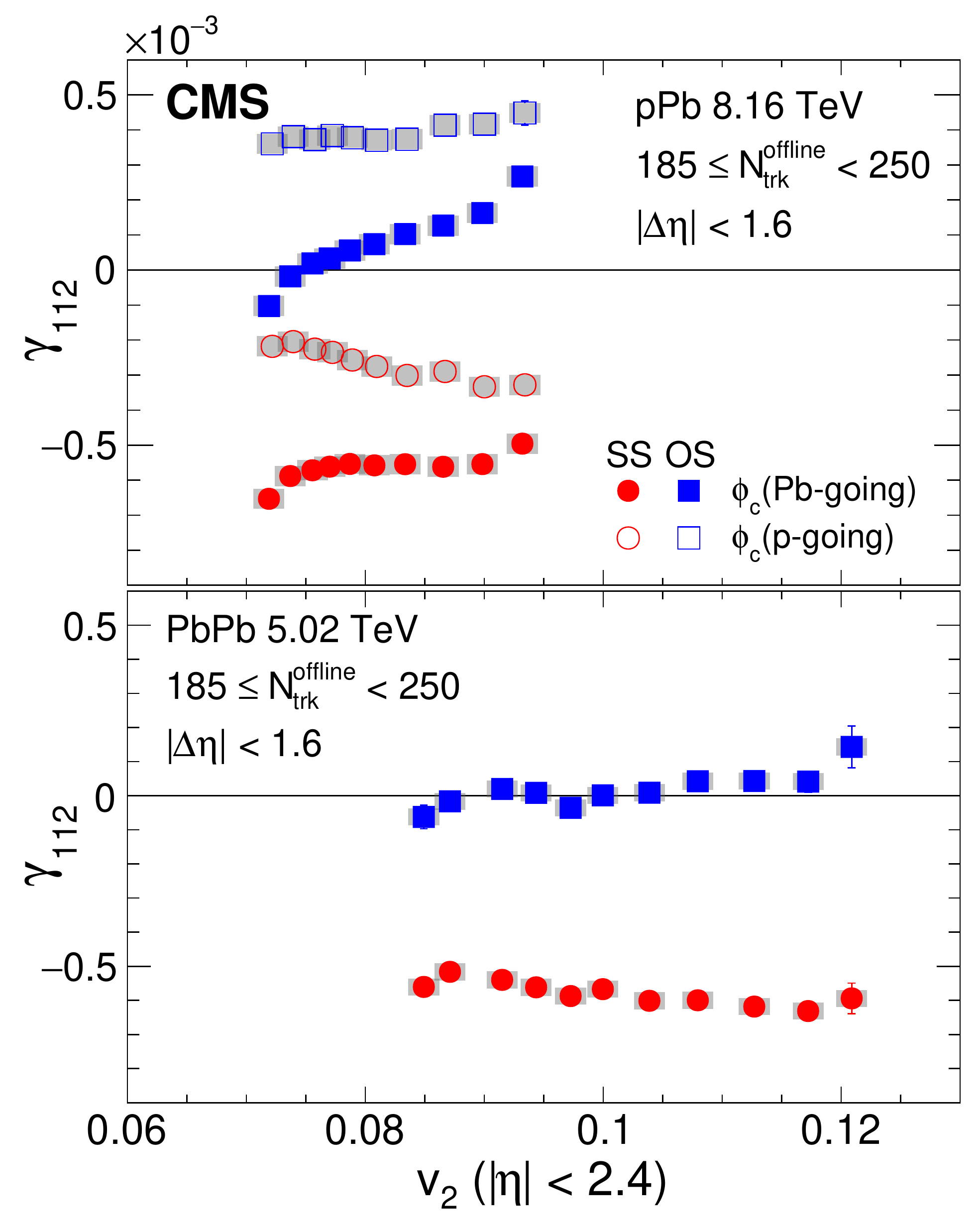}
  \caption{ \label{fig7} The SS and OS three-particle
  correlators, $\gamma_{112}$, averaged over $\abs{\deta}<1.6$
  as a function of $v_2$ (evaluated as the average $v_2$ value for each corresponding $q_2$ event class), for the multiplicity range
  $185 \leq \noff < 250$ in \pPb collisions at $\rootsNN = 8.16$\TeV (upper) and
  \PbPb collisions at 5.02\TeV (lower). The \pPb results are obtained
  with particle $c$ from \Pb- and \Pp-going sides separately. Statistical and systematic
  uncertainties are indicated by the error bars and shaded regions, respectively.  }
\end{figure}

The SS and OS three-particle correlators,
$\gamma_{112}$, averaged over $\abs{\deta}<1.6$,
are shown as a function of $v_2$ (evaluated as the average $v_2$ value for each corresponding $q_2$ event class in Fig.~\ref{fig7}), for the multiplicity range
$185 \leq \noff < 250$ in \pPb collisions at $\rootsNN = 8.16$\TeV (upper) and
\PbPb collisions at 5.02\TeV (lower). The \pPb results are obtained
with particle $c$ from the \Pb- and \Pp-going sides separately.

\begin{figure}[thbp]
\centering
  \includegraphics[width=\cmsFigWidth]{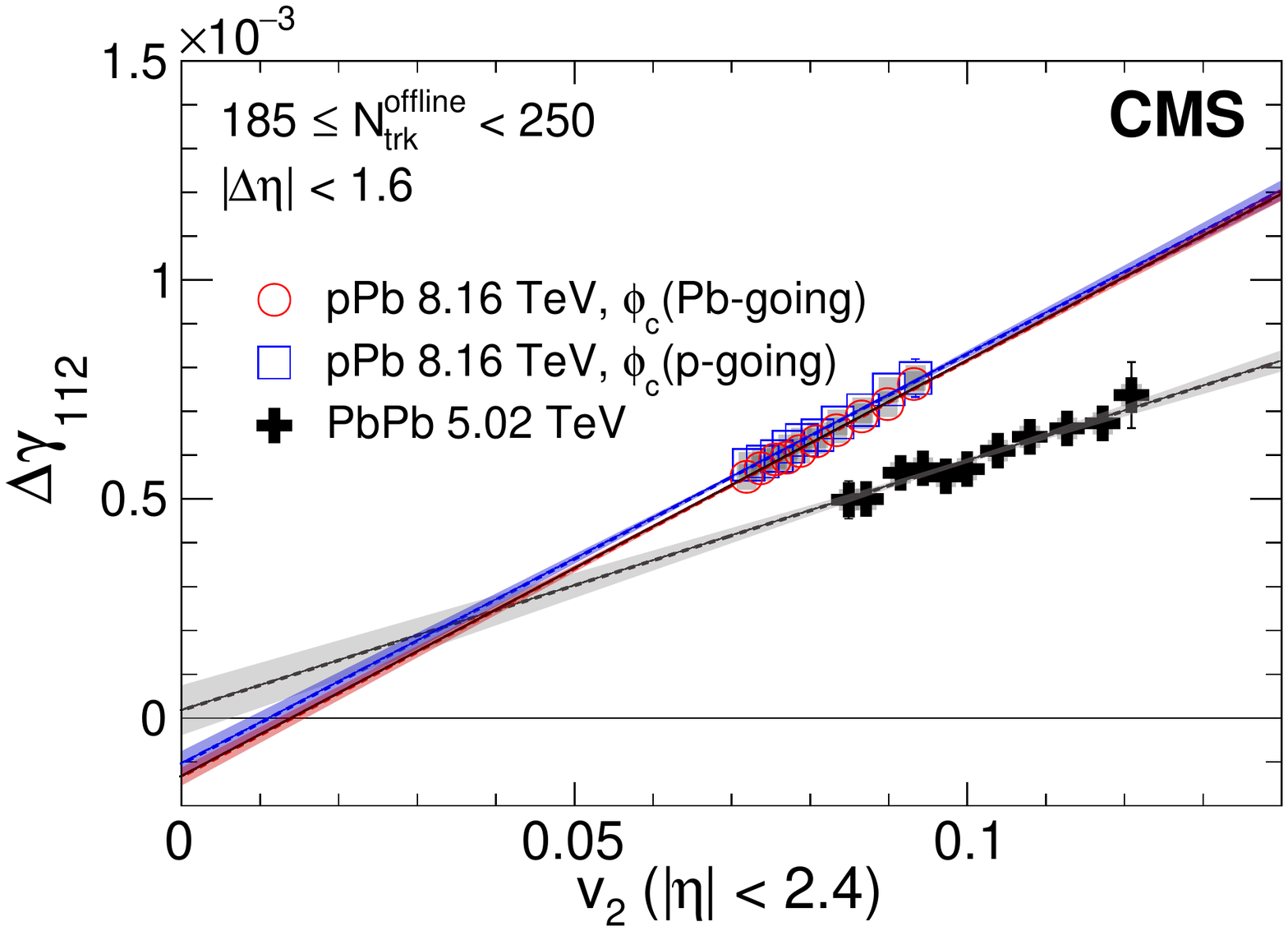}
  \includegraphics[width=\cmsFigWidth]{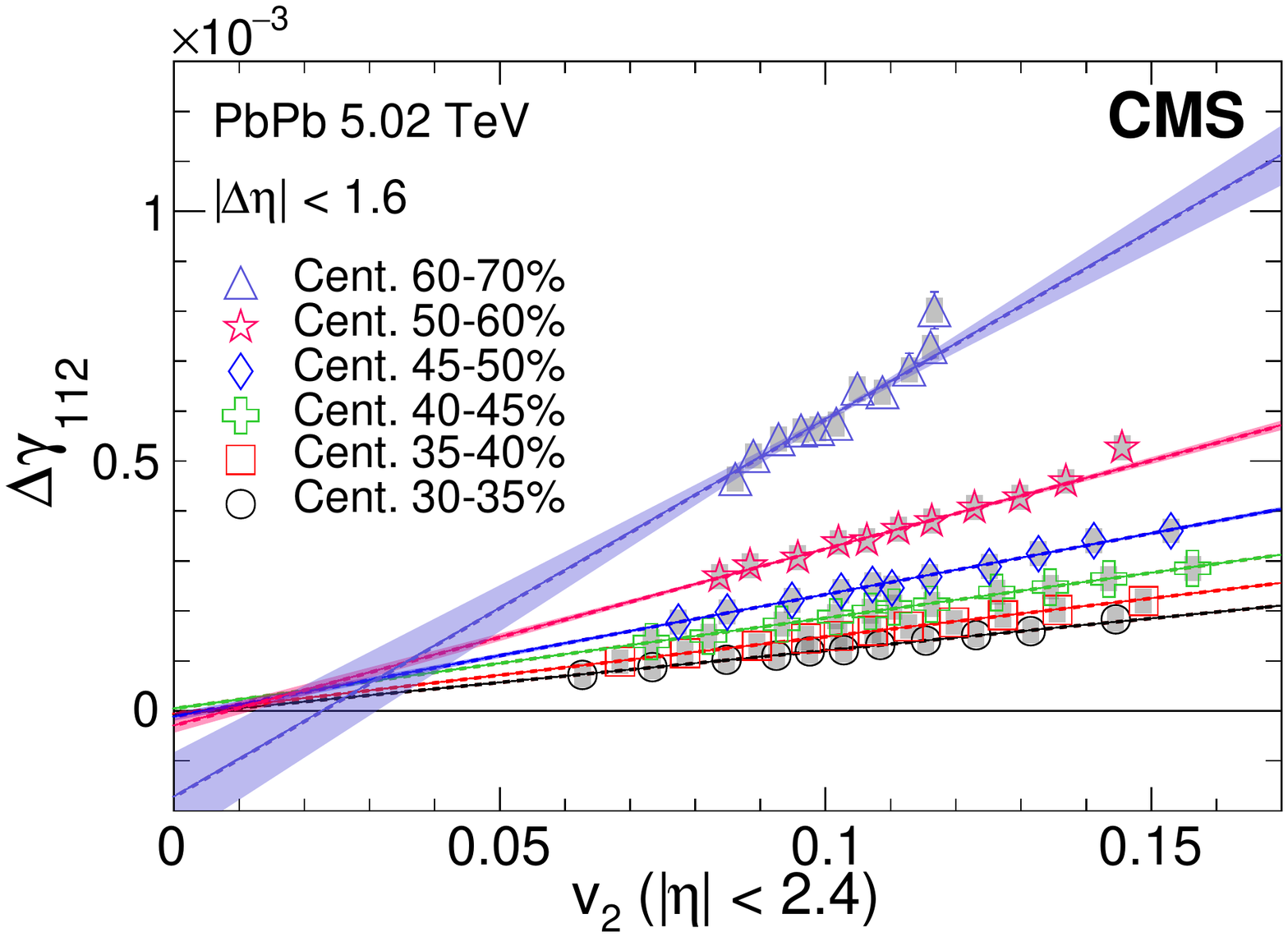}
  \caption{ \label{fig8} The difference of the OS and SS
  three-particle correlators, $\gamma_{112}$, averaged over $\abs{\deta}<1.6$
  as a function of $v_2$ evaluated in each $q_2$ class, for the multiplicity range
  $185 \leq \noff < 250$ in \pPb collisions at $\rootsNN = 8.16$\TeV and
  \PbPb collisions at 5.02\TeV (upper), and for different centrality
  classes in \PbPb collisions at 5.02\TeV (lower). Statistical and systematic
  uncertainties are indicated by the error bars and shaded regions, respectively. A one standard deviation uncertainty from the fit is also shown.
  }
\end{figure}

Both SS and OS $\gamma_{112}$ correlators in both \pPb (both
beam directions for particle $c$) and \PbPb collisions show a dependence on $v_2$.
A clear linear dependence on the $v_2$ value is not seen for any of the SS and
OS correlators studied.

\begin{figure}[thbp]
\centering
  \includegraphics[width=\cmsFigWidth]{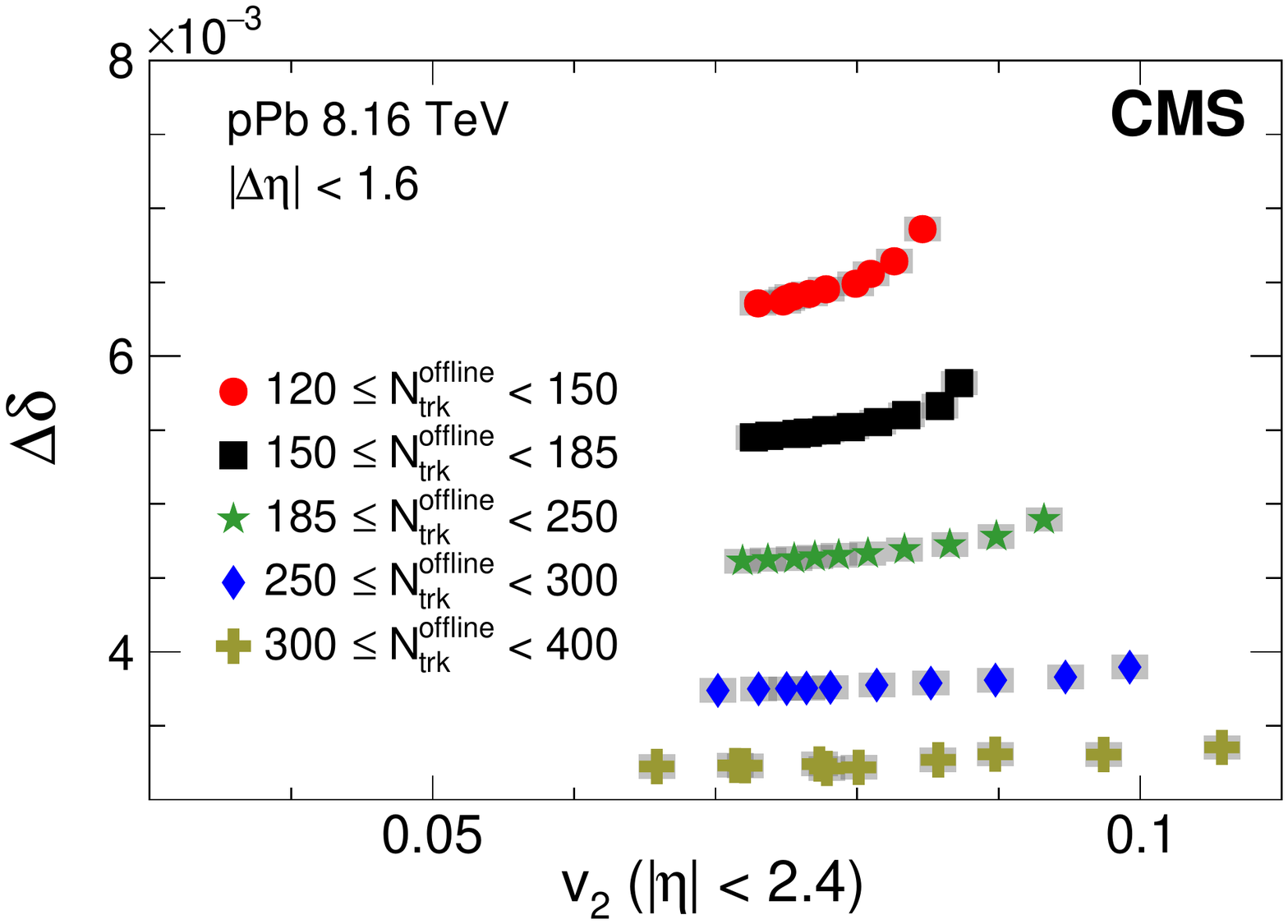}
  \includegraphics[width=\cmsFigWidth]{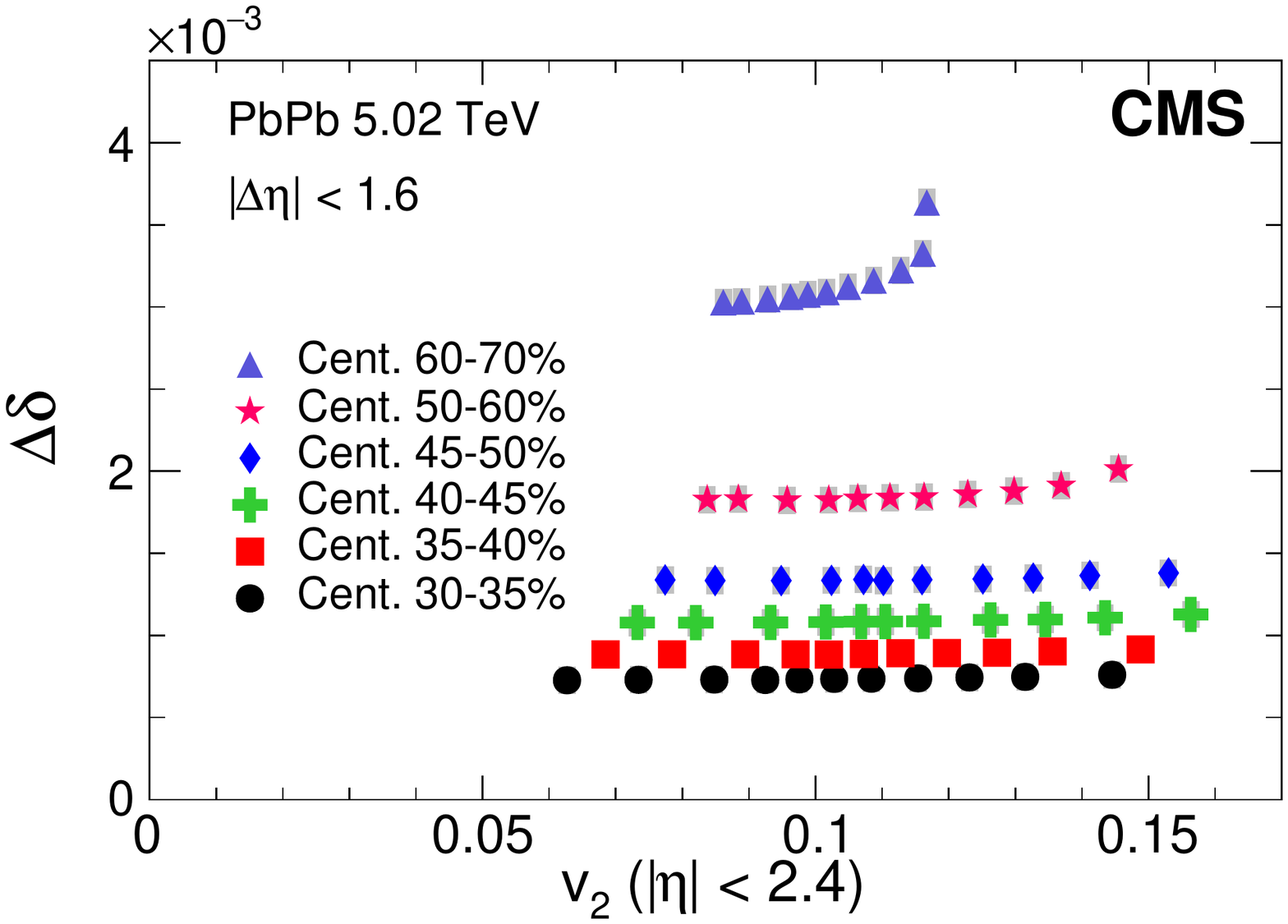}
  \caption{ \label{fig9} The difference of the OS and SS
  two-particle correlators, $\delta$, averaged over $\abs{\deta}<1.6$
  as a function of $v_2$ evaluated in each $q_2$ class, for different multiplicity ranges in \pPb collisions at $\rootsNN = 8.16$\TeV (upper), and for different centrality
  classes in \PbPb collisions at 5.02\TeV (lower). Statistical and systematic
  uncertainties are indicated by the error bars and shaded regions, respectively.
  }
\end{figure}

Similar to the analysis in Section~\ref{subsec:results_mixed}, the difference
between OS and SS correlators is taken in order to eliminate the charge-independent
sources of the correlators. The results, averaged over $\abs{\deta}<1.6$,
are shown in Fig.~\ref{fig8} (upper) , as a function of $v_2$ evaluated in each $q_2$ class,
for the multiplicity range $185 \leq \noff < 250$ in \pPb collisions at $\rootsNN = 8.16$\TeV\
and \PbPb collisions at 5.02\TeV. The results obtained in each centrality
class of \PbPb collisions at 5.02\TeV are also presented in Fig.~\ref{fig8} (lower).
The lines are linear fits to the data,
\begin{linenomath}
\begin{equation}
\label{ese_fit}
\Delta\gamma_{112} = a \, v_2 + b,
\end{equation}
\end{linenomath}
where the first term corresponds to the $v_2$-dependent background contribution
with the slope parameter $a$ equal to $\kappa_{2}\Delta\delta$ (from Eq.~(\ref{eq:lcc})), which is assumed to be $v_2$ independent. The intercept parameter $b$ denotes the
$v_{\mathrm 2}$-independent contribution (when linearly extrapolating to $v_{\mathrm 2}=0$)
in the $\gamma_{112}$ correlator. In particular, as the CME contribution to the
$\Delta\gamma_{112}$ is expected to be largely $v_2$-independent within narrow multiplicity (centrality) ranges, the $b$ parameter may
provide an indication to a possible observation of the CME, or set an upper limit on
the CME contribution.

\begin{figure}[thbp]
\centering
  \includegraphics[width=\cmsFigWidth]{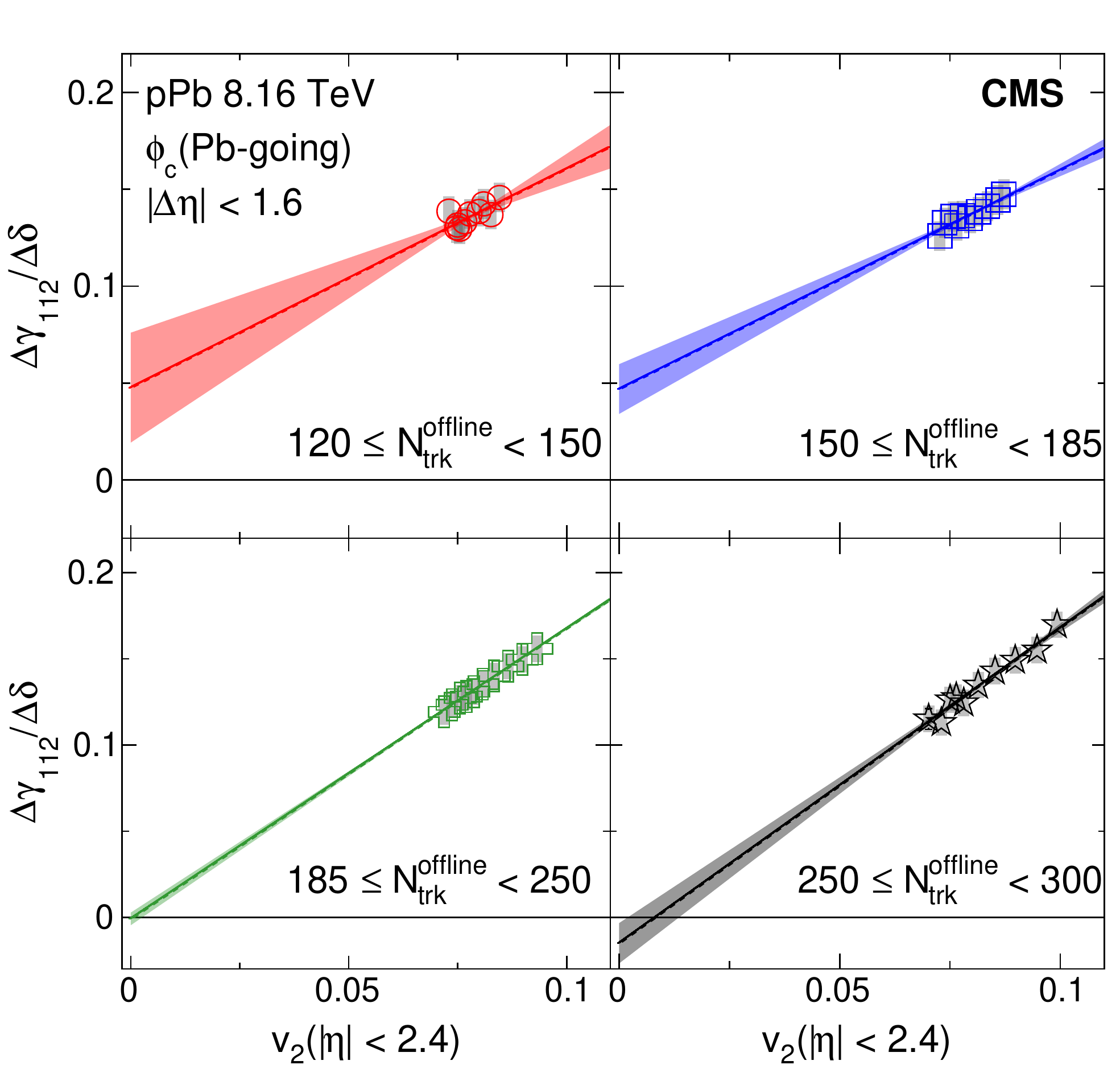}
  \includegraphics[width=\cmsFigWidth]{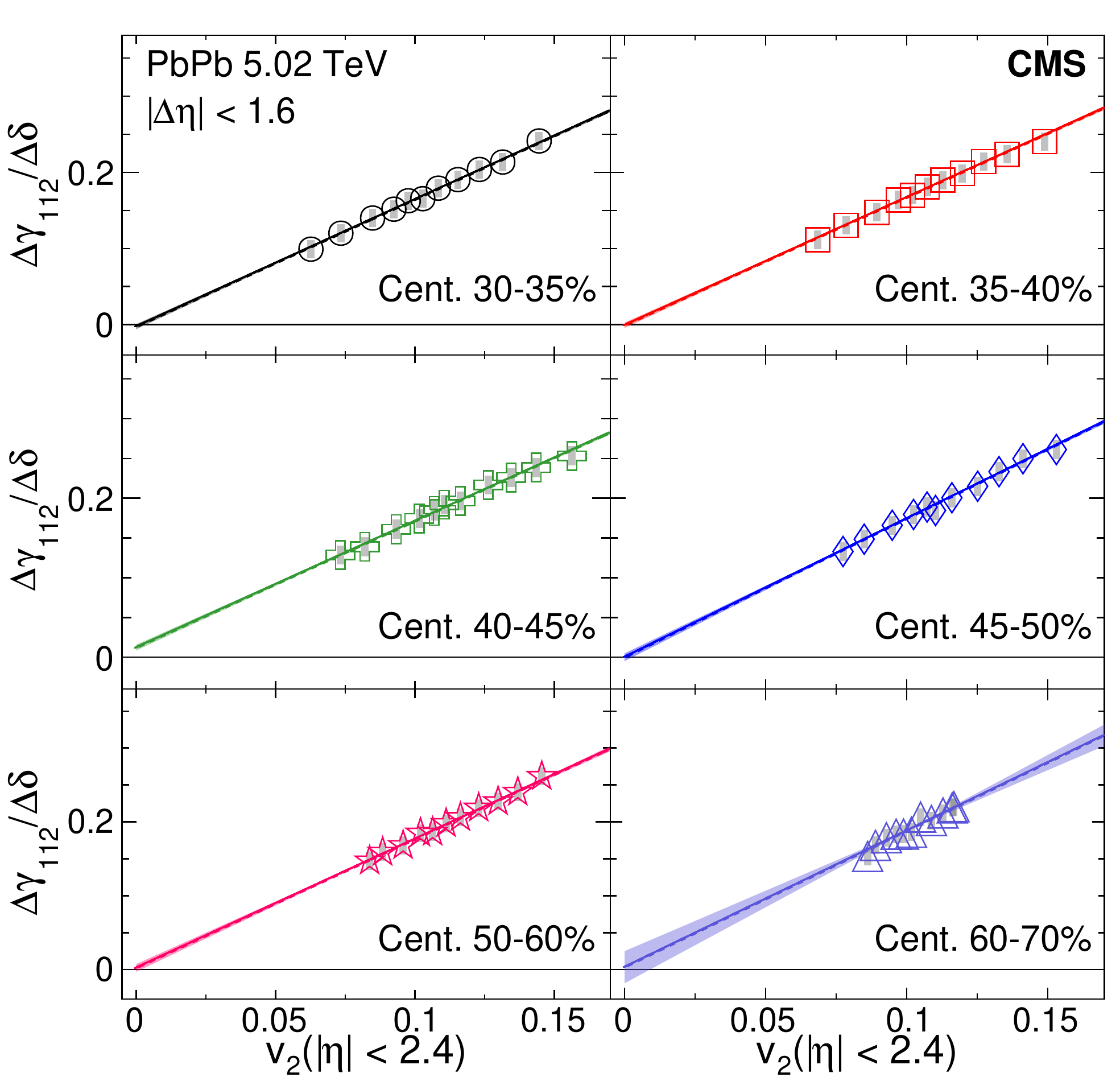}
  \caption{ \label{fig10} The ratio between the difference of the OS and SS
  three-particle correlators and the difference of OS and SS in $\delta$ correlators, $\Delta\gamma_{112}/\Delta\delta$, averaged over $\abs{\deta}<1.6$
  as a function of $v_2$ evaluated in each $q_2$ class, for different multiplicity ranges
  in \pPb collisions at $\rootsNN = 8.16$\TeV (upper), and for different centrality
  classes in \PbPb collisions at 5.02\TeV (lower). Statistical and systematic
  uncertainties are indicated by the error bars and shaded regions, respectively.
  A one standard deviation uncertainty from the fit is also shown.
  }
\end{figure}

As shown in Fig.~\ref{fig8}, for both \pPb and \PbPb collisions in each multiplicity
or centrality range, a clear linear dependence of the $\Delta\gamma_{112}$ correlator as
a function of $v_2$ is observed. Fitted by a linear function, the intercept parameter, $b$,
can be extracted. A one standard deviation uncertainty band
is also shown for the linear fit. Taking the statistical uncertainties into account,
the values of $b$ are found to be nonzero for multiplicity range $185 \leq \noff < 250$ in \pPb and 60--70\% centrality in \PbPb collisions.

\begin{figure}[thb]
\centering
  \includegraphics[width=\cmsFigWidth]{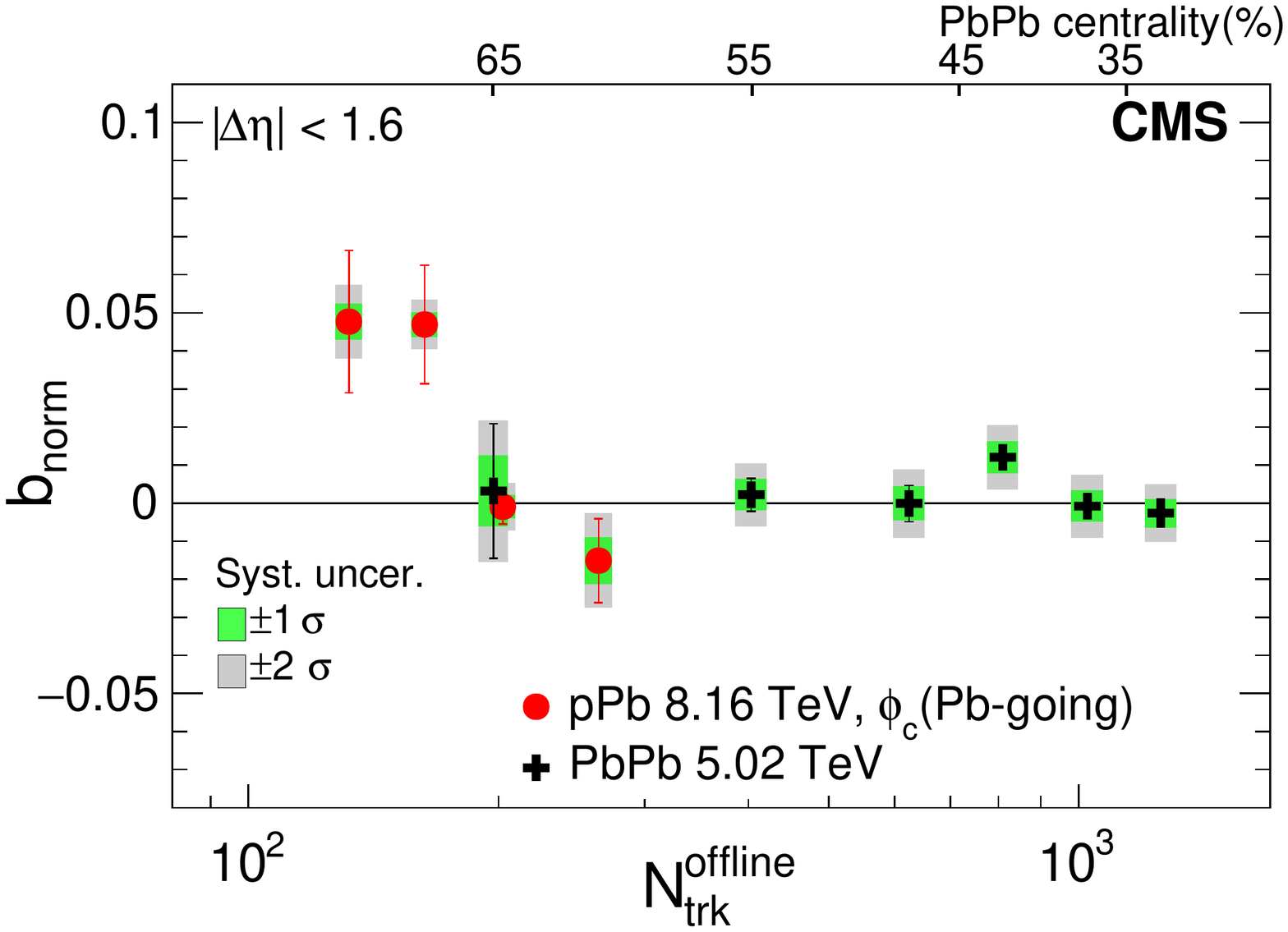}
  \includegraphics[width=\cmsFigWidth]{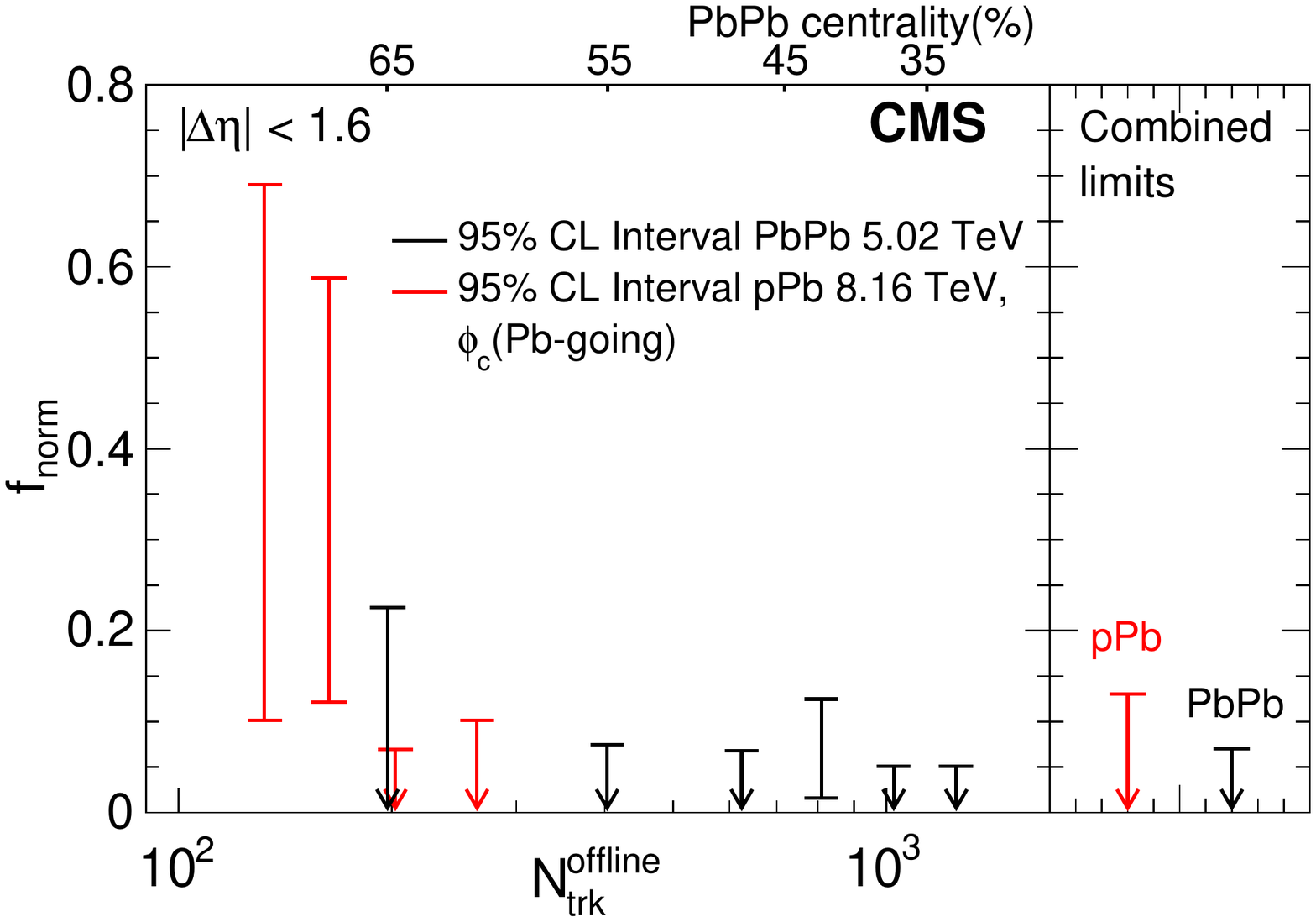}
  \caption{ \label{fig12} Extracted intercept parameter $b_{\text{norm}}$ (upper) and
  corresponding upper limit of the fraction of $v_2$-independent $\gamma_{112}$
  correlator component (lower), averaged over $\abs{\deta}<1.6$,
  as a function of \noff in \pPb collisions at $\rootsNN = 8.16$\TeV and
  \PbPb collisions at 5.02\TeV. Statistical and systematic uncertainties
are indicated by the error bars and shaded regions in the top panel, respectively.}
\end{figure}

Observing a nonzero intercept $b$ from Fig.~\ref{fig8} may or may not lead to a conclusion
of a finite CME signal, as an assumption is made for the background contribution term,
namely that $\Delta\delta$ is independent of $v_2$. To check this assumption explicitly,
the $\Delta\delta$ correlator is shown in Fig.~\ref{fig9} as a function of $v_{\mathrm 2}$
in different multiplicity and centrality ranges in \pPb (upper)  and \PbPb (lower)
collisions. It is observed that the value of $\Delta\delta$
remains largely constant as a function of $v_2$ in low- or intermediate-$q_{2}$ classes,
but starts rising as $v_{\mathrm 2}$ increases in high-$q_{2}$ classes. The multiplicity, within a centrality or multiplicity range, decreases slightly with increasing $q_{2}$, which qualitatively could contribute to the rising $\Delta\delta$ due to a multiplicity dilution effect. However, this is only found to be true for \PbPb collisions, but not for \pPb collisions. The other reason may be related to larger jet-like correlations selected by requiring
large $q_2$ values. Events with higher multiplicities
show a weaker dependence on $v_{\mathrm 2}$ than those with lower multiplicities, which is consistent with the expectation that short-range jet-like correlations are stronger in peripheral events.
Because of the possible bias towards larger jet-like correlations at higher $q_2$ from the ESE technique, the $v_{\mathrm 2}$
dependence of $\Delta\delta$ is hard to completely eliminate. This presents
a challenge to the interpretation of the intercept values from the linear fits in Fig.~\ref{fig8}.

In order to avoid the issue of $\Delta\delta$ being dependent on $v_{\mathrm 2}$,
the ratio $\Delta\gamma_{112}/\Delta\delta$ as function of $v_{\mathrm 2}$ is
shown in Fig.~\ref{fig10} for different multiplicity ranges in \pPb collisions
at $\rootsNN = 8.16$\TeV (upper) and for different centrality classes in \PbPb collisions at 5.02\TeV (lower).
Particularly in the scenario of a pure $v_{2}$-dependent background,
the ratio $\Delta\gamma_{112}/\Delta\delta$ is expected to be proportional
to $v_2$. A linear function is fitted again using
\begin{linenomath}
\begin{equation}
\label{ese_fit2}
\frac{\Delta\gamma_{112}}{\Delta\delta} = a_{\text{norm}} \, v_2 + b_{\text{norm}}.
\end{equation}
\end{linenomath}
Here, comparing to the intercept parameter $b$ in
Eq.~(\ref{ese_fit}), the $b_{\text{norm}}$ parameter is equivalent to
$b$ scaled by the $\Delta\delta$ factor. The fitted linear slope and intercept parameters,
$a_{\text{norm}}$ and $b_{\text{norm}}$, are summarized
in Tables~\ref{tab:pPb_Ntrk} and ~\ref{tab:PbPb_Cent} in \noff and centrality
classes for \pPb and \PbPb collisions, respectively.

\begin{table*}
\centering
\topcaption{ \label{tab:pPb_Ntrk} The summary of slope and intercept
parameter $a_{\text{norm}}$ and $b_{\text{norm}}$ for different \noff classes
in \pPb collisions, and the goodness of fit $\chi^{2}$ per degree of freedom (ndf). The statistical and systematic uncertainties are shown after the central values, respectively.}
\newcolumntype{x}{D{,}{\,\pm\,}{5.13}}
\begin{scotch}{lcxr}
 \noff &  $a_{\text{norm}}$ & \multicolumn{1}{c}{$b_{\text{norm}}$} & $\chi^{2}/\text{ndf}$   \\
\hline
120--150 & 1.13 $\pm$ 0.24 $\pm$ 0.14 & 0.048 , 0.019 \pm 0.012 & 16.3/8 \\
150--185 & 1.13 $\pm$ 0.19 $\pm$ 0.04 & 0.047 , 0.016 \pm 0.008 & 4.9/8 \\
185--250 & 1.69 $\pm$ 0.06 $\pm$ 0.01 & -0.0009 , 0.0050 \pm 0.0078 & 4.5/8 \\
250--300 & 1.83 $\pm$ 0.13 $\pm$ 0.15 & -0.015 , 0.011 \pm 0.016 & 8.1/8 \\
\end{scotch}
\end{table*}

\begin{table*}
\centering
\caption{ \label{tab:PbPb_Cent} The summary of slope and intercept
parameter $a_{\text{norm}}$ and $b_{\text{norm}}$ for different centrality classes
in \PbPb collisions, and the goodness of fit $\chi^{2}$ per degree of freedom (ndf). The statistical and systematic uncertainties are shown after the central values, respectively.}
\newcolumntype{x}{D{,}{\,\pm\,}{5.13}}
\begin{scotch}{lcxr}
 Centrality &  $a_{\text{norm}}$ & \multicolumn{1}{c}{$b_{\text{norm}}$} & $\chi^{2}/\text{ndf}$   \\
\hline
60--70\% & 1.85 $\pm$ 0.17 $\pm$ 0.21 & 0.003 , 0.017 \pm 0.023 & 12.3/9 \\
50--60\% & 1.75 $\pm$ 0.04 $\pm$ 0.01 & 0.002 , 0.004 \pm 0.010 & 11.8/9 \\
45--50\% & 1.74 $\pm$ 0.04 $\pm$ 0.03 & 0.000 , 0.005 \pm 0.011  & 8.4/9 \\
40--45\% & 1.59 $\pm$ 0.03 $\pm$ 0.01 & 0.012 , 0.003 \pm 0.011  & 9.1/9 \\
35--40\% & 1.68 $\pm$ 0.03 $\pm$ 0.01 & -0.001 , 0.003 \pm 0.010 & 15.1/9 \\
30--35\% & 1.67 $\pm$ 0.04 $\pm$ 0.01 & -0.0026 , 0.0036 \pm 0.0095  & 6.9/9 \\
\end{scotch}
\end{table*}

The values of the intercept parameter $b_{\text{norm}}$ are shown as a function of
event multiplicity in Fig.~\ref{fig12} (upper) , for both \pPb and \PbPb collisions. The $\pm$1$\sigma$ and $\pm$2$\sigma$ systematic uncertainty is shown, which correspond to a 68\% and 95\% confidence level (\CL), respectively.
Within statistical and systematic uncertainties, no significant positive value for $b_{\text{norm}}$ is observed for most multiplicities in \pPb or centralities in \PbPb collisions. For multiplicity ranges $120 \leq \noff < 150$ and $150 \leq \noff < 185$ in \pPb collisions, an indication of positive values with significances of more than two standard deviations is seen. However, results in these multiplicity ranges are likely to be highly sensitive to the very limited $v_2$ coverage using the ESE technique, as shown in the upper panel of Fig.~\ref{fig10}. Overall, the result suggests that the $v_2$-independent
contribution to the $\Delta\gamma_{112}$ correlator is consistent with zero, and
correlation data are consistent with the background-only scenario of charge-dependent
two-particle correlations plus an anisotropic flow, $v_n$. This conclusion is consistent
with that drawn from the study of higher-order harmonic three-particle correlators discussed earlier.

Based on the assumption of a nonnegative CME signal, the upper limit of the $v_2$-independent fraction in the $\Delta\gamma_{112}$ correlator is obtained from the Feldman--Cousins approach~\cite{Feldman:1997qc} with the measured statistical and systematic uncertainties. In Fig.~\ref{fig12} (lower), the upper limit of the fraction $f_{\text{norm}}$, where $f_{\text{norm}}$ is the ratio of the $b_{\text{norm}}$ value to
the value of $\left\langle \Delta\gamma_{112}\right\rangle$/$\left\langle \Delta\delta \right\rangle$, is presented at 95\% \CL
as a function of event multiplicity. The $v_2$-independent
component of the $\Delta\gamma_{112}$ correlator is less than 8--15\% for most of the multiplicity or centrality range. The combined limits from all presented multiplicities and centralities are also shown in \pPb and \PbPb collisions. An upper limit on the $v_2$-independent fraction of the three-particle correlator, or possibly the CME signal contribution, is
estimated to be 13\% in \pPb and 7\% in \PbPb collisions, at 95\% \CLp.
Note that the conclusion here is based on the assumption of a CME signal independent of $v_2$ in a narrow multiplicity or centrality range.
As pointed out in a study by the ALICE collaboration after this manuscript was submitted~\cite{Acharya:2017fau},
the observed CME signal may be reduced as $v_2$ decreases for small $v_2$ values (e.g., $<$6\%),
due to a weaker correlation between magnetic field and event-plane orientations as a result of initial-state fluctuations.
Depending on specific models of initial-state fluctuations, the upper limits obtained in this paper may increase relatively by about 20\%, although still well within a few \% level.
On the other hand, covering a wide range of $v_2$ values in this analysis (6--15\%), the $v_2$ dependence of the observed CME signal
is minimized to the largest extent, especially for more central events. The data also
rule out any significant nonlinear $v_2$ dependence of the observed CME signal, as suggested by Ref.~\cite{Acharya:2017fau}.
Therefore, the high-precision data presented in this paper indicate that the charge-dependent
three-particle azimuthal correlations in \pPb and \PbPb collisions are consistent
with a $v_2$-dependent background-only scenario, posing a significant challenge to the search
for the CME in heavy ion collisions using three-particle azimuthal correlations.

\section{Summary}
\label{sec:conclusion}

Charge-dependent azimuthal correlations of same- and opposite-sign (SS and OS)
pairs with respect to the second- and third-order event planes
have been studied in \pPb collisions at \rootsNN = 8.16\TeV and \PbPb\
collisions at 5.02\TeV by the CMS experiment at the LHC.
The correlations are extracted via three-particle correlators
as functions of pseudorapidity difference, transverse momentum difference, and \pt average of
SS and OS particle pairs, in various multiplicity or centrality ranges of the collisions.
The differences in correlations between OS and SS particles with
respect to both second- and third-order event planes as functions of \deta\ and multiplicity are found to agree for \pPb and \PbPb collisions, indicating
a common underlying mechanism for the two systems.
Dividing the OS and SS difference of the three-particle correlator
by the product of the $v_n$ harmonic of the corresponding order and the difference of the
two-particle correlator, the ratios are found to be similar for the second-
and third-order event planes, and show a weak dependence on event multiplicity.
These observations support a scenario in which the charge-dependent three-particle
correlator is predominantly a consequence of charge-dependent two-particle correlations coupled to an anisotropic flow signal.

To establish the relation between the three-particle correlator and anisotropic
flow harmonic in detail, an event shape engineering technique is applied. A linear relation for the ratio of three- to two-particle correlator difference as a function of
$v_2$ is observed, which extrapolates to an intercept that is consistent with zero within uncertainties for most of multiplicities.
An upper limit on the $v_2$-independent fraction of the three-particle correlator,
or the possible CME signal contribution (assumed independent of $v_2$ within the same narrow multiplicity or centrality range),
is estimated to be 13\% for \pPb data and 7\% for \PbPb data at a 95\% confidence level.
The data presented in this paper provide new stringent constraints on
the nature of the background contribution to the charge-dependent azimuthal
correlations, and establish a new baseline for the search for the chiral magnetic effect
in heavy ion collisions.

\ifthenelse{\boolean{cms@external}}{}{\clearpage}
\begin{acknowledgments}
We congratulate our colleagues in the CERN accelerator departments for the excellent performance of the LHC and thank the technical and administrative staffs at CERN and at other CMS institutes for their contributions to the success of the CMS effort. In addition, we gratefully acknowledge the computing centers and personnel of the Worldwide LHC Computing Grid for delivering so effectively the computing infrastructure essential to our analyses. Finally, we acknowledge the enduring support for the construction and operation of the LHC and the CMS detector provided by the following funding agencies: BMWFW and FWF (Austria); FNRS and FWO (Belgium); CNPq, CAPES, FAPERJ, and FAPESP (Brazil); MES (Bulgaria); CERN; CAS, MoST, and NSFC (China); COLCIENCIAS (Colombia); MSES and CSF (Croatia); RPF (Cyprus); SENESCYT (Ecuador); MoER, ERC IUT, and ERDF (Estonia); Academy of Finland, MEC, and HIP (Finland); CEA and CNRS/IN2P3 (France); BMBF, DFG, and HGF (Germany); GSRT (Greece); OTKA and NIH (Hungary); DAE and DST (India); IPM (Iran); SFI (Ireland); INFN (Italy); MSIP and NRF (Republic of Korea); LAS (Lithuania); MOE and UM (Malaysia); BUAP, CINVESTAV, CONACYT, LNS, SEP, and UASLP-FAI (Mexico); MBIE (New Zealand); PAEC (Pakistan); MSHE and NSC (Poland); FCT (Portugal); JINR (Dubna); MON, RosAtom, RAS, RFBR and RAEP (Russia); MESTD (Serbia); SEIDI, CPAN, PCTI and FEDER (Spain); Swiss Funding Agencies (Switzerland); MST (Taipei); ThEPCenter, IPST, STAR, and NSTDA (Thailand); TUBITAK and TAEK (Turkey); NASU and SFFR (Ukraine); STFC (United Kingdom); DOE and NSF (USA).

\hyphenation{Rachada-pisek} Individuals have received support from the Marie-Curie program and the European Research Council and Horizon 2020 Grant, contract No. 675440 (European Union); the Leventis Foundation; the A. P. Sloan Foundation; the Alexander von Humboldt Foundation; the Belgian Federal Science Policy Office; the Fonds pour la Formation \`a la Recherche dans l'Industrie et dans l'Agriculture (FRIA-Belgium); the Agentschap voor Innovatie door Wetenschap en Technologie (IWT-Belgium); the Ministry of Education, Youth and Sports (MEYS) of the Czech Republic; the Council of Science and Industrial Research, India; the HOMING PLUS program of the Foundation for Polish Science, cofinanced from European Union, Regional Development Fund, the Mobility Plus program of the Ministry of Science and Higher Education, the National Science Center (Poland), contracts Harmonia 2014/14/M/ST2/00428, Opus 2014/13/B/ST2/02543, 2014/15/B/ST2/03998, and 2015/19/B/ST2/02861, Sonata-bis 2012/07/E/ST2/01406; the National Priorities Research Program by Qatar National Research Fund; the Programa Clar\'in-COFUND del Principado de Asturias; the Thalis and Aristeia programs cofinanced by EU-ESF and the Greek NSRF; the Rachadapisek Sompot Fund for Postdoctoral Fellowship, Chulalongkorn University and the Chulalongkorn Academic into Its 2nd Century Project Advancement Project (Thailand); and the Welch Foundation, contract C-1845.
\end{acknowledgments}

\section*{Appendices}

\appendix

\section{General relation of \texorpdfstring{$v_n$}{v[n]} harmonics, two- and three-particle azimuthal correlations}
\label{app:a}

In Section~\ref{sec:intro}, Eq.~(\ref{eq:lcc123}) can be derived in
a way similar to Eq.~(\ref{eq:lcc}), with details which can be found in Ref.~\cite{Bzdak:2012ia}.
Here, a general derivation of Eq.~(\ref{eq:lcc123}) for all higher-order-harmonic correlators is given.

Similar to Eq. (40) in Ref.~\cite{Bzdak:2012ia}, the general relation between
the $n$th order anisotropy harmonic $v_{n}$ and the three-particle correlator
with respect to the $n$th order event plane can be derived starting from,
\begin{linenomath}
\begin{multline}
\label{eq:lcc123_1}
\gamma_{1,n-1;n} \equiv \left\langle \cos(\phi_{\alpha} + (n-1)\phi_{\beta} - n\Psi_\mathrm{n}) \right\rangle  \\
= \frac{\int \rho_2\cos{(\phi_{\alpha}+(n-1)\phi_{\beta}-n\Psi_n) \rd\phi_{\alpha}\,\rd\phi_{\beta}\,\rd x_{\alpha}\,\rd x_{\beta}}}{\int \rho_2 \rd\phi_{\alpha}\,\rd\phi_{\beta}\,\rd x_{\alpha}\,\rd x_{\beta}} \\
= \frac{\int \rho_2\cos{\left(\phi_{\alpha}-\phi_{\beta}+n(\phi_{\beta}-\Psi_n)\right) \rd\phi_{\alpha}\,\rd\phi_{\beta}\,\rd x_{\alpha}\,\rd x_{\beta}}}{\int \rho_2 \rd\phi_{\alpha}\,\rd\phi_{\beta}\,\rd x_{\alpha}\,\rd x_{\beta}},
\end{multline}
\end{linenomath}
where $x$ denotes $(\pt,\eta)$ and $\rd x = \pt\, \rd\pt\, \rd\eta$. $\rho_{2}$ is the two-particle pair density
distribution, which can be expressed in terms of the single-particle density distribution and its underlying
two-particle correlation function (see Section 2 in Ref.~\cite{Bzdak:2012ia}),
\begin{linenomath}
\begin{equation}
\label{eq:lcc123_2}
\rho_2 = \rho(\phi_{\alpha},x_{\alpha})\rho(\phi_{\beta},x_{\beta})\left [ 1 + C(\phi_{\alpha},\phi_{\beta},x_{\alpha},x_{\beta}) \right ].
\end{equation}
\end{linenomath}

In presence of collective anisotropy flow, the single-particle azimuthal distribution can be
expressed in terms of a Fourier series with respect to the event plane of the corresponding order,
\begin{linenomath}
\begin{equation}
\label{eq:lcc123_3}
\rho(\phi,x) = \frac{\rho_0(x)}{2\pi}\left [ 1 + \sum_{n=1}^{\infty} nv_\mathrm{n}(x)\cos{n(\phi - \Psi_n)} \right ],
\end{equation}
\end{linenomath}
where $\rho_0(x)$ depends on \pt and $\eta$ only.

The two-particle correlation function $C$
describes intrinsic correlations that are insensitive to the event plane $\Psi_n$, but only involve azimuthal angle difference $\Delta\phi = \phi_{\alpha}-\phi_{\beta}$. It can be also expanded
in Fourier series~\cite{Bzdak:2012ia},
\begin{linenomath}
\begin{equation}
\label{eq:lcc123_4}
C(\Delta\phi,x_{\alpha},x_{\beta}) = \sum^{\infty}_{n=1} a_{n}(x_{\alpha},x_{\beta})\cos{(n\Delta\phi)},
\end{equation}
\end{linenomath}
where $a_{n}(x_{\alpha},x_{\beta})$ is the two-particle Fourier coefficient. By definition,
$a_{1}(x_{\alpha},x_{\beta})$ is equal to the two-particle
correlator $\delta (x_{\alpha},x_{\beta})$, introduced in Section~\ref{sec:intro},
as a function of $x_{\alpha}$ and $x_{\beta}$ (i.e., \pt and $\eta$ of both particles).

Therefore,
we substitute Eqs.~(\ref{eq:lcc123_4}) and~(\ref{eq:lcc123_2}) into~(\ref{eq:lcc123_1}) and obtain,
\begin{linenomath}
\begin{multline}
\label{eq:lcc123_5}
\gamma_{1,n-1;n} = \frac{1}{2N^{2}}\int \rho_0(x_{\alpha})\rho_0(x_{\beta})a_1(x_{\alpha},x_{\beta}) \\
\left [ v_{n}(x_{\alpha})
+v_{n}(x_{\beta}) \right ]\rd x_{\alpha}\,\rd x_{\beta} \\
= \frac{1}{2N^{2}}\int \rho_0(x_{\alpha})\rho_0(x_{\beta})\delta(x_{\alpha},x_{\beta}) \\
\left [ v_{n}(x_{\alpha}) +v_{n}(x_{\beta}) \right ]\rd x_{\alpha}\,\rd x_{\beta}
\end{multline}
\end{linenomath}
where $N = \int \rho_0(x) dx$. This is the general equation explaining
why a nonzero two-particle correlation $\delta(x_{\alpha},x_{\beta})$ plus
an anisotropy flow of $n$th order $v_n(x)$ contribute to the three-particle correlator, $\gamma_{1,n-1;n}$.

Therefore, this general form of $\gamma_{1,n-1;n}$ can be applied to any order $n$ and decomposed into the two-particle correlator $\delta$ and the $n$th order harmonic $v_{n}$, where $n=2$ and 3 are studied in detail in Section~\ref{subsec:results_mixed}.

\section{Supporting results of the event shape engineering method}
\label{app:b}

\begin{figure}[thb]
\centering
  \includegraphics[width=\cmsFigWidth]{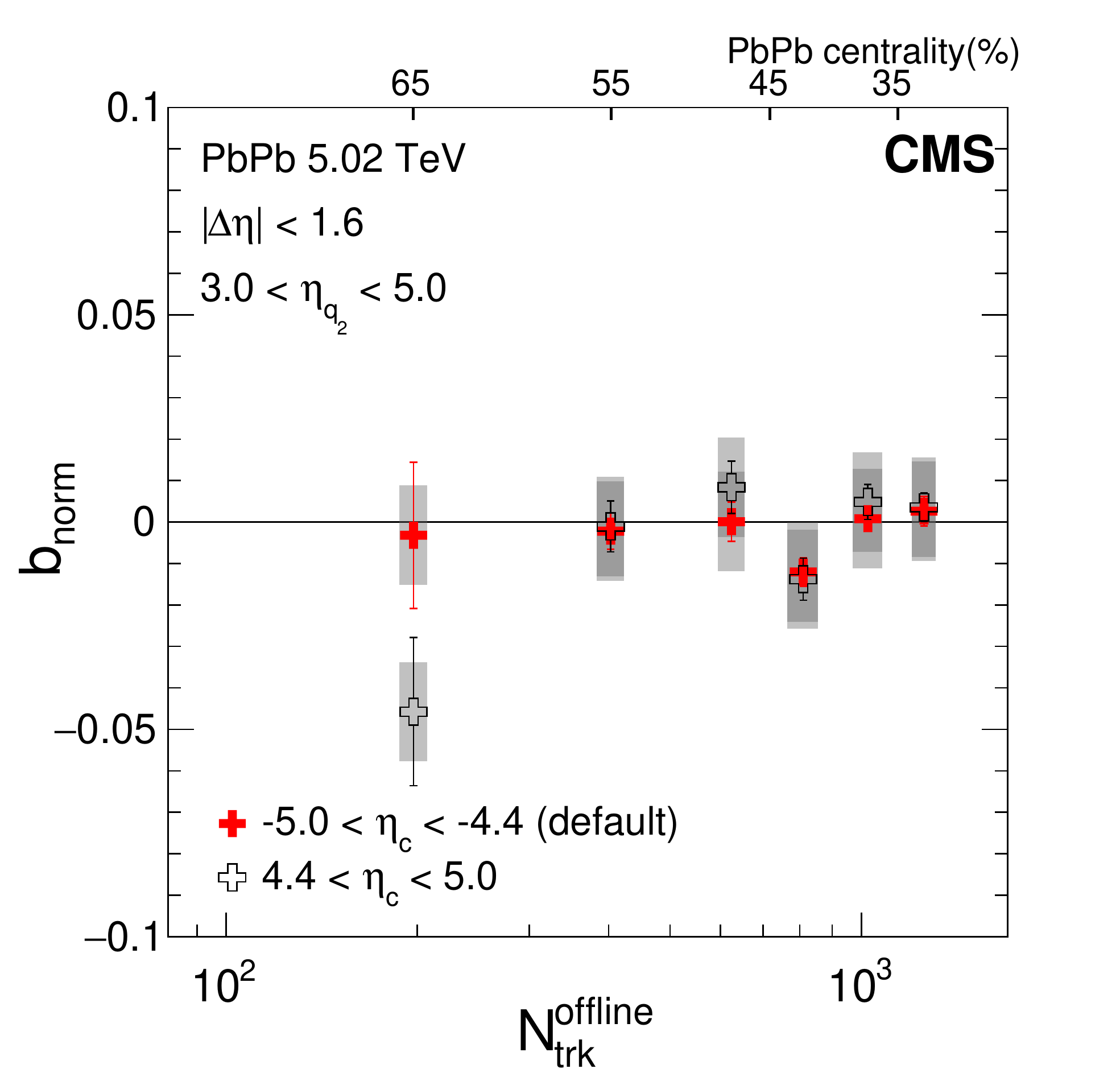}
  \caption{ \label{fig11_compare} The intercepts $b_{\text{norm}}$ of $v_2$-independent $\gamma_{112}$
  correlator component using particle $c$ from HF$+$ and HF$-$ data, averaged over $|\Delta\eta|<1.6$, are shown
  as a function of \noff in \PbPb collisions at $\rootsNN = 5.02$\TeV. Statistical and systematic uncertainties
are indicated by the error bars and shaded regions, respectively. }
\end{figure}

As stated in Section~\ref{subsec:analysis_ese}, the $Q_{2}$ vector is calculated using one side of the HF detector within the $\eta$ range of 3 to 5 units. The default result in Section~\ref{subsec:results_ese} presents the $\Delta\gamma_{112}$ as a function of $v_2$, where the particle $c$ in the $\gamma_{112}$ correlator corresponds to the $\eta$ range $-$5.0 to $-$4.4. However, the results are found to be independent of where the particle $c$ is reconstructed, as it is shown in Fig.~\ref{fig11_compare}.

\begin{figure}[thb]
\centering
  \includegraphics[width=\cmsFigWidth]{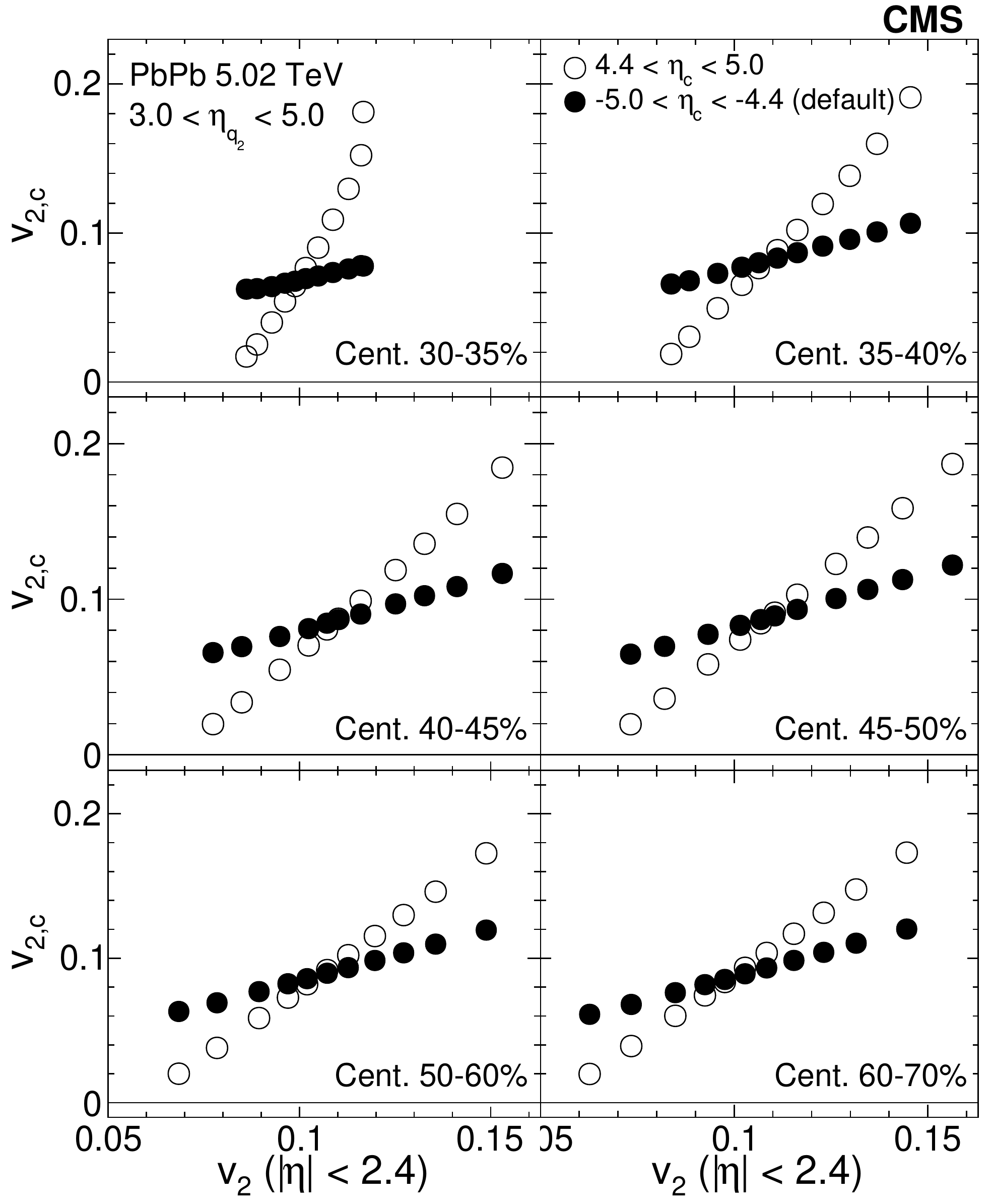}
  \caption{ \label{fig13_a} The $v_{2,c}$ using particle $c$ from HF$+$ and HF$-$ data
  are shown as a function of $v_{2}$ in the tracker region ($|\eta|<2.4$)
  in \PbPb collisions at $\rootsNN = 5.02$\TeV. }
\end{figure}

\begin{figure}[thb]
\centering
  \includegraphics[width=\cmsFigWidth]{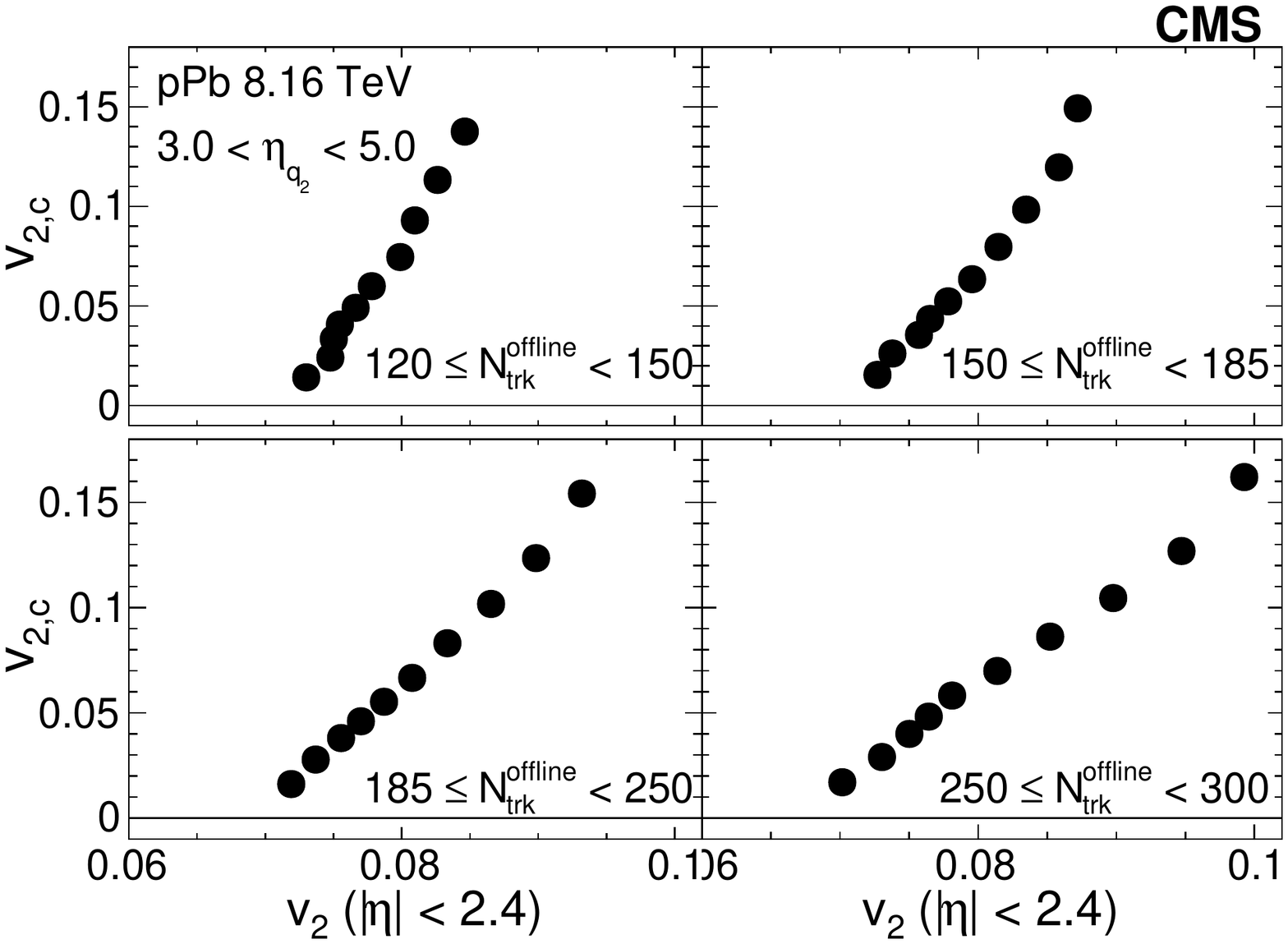}
  \caption{ \label{fig13_b} The $v_{2,c}$ using particle $c$ from the \Pb-going side of the HF ($4.4 < \eta < 5.0$) data are shown as a function of $v_{2}$ in the tracker region ($|\eta|<2.4$)
  in \pPb collisions at $\rootsNN = 8.16$\TeV. }
\end{figure}

In Figs.~\ref{fig13_a} and~\ref{fig13_b}, the denominators of Eq.~(\ref{3pcorrelatorQVector}), $v_{2,c}$, for different $Q_{2}$ classes with respect to HF$+$ and HF$-$ in \PbPb collisions at
$\rootsNN = 5.02$\TeV, and the \Pb-going side of the HF in \pPb collisions at
8.16\TeV, are shown as a function of $v_{2}$ in the tracker region. Here $v_{2,c}$ is
a measure of elliptic anisotropy of the transverse energy registered in the HF detectors without
being corrected to the particle-level elliptic flow. It serves as the resolution correction factor
when deriving the three-particle correlators or the $v_2$ values in the tracker region using the
scalar-product method.

\begin{figure}[thb]
\centering
  \includegraphics[width=\cmsFigWidth]{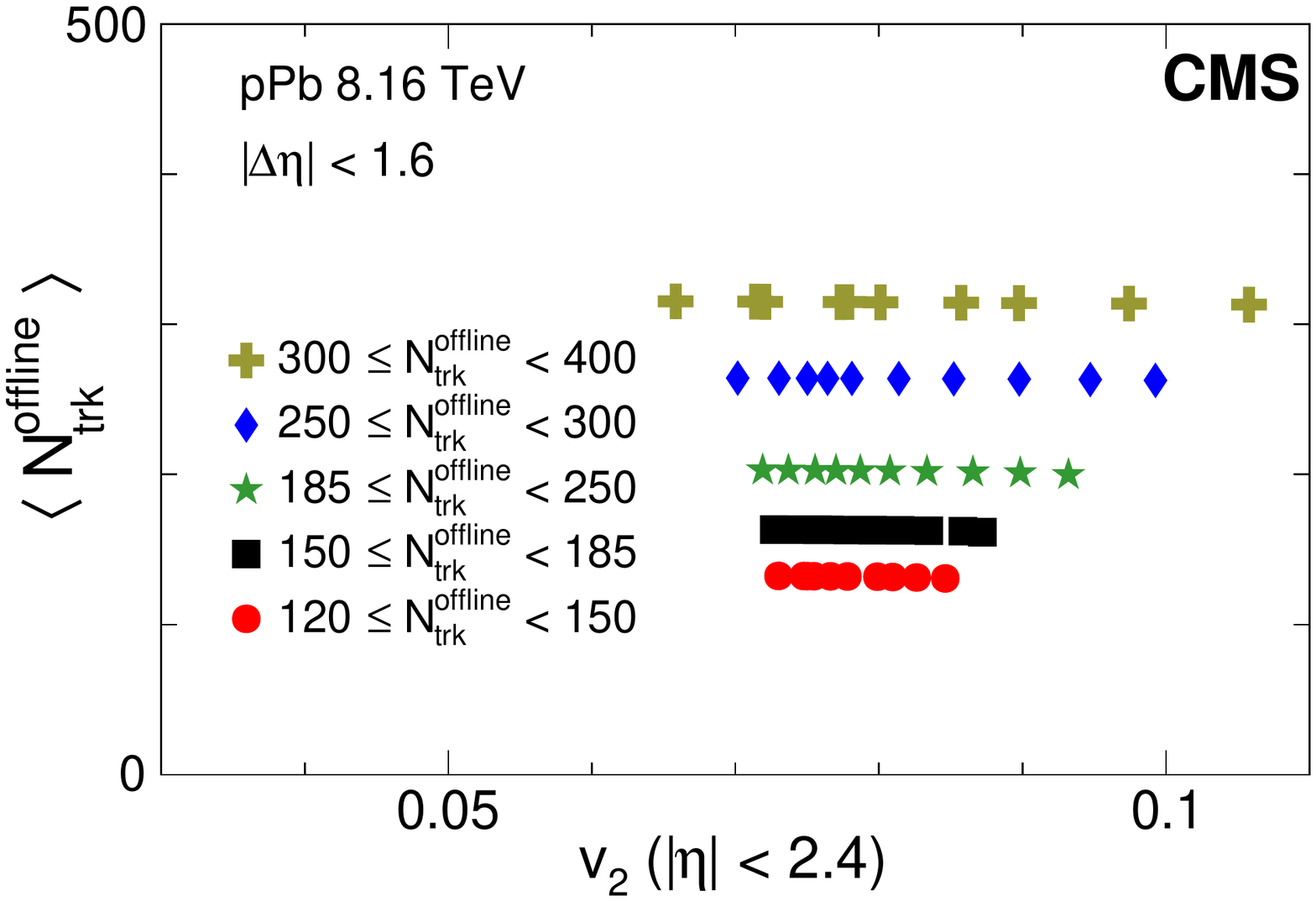}
  \includegraphics[width=\cmsFigWidth]{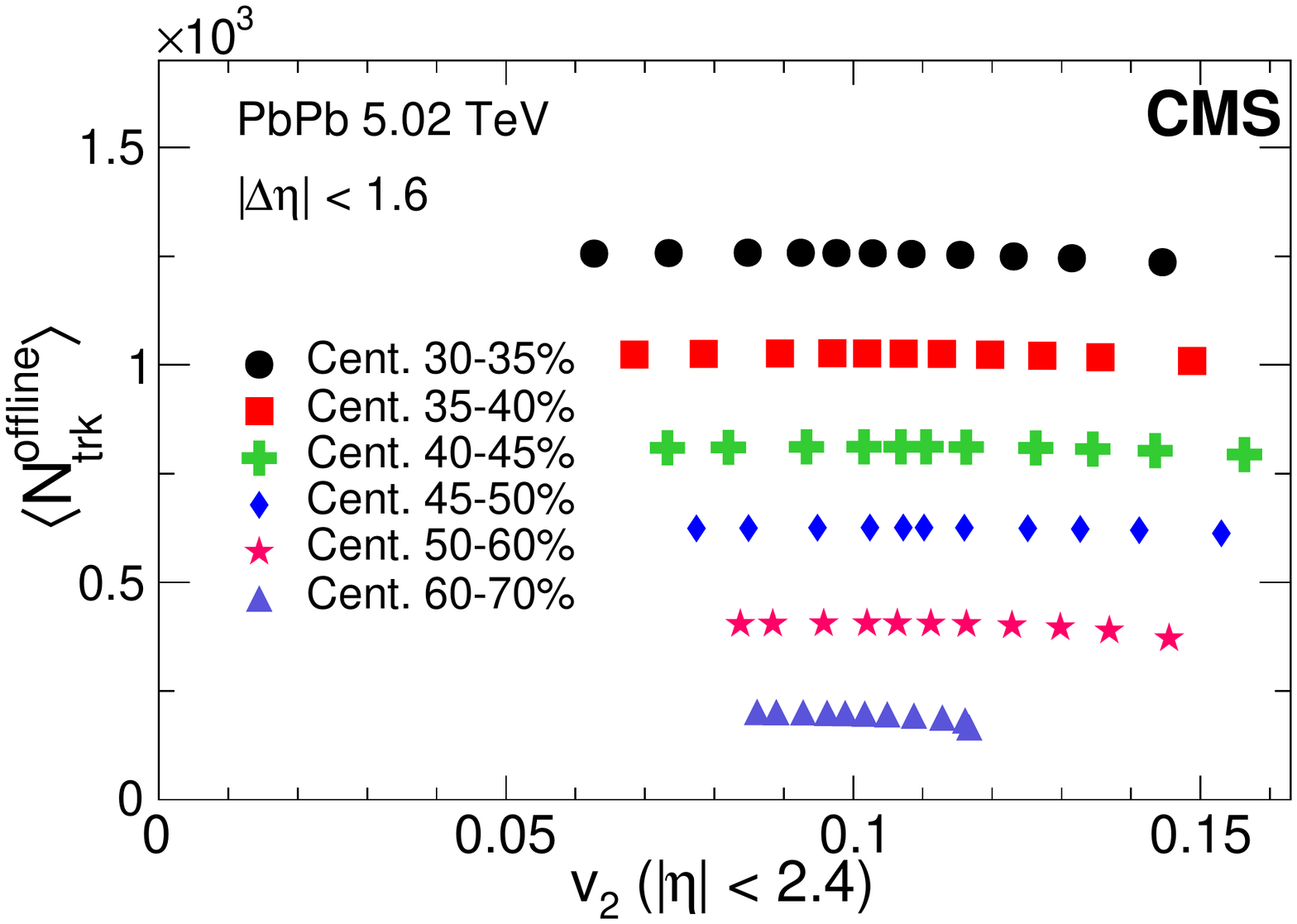}
  \caption{ \label{fig9_ntrk} The average multiplicity \noff as a function of $v_2$ evaluated in each $q_2$ class, for different multiplicity ranges
  in \pPb collisions at $\rootsNN = 8.16$\TeV (upper), and for different centrality
  classes in \PbPb collisions at 5.02\TeV (lower). Statistical uncertainties are invisible on the current scale.
  }
\end{figure}

In Fig.~\ref{fig9_ntrk}, the average \noff is shown as a function of $v_{\mathrm 2}$ in different multiplicity and centrality ranges in \pPb (upper)  and \PbPb collisions (lower), respectively. The average \noff is found to be weakly dependent on $v_2$, but with a slight decreasing trend as $v_2$ increases. Similar to Fig.~\ref{fig9}, the effect at low multiplicities is stronger than that at high multiplicities. Overall, this effect is negligible for the results shown in Section~\ref{subsec:results_ese}.

\section{Three- and two-particle correlator as functions of differential variables in different multiplicity and centrality classes}
\label{app:c}

\begin{figure*}[thb]
\centering
  \includegraphics[width=0.48\textwidth]{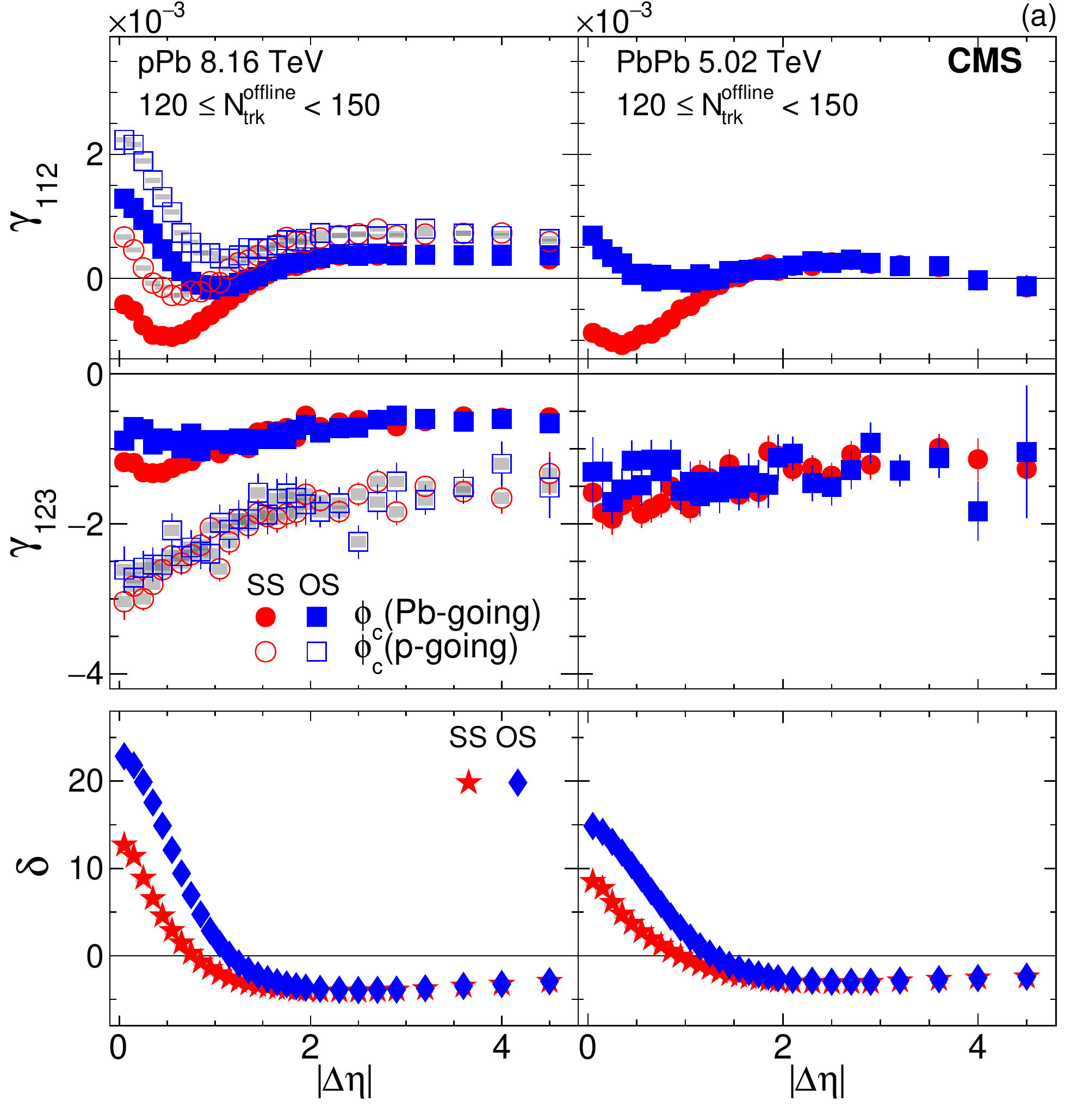}
  \includegraphics[width=0.48\textwidth]{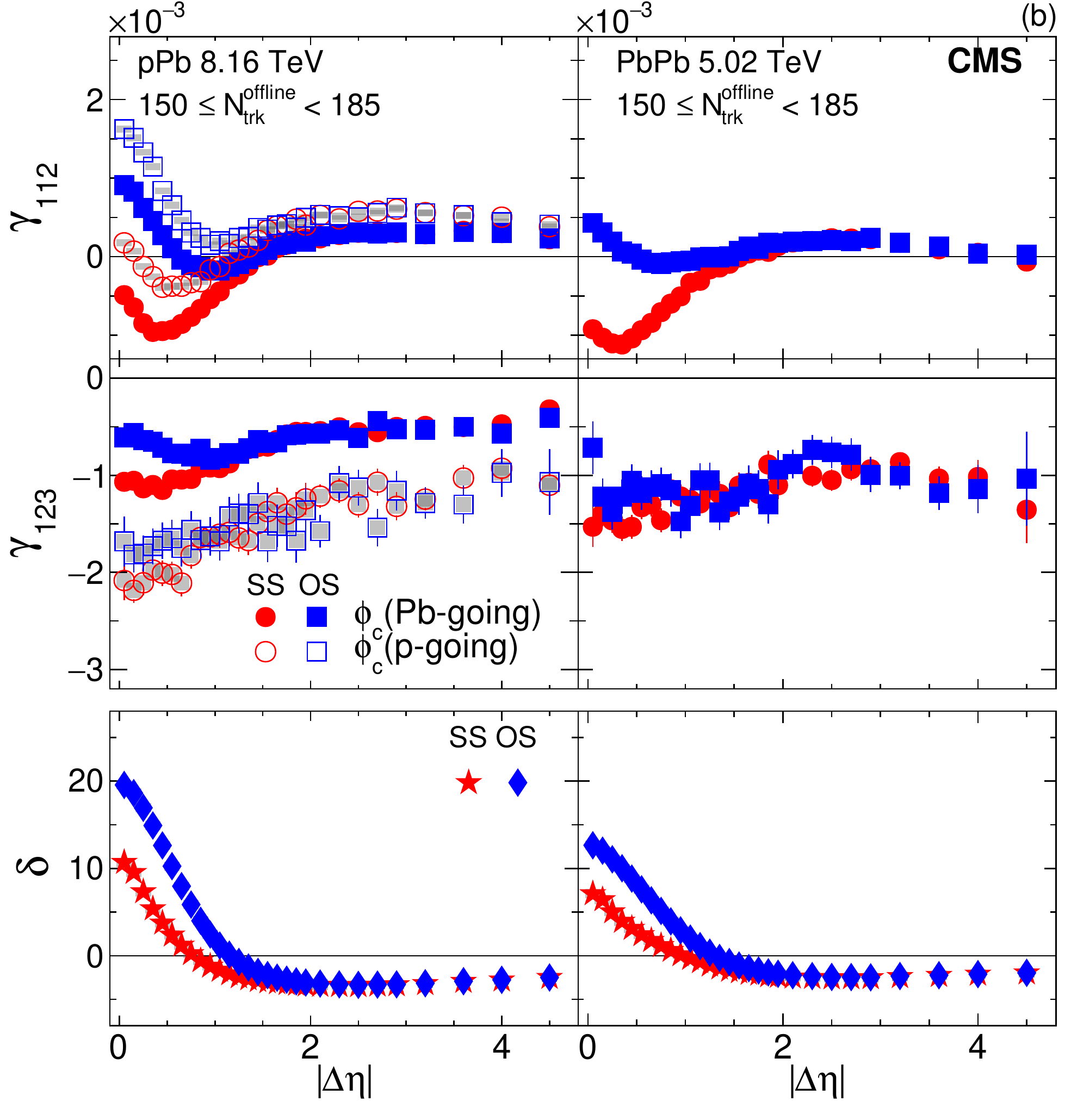}
  \includegraphics[width=0.48\textwidth]{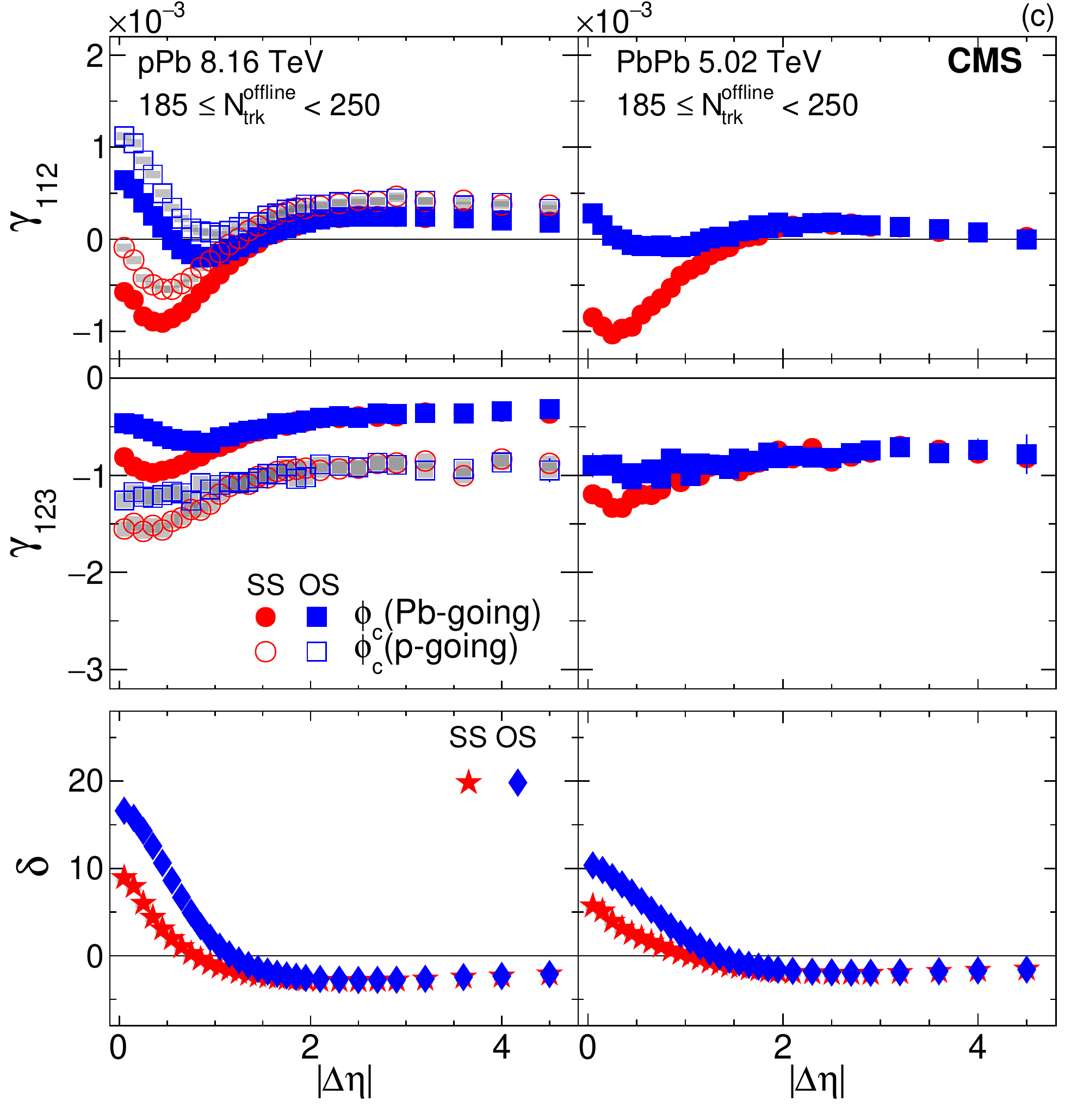}
  \includegraphics[width=0.48\textwidth]{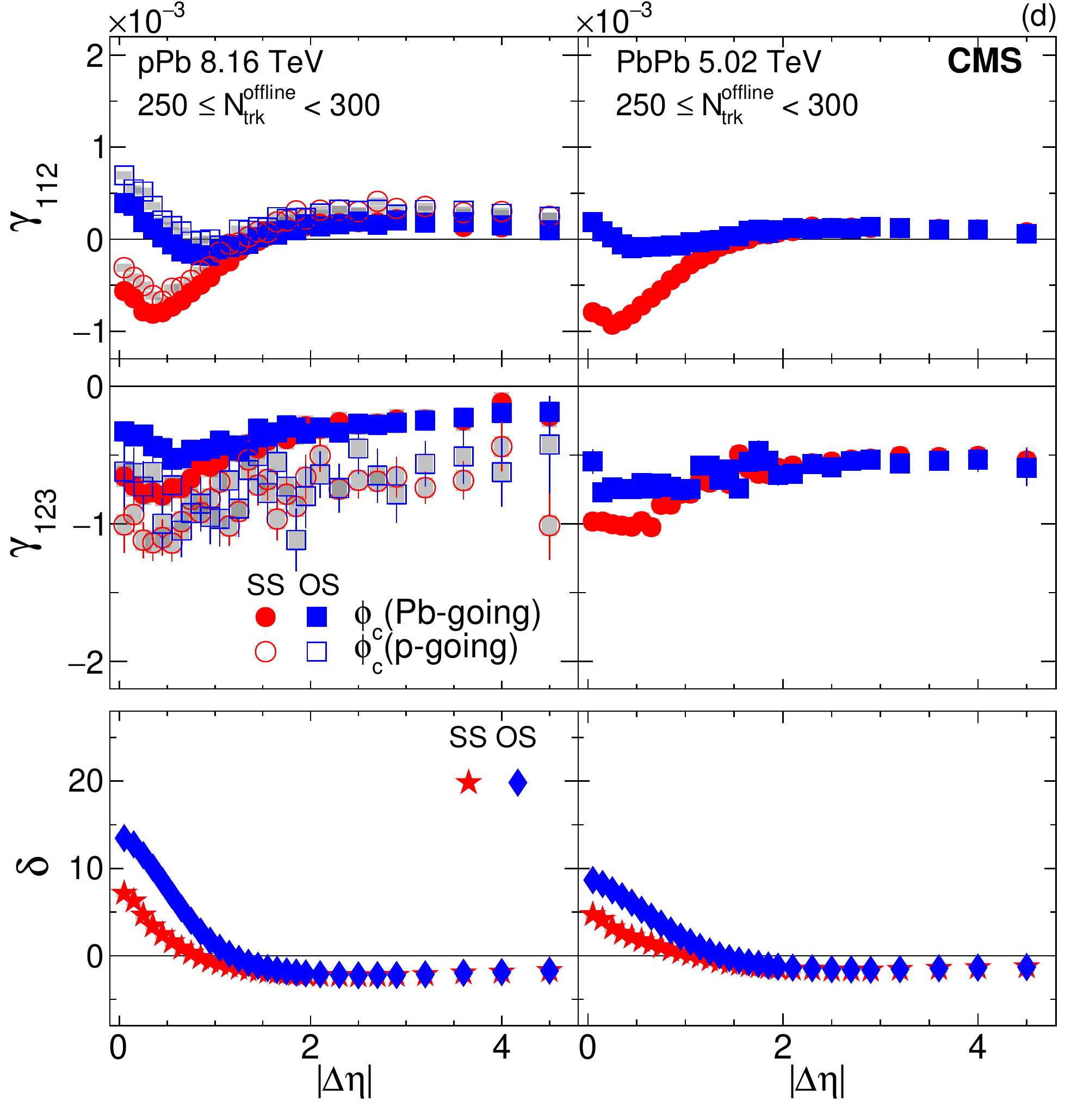}
  \caption{ \label{fig1_appendix_a}
The SS and OS three-particle correlators,
$\gamma_{112}$ (upper) and $\gamma_{123}$ (middle), and two-particle
correlator, $\delta$ (lower), as a function of $\abs{\deta}$
for four multiplicity ranges in \pPb collisions at $\rootsNN = 8.16$\TeV~(left) and
\PbPb collisions at 5.02\TeV~(right). The \pPb results obtained with particle $c$
in \Pb-going (solid markers) and \Pp-going (open markers) sides are
shown separately. The SS and OS two-particle correlators are denoted by different markers for both \pPb and \PbPb collisions. Statistical and systematic uncertainties
are indicated by the error bars and shaded regions, respectively.
   }
\end{figure*}

\begin{figure*}[thb]
\centering
  \includegraphics[width=0.48\textwidth]{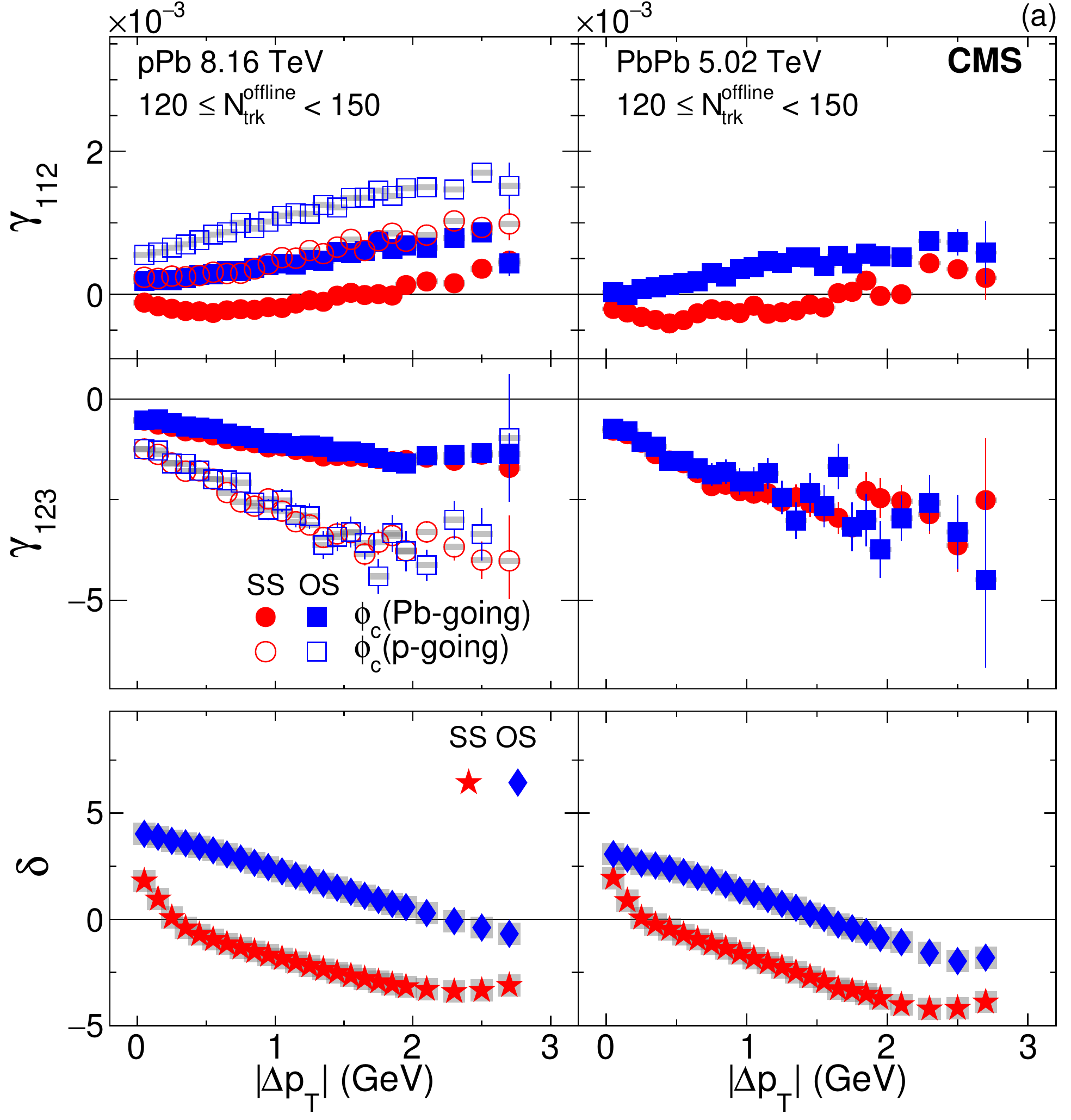}
  \includegraphics[width=0.48\textwidth]{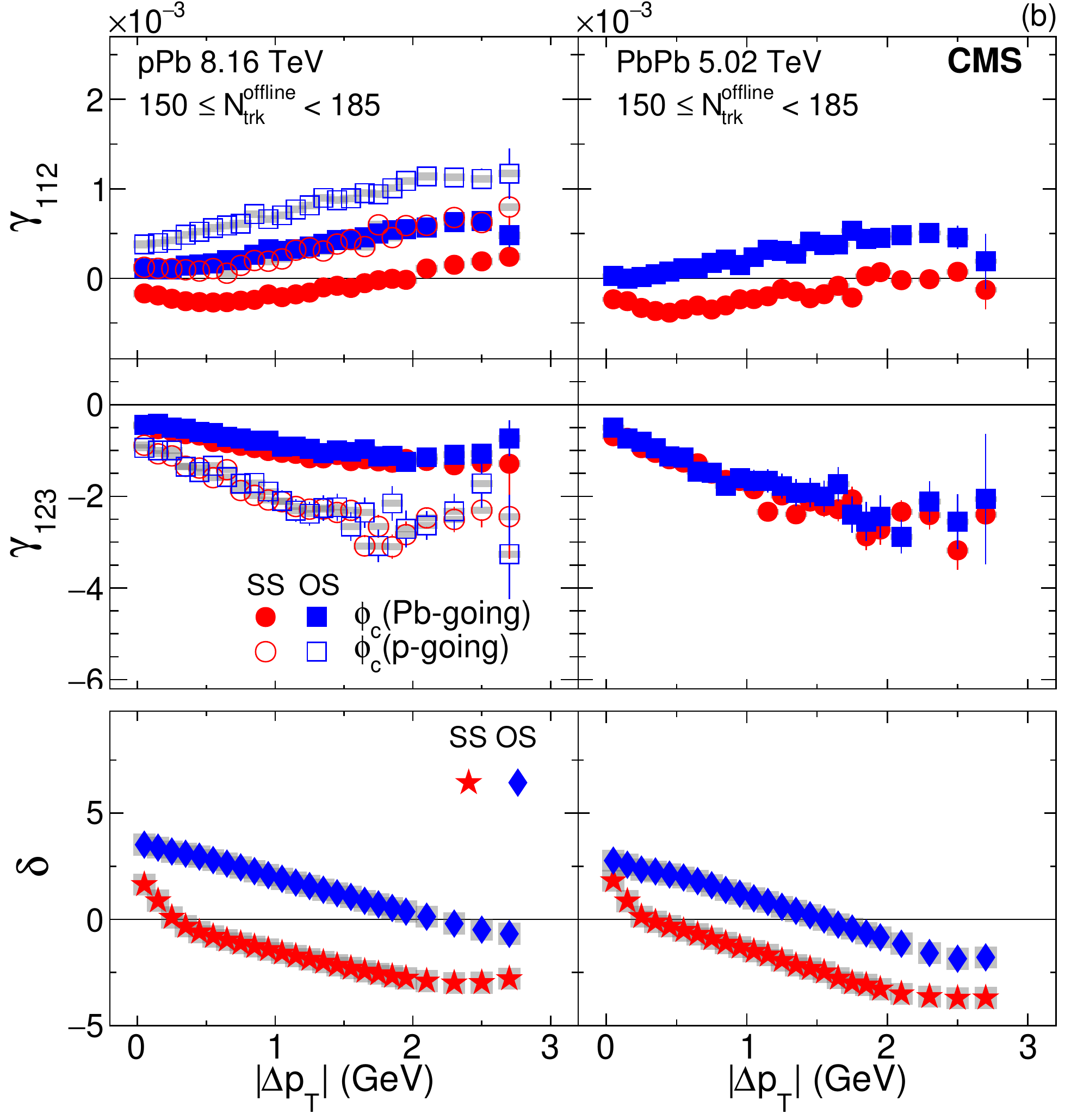}
  \includegraphics[width=0.48\textwidth]{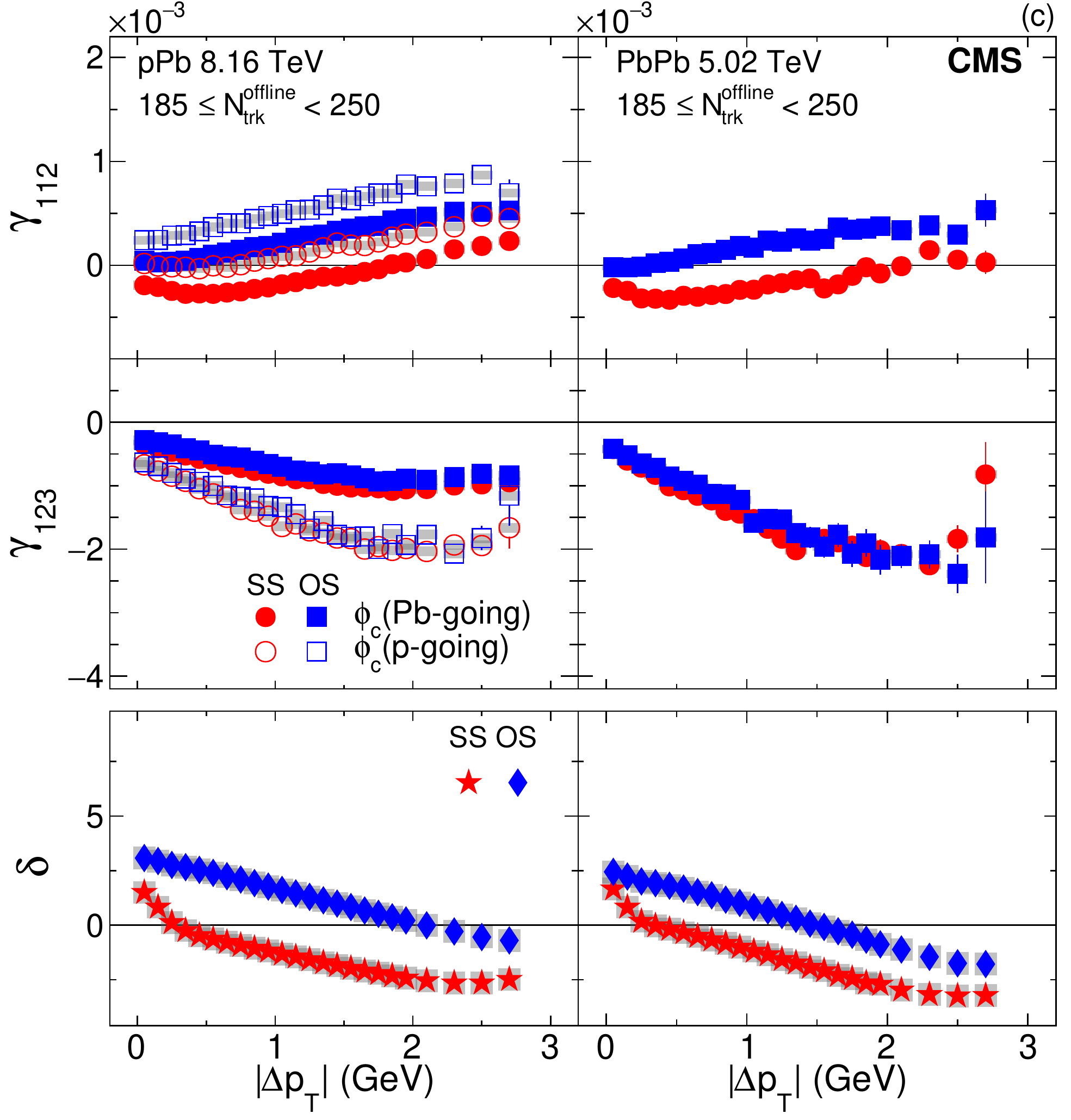}
  \includegraphics[width=0.48\textwidth]{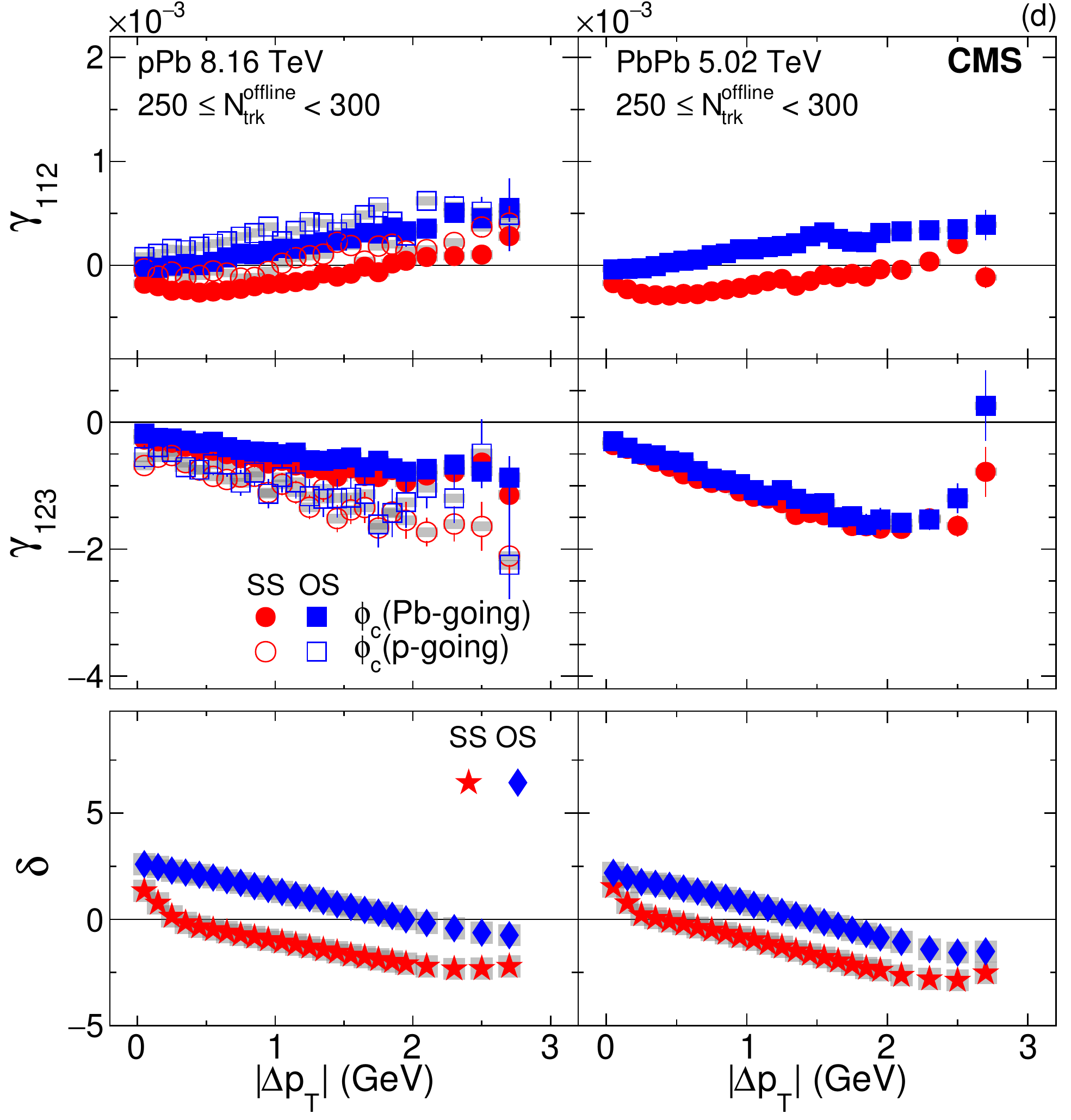}
\caption{ \label{fig1_appendix_b}
The SS and OS three-particle correlators,
$\gamma_{112}$ (upper) and $\gamma_{123}$ (middle), and two-particle
correlator, $\delta$ (lower), as a function of $\abs{ \Delta\pt } $
for four multiplicity ranges in \pPb collisions at $\rootsNN = 8.16$\TeV~(left)
and \PbPb collisions at 5.02\TeV~(right) collisions. The \pPb results
obtained with particle $c$ in \Pb-going (solid markers) and \Pp-going
(open markers) sides are shown separately. The SS and OS two-particle correlators are denoted by different markers for both \pPb and \PbPb collisions. Statistical and
systematic uncertainties are indicated by the error bars and
shaded regions, respectively.
}

\end{figure*}

\begin{figure*}[thb]
\centering
  \includegraphics[width=0.48\textwidth]{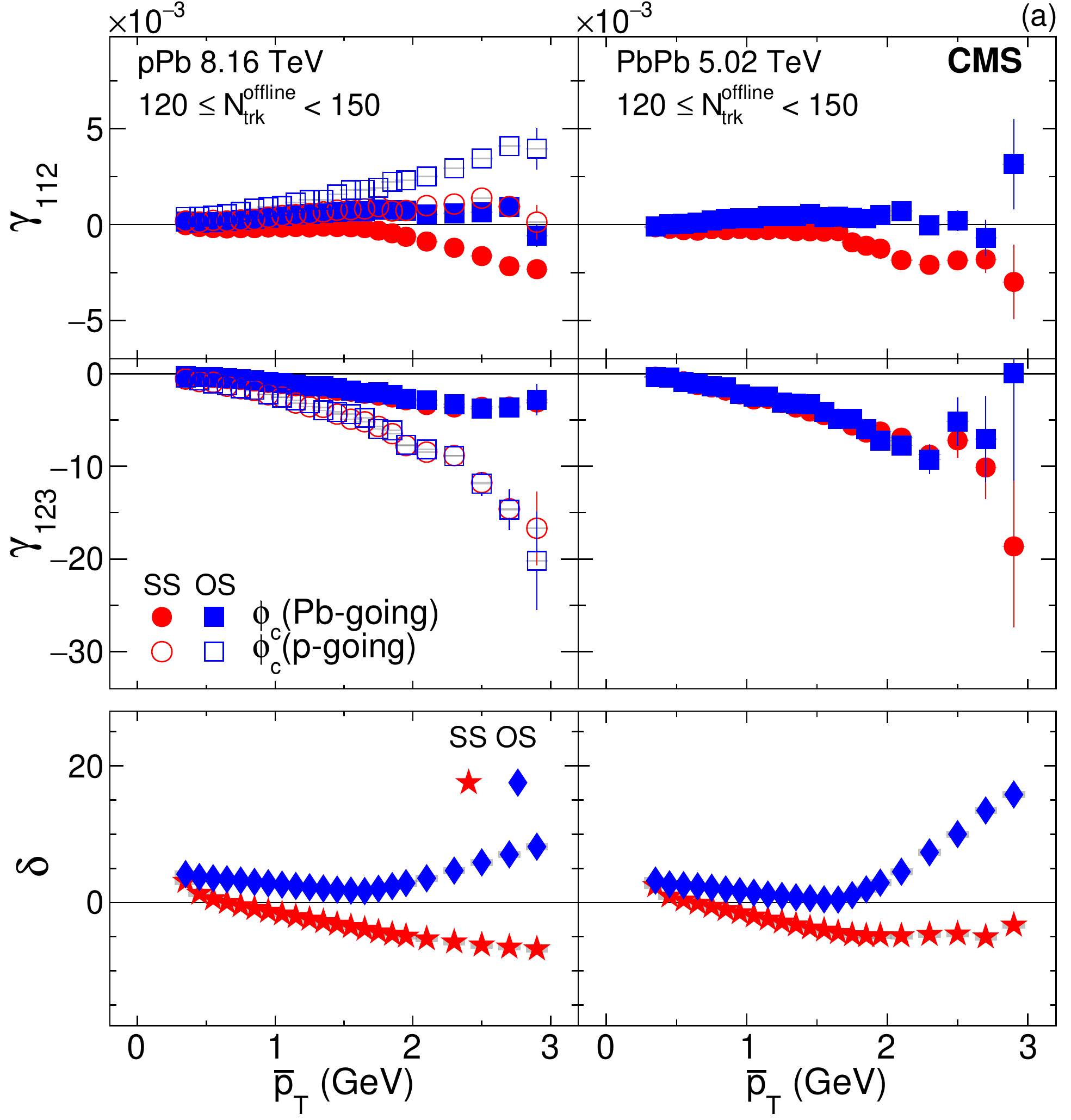}
  \includegraphics[width=0.48\textwidth]{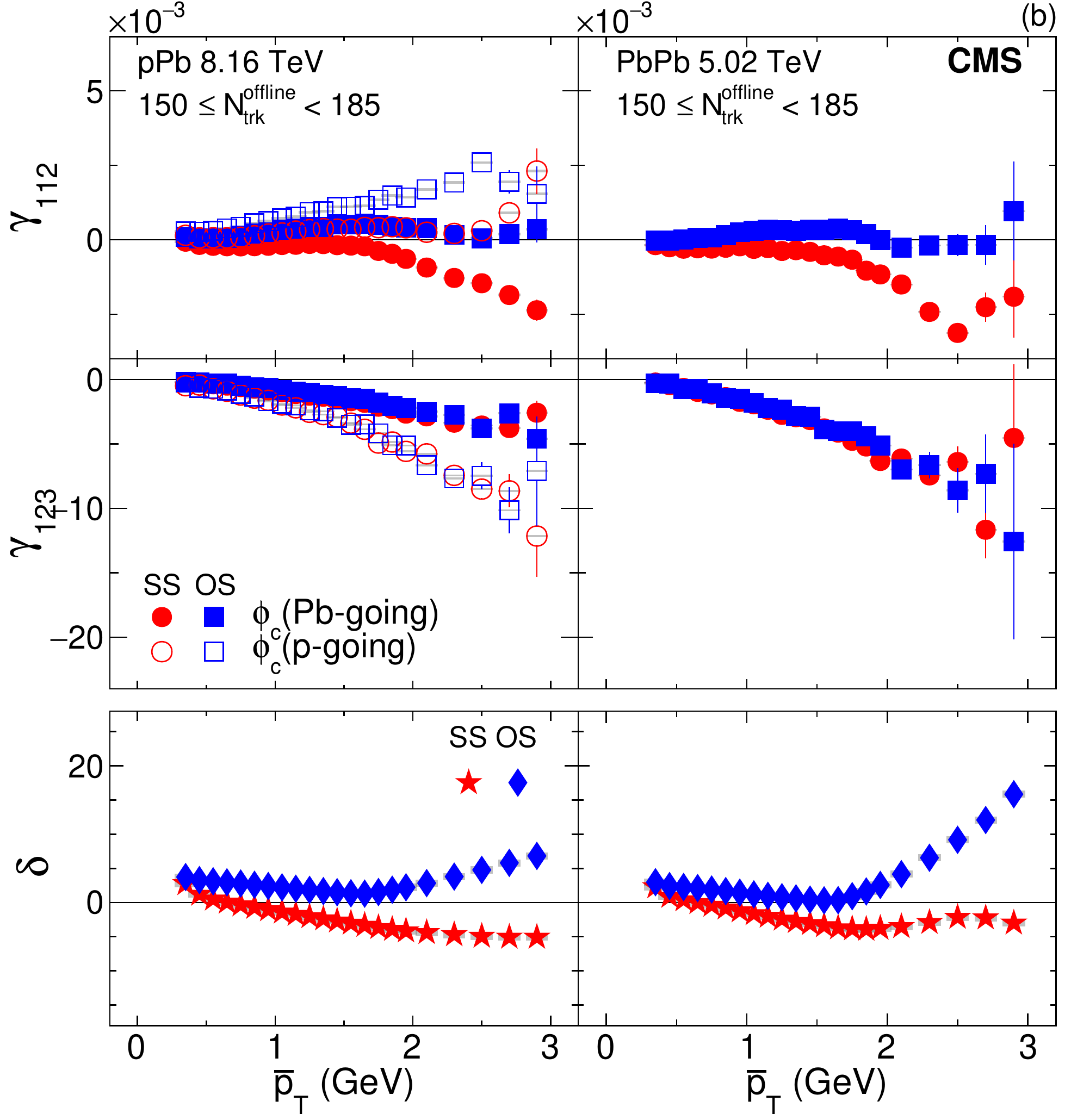}
  \includegraphics[width=0.48\textwidth]{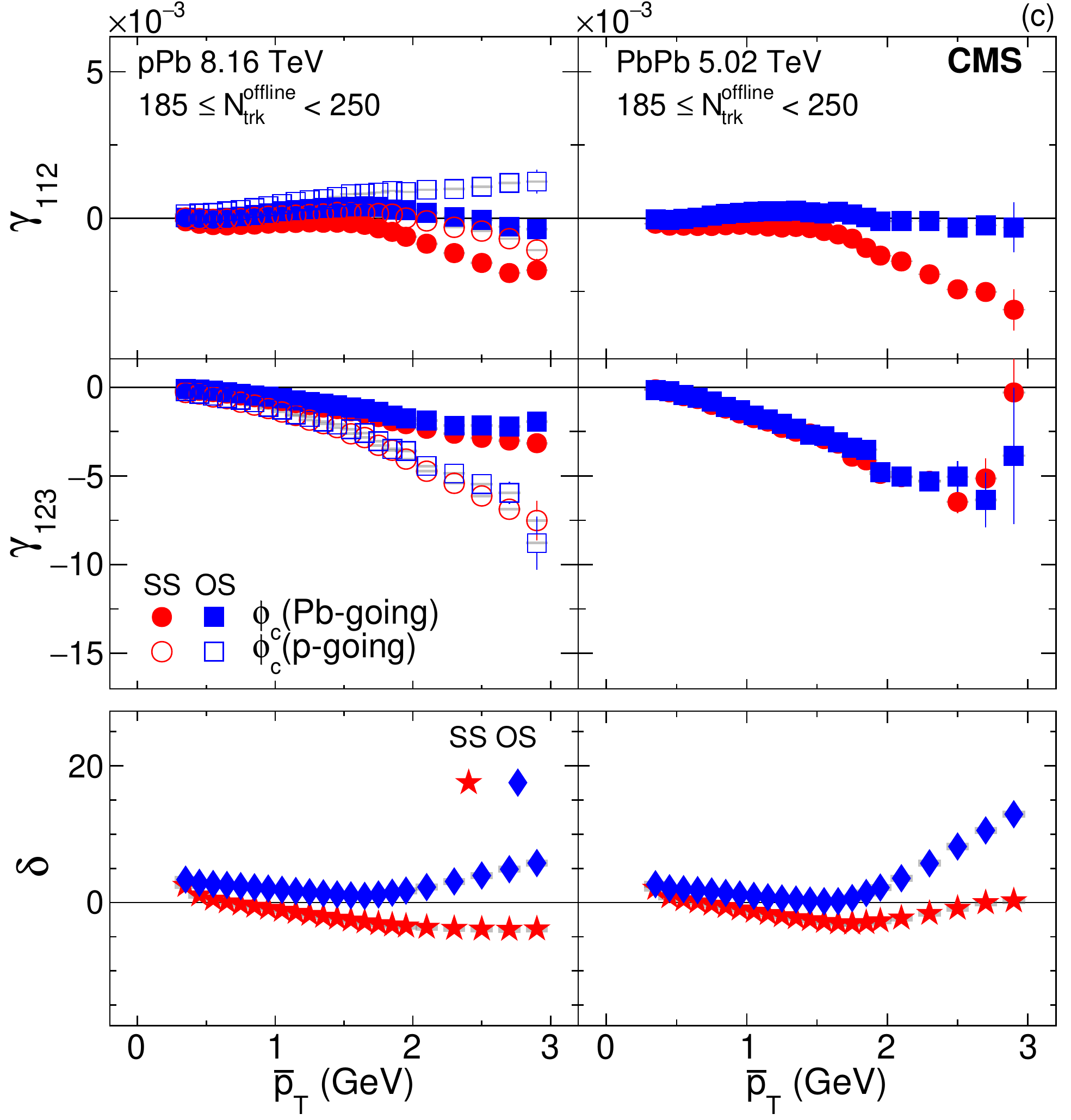}
  \includegraphics[width=0.48\textwidth]{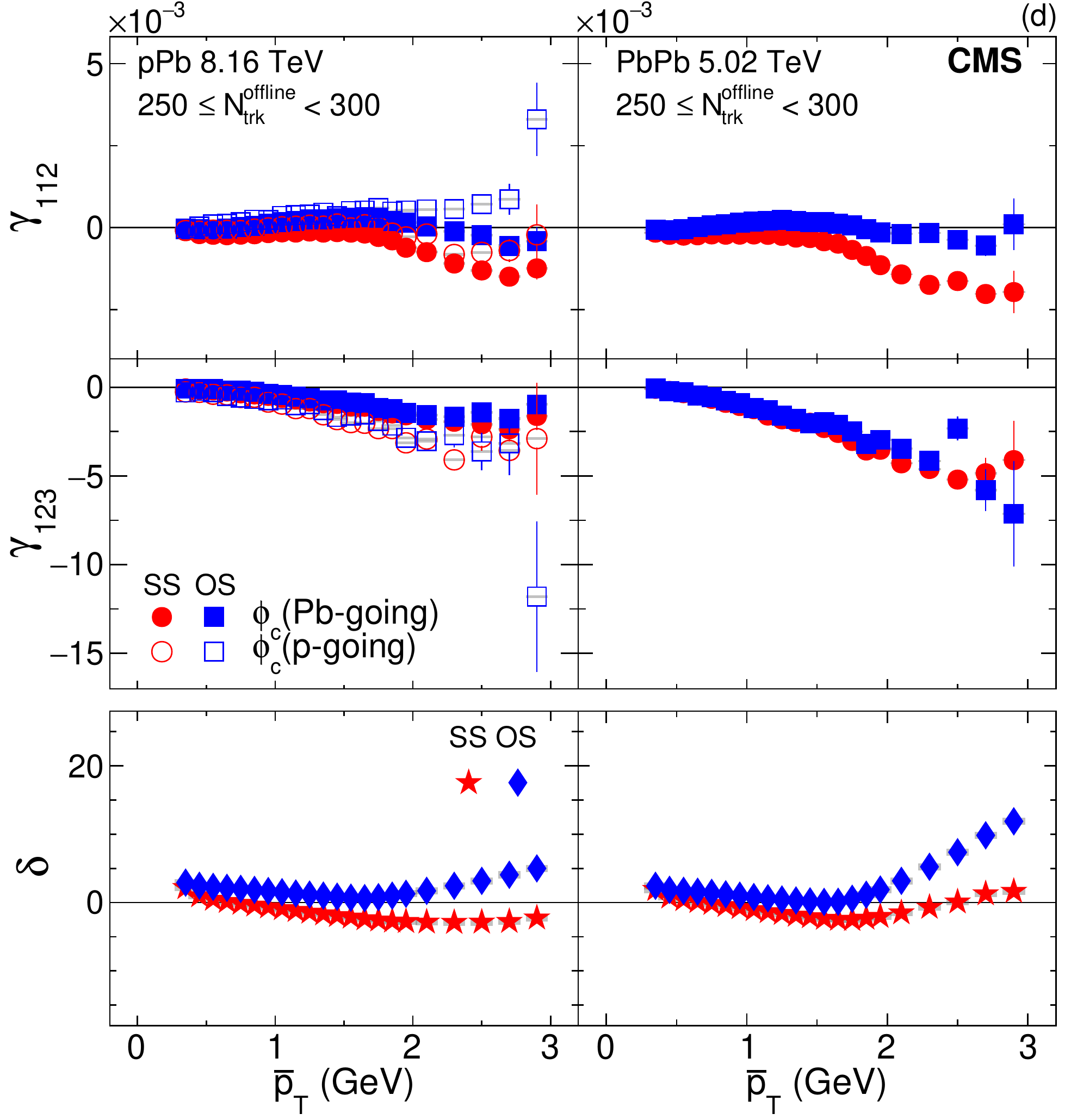}
  \caption{
  \label{fig1_appendix_c}
The SS and OS three-particle correlators,
$\gamma_{112}$ (upper) and $\gamma_{123}$ (middle), and two-particle
correlator, $\delta$ (lower), as a function of $\ptbar$
for four multiplicity ranges in \pPb collisions at $\rootsNN = 8.16$\TeV~(left) and
\PbPb collisions at $\rootsNN = 5.02$\TeV~(right). The \pPb results
obtained with particle $c$ in \Pb-going (solid markers) and \Pp-going (open markers)
sides are shown separately. The SS and OS two-particle correlators are denoted by different markers for both \pPb and \PbPb collisions. Statistical and systematic uncertainties
are indicated by the error bars and shaded regions, respectively.
   }
\end{figure*}

\begin{figure*}[thb]
\centering
  \includegraphics[width=0.32\textwidth]{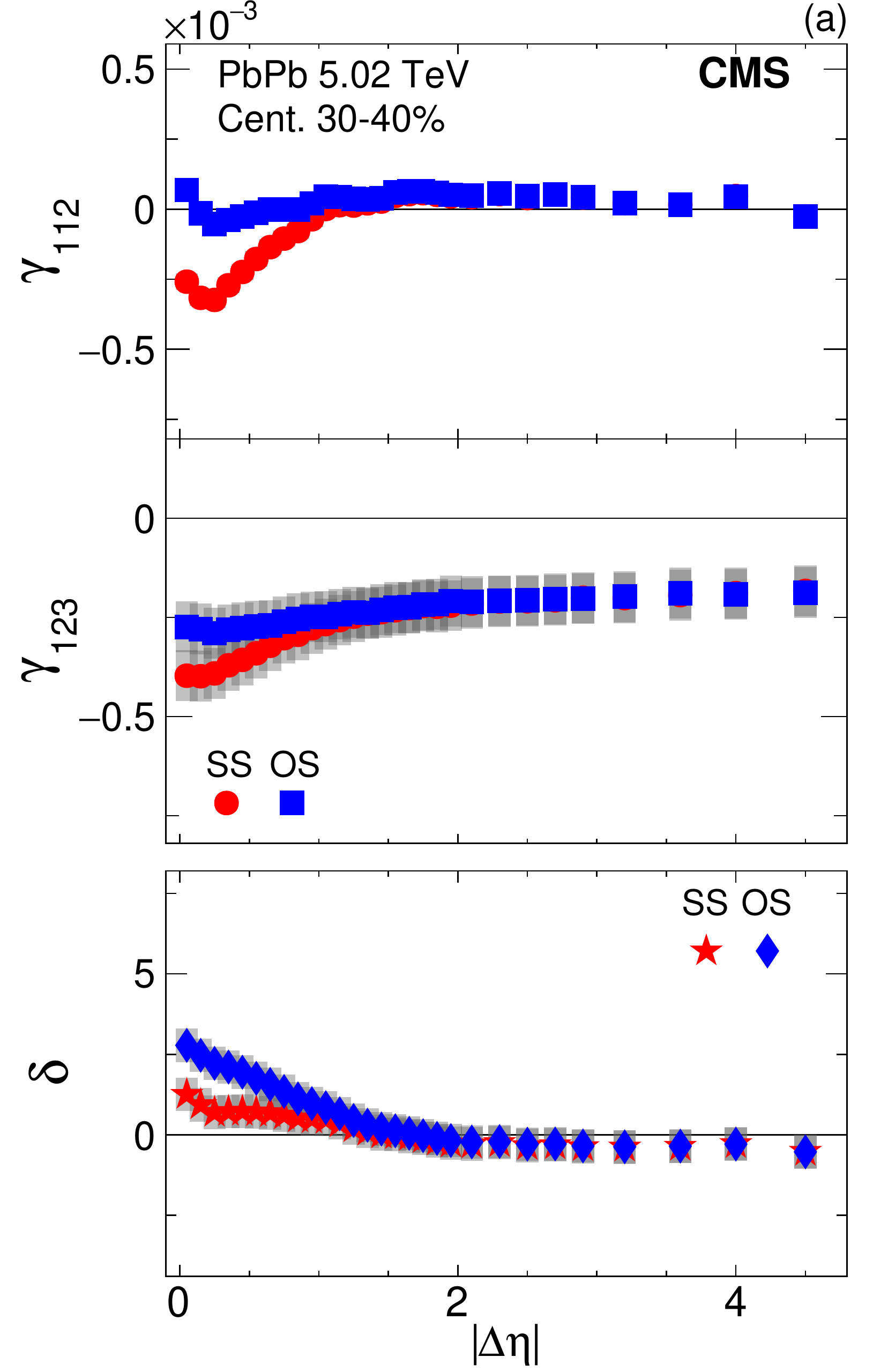}
  \includegraphics[width=0.32\textwidth]{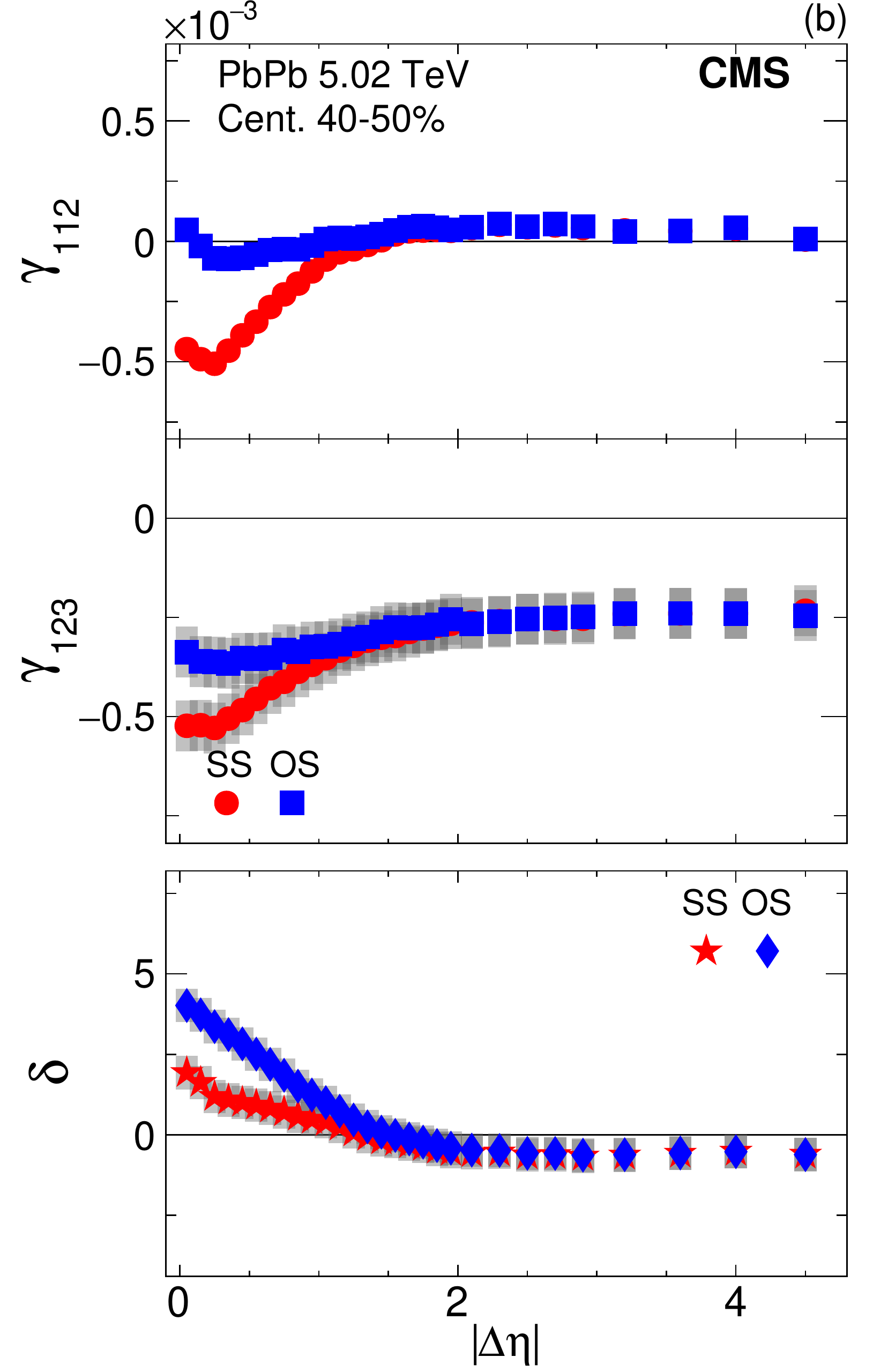}
  \includegraphics[width=0.32\textwidth]{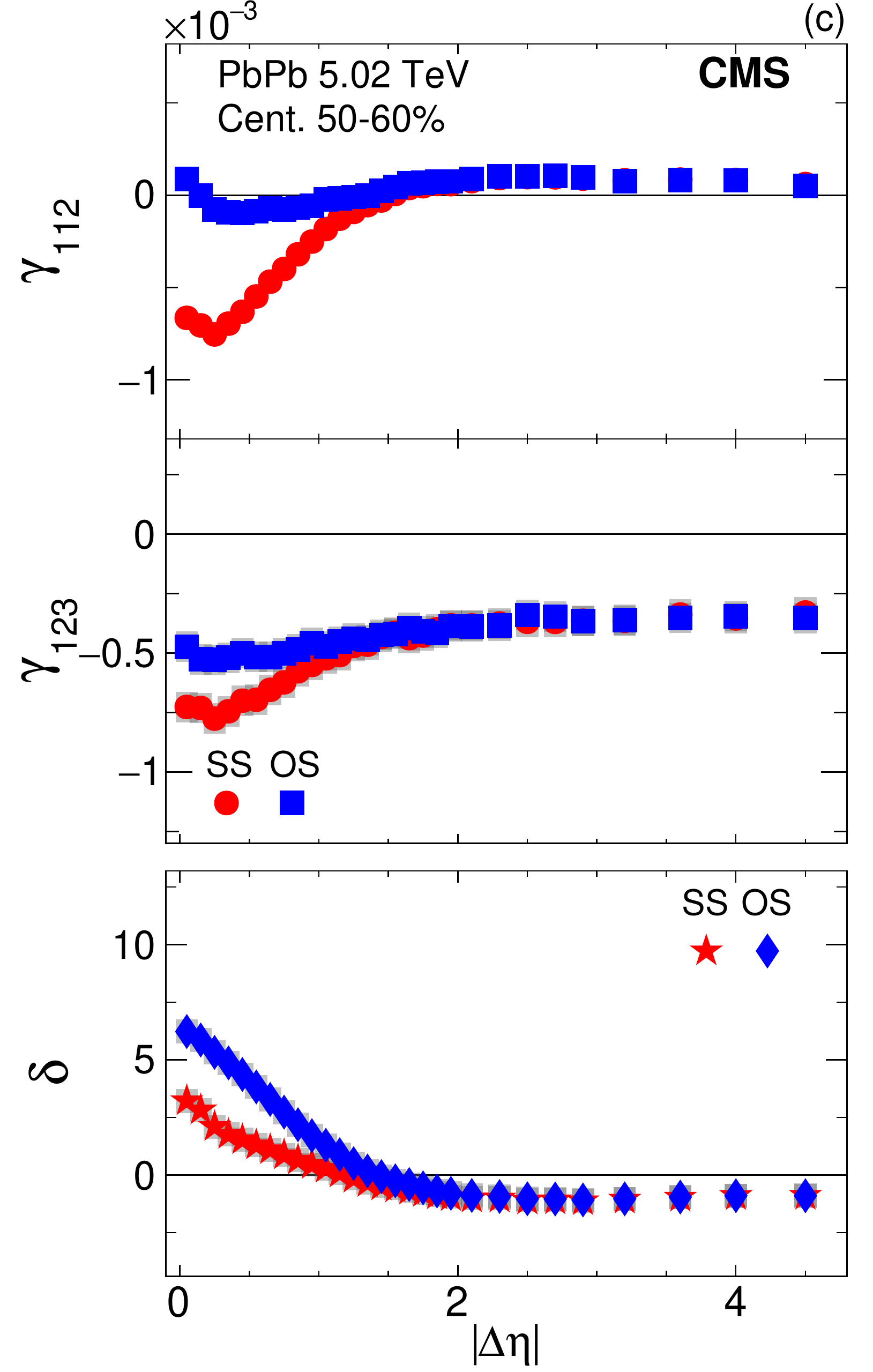}
  \includegraphics[width=0.32\textwidth]{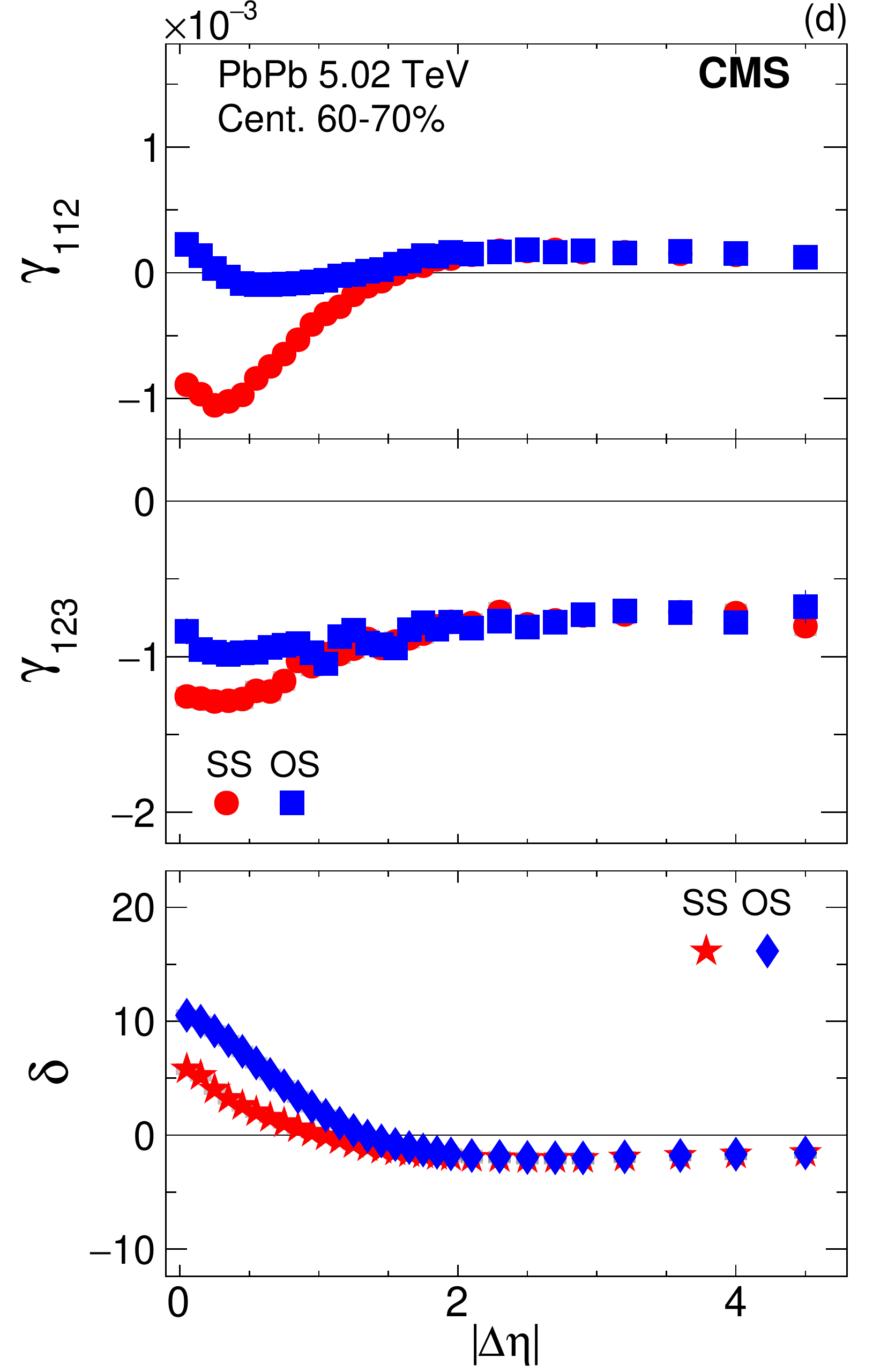}
  \includegraphics[width=0.32\textwidth]{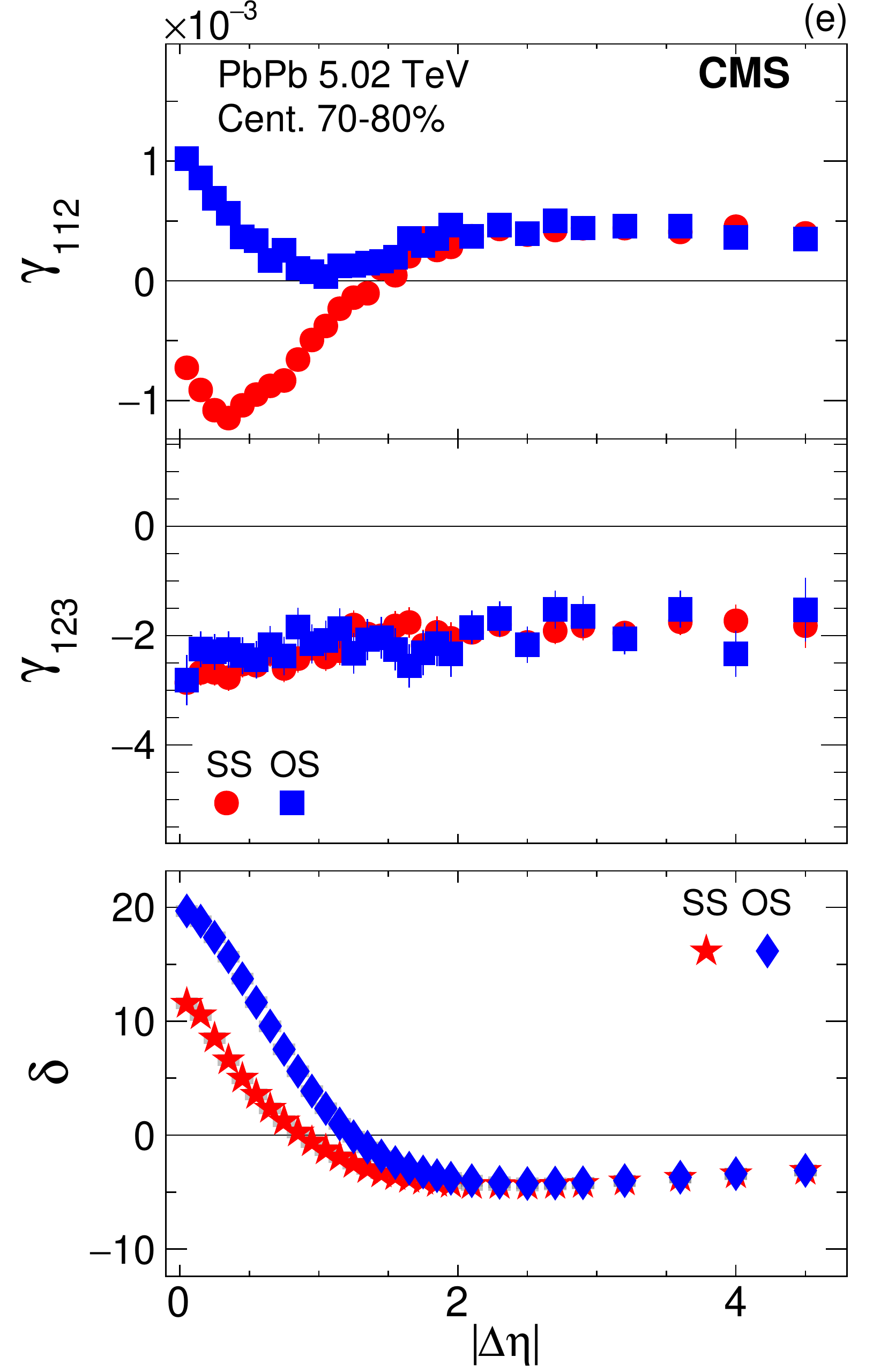}
  \caption{ \label{fig1_Cent_appendix_a}
The SS and OS three-particle correlators,
$\gamma_{112}$ (upper)  and $\gamma_{123}$ (middle), and two-particle
correlator, $\delta$ (lower), as a function of $\abs{\deta}$
for five centrality classes in \PbPb collisions at 5.02\TeV. The SS and OS two-particle correlators are denoted by different markers. Statistical and systematic uncertainties
are indicated by the error bars and shaded regions, respectively.
   }
\end{figure*}

\begin{figure*}[thb]
\centering
  \includegraphics[width=0.32\textwidth]{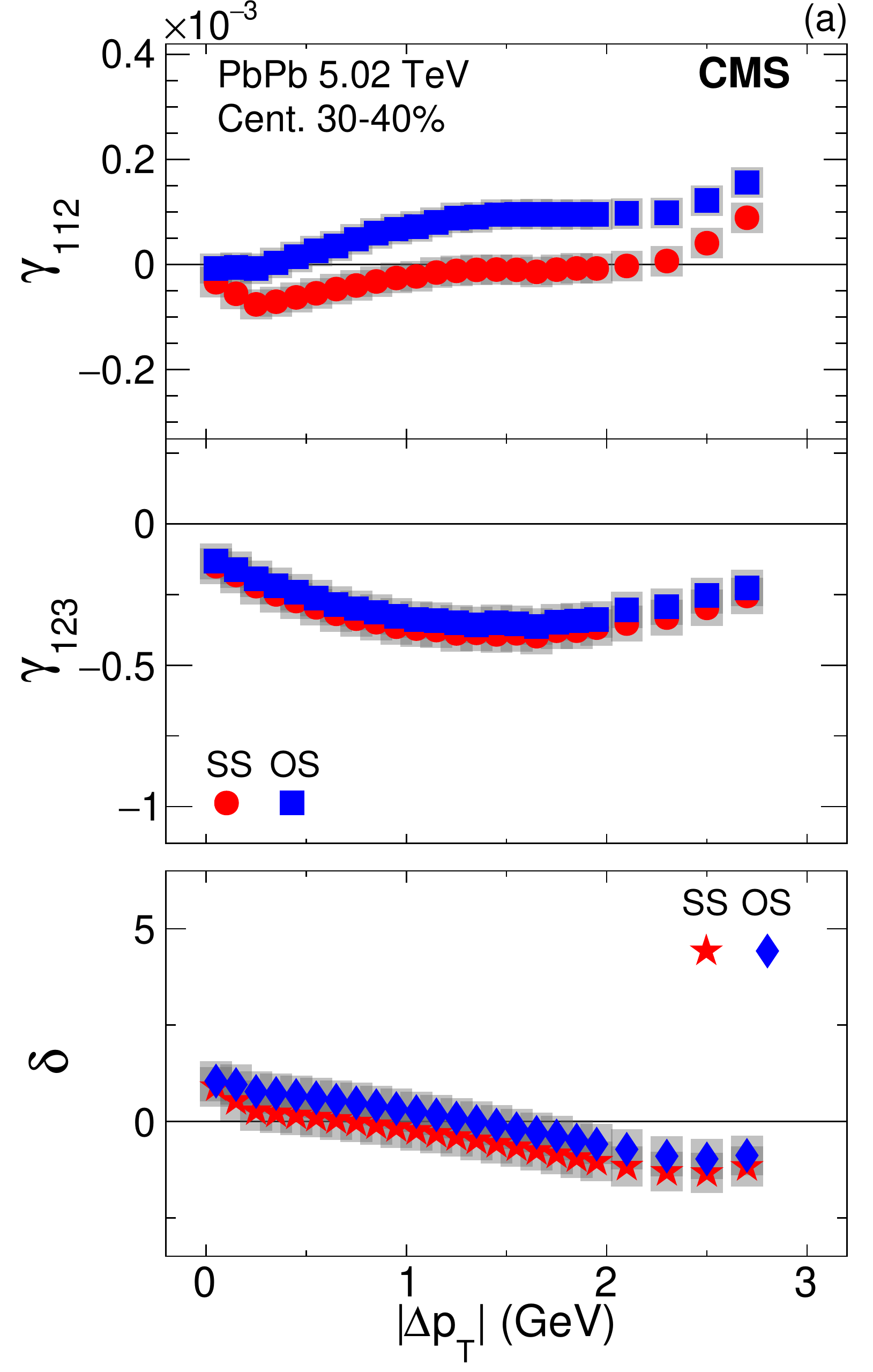}
  \includegraphics[width=0.32\textwidth]{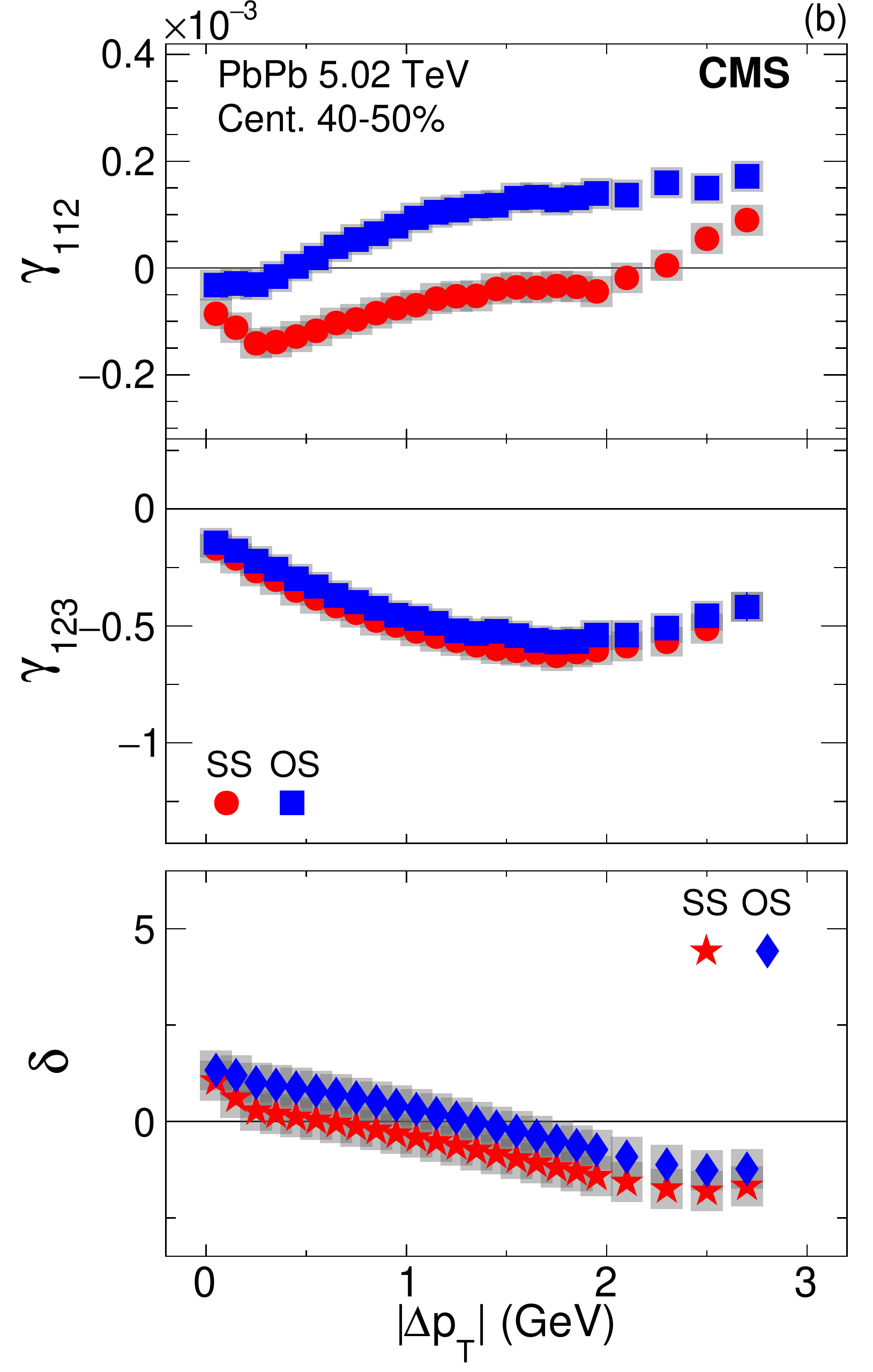}
  \includegraphics[width=0.32\textwidth]{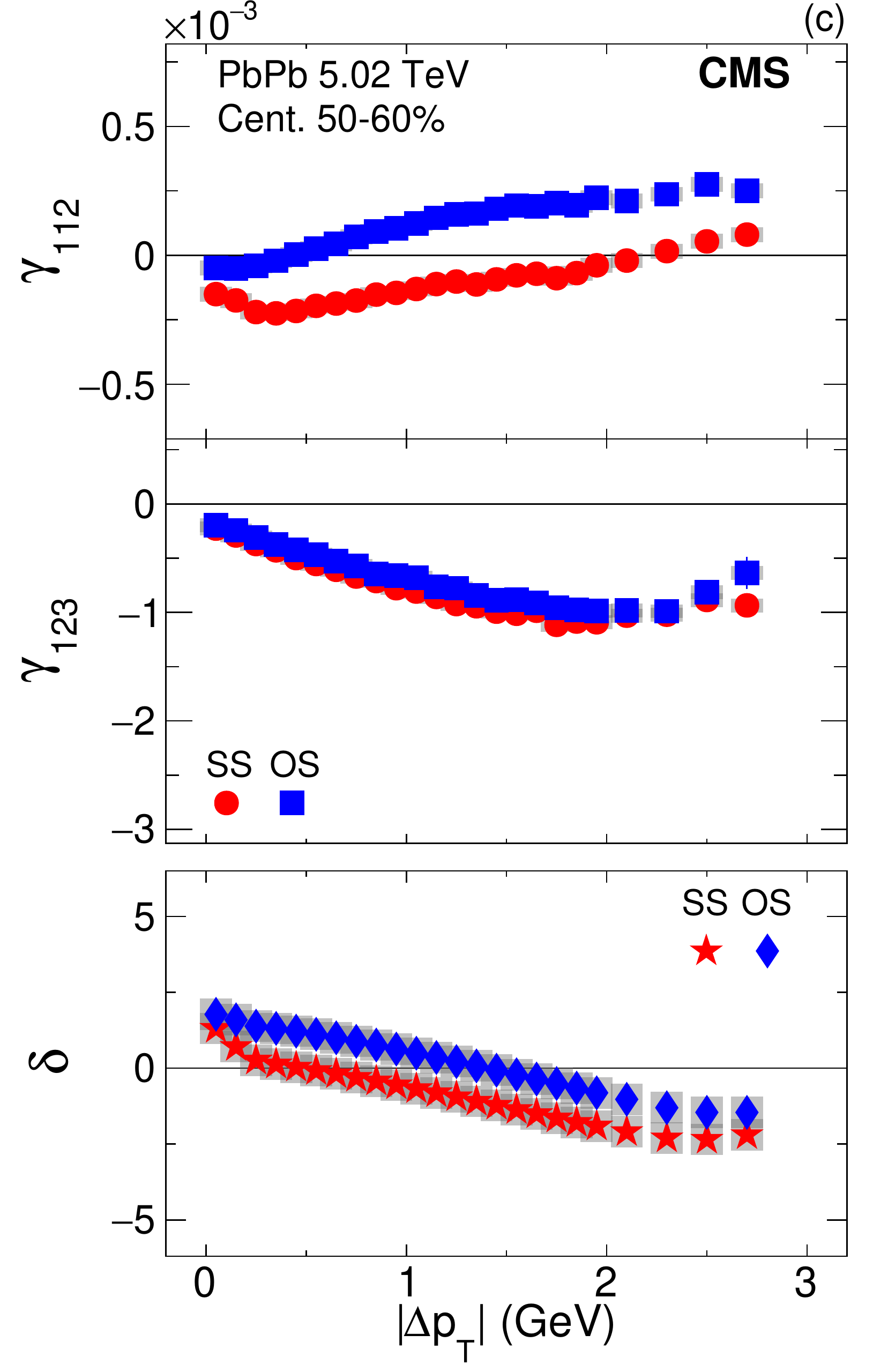}
  \includegraphics[width=0.32\textwidth]{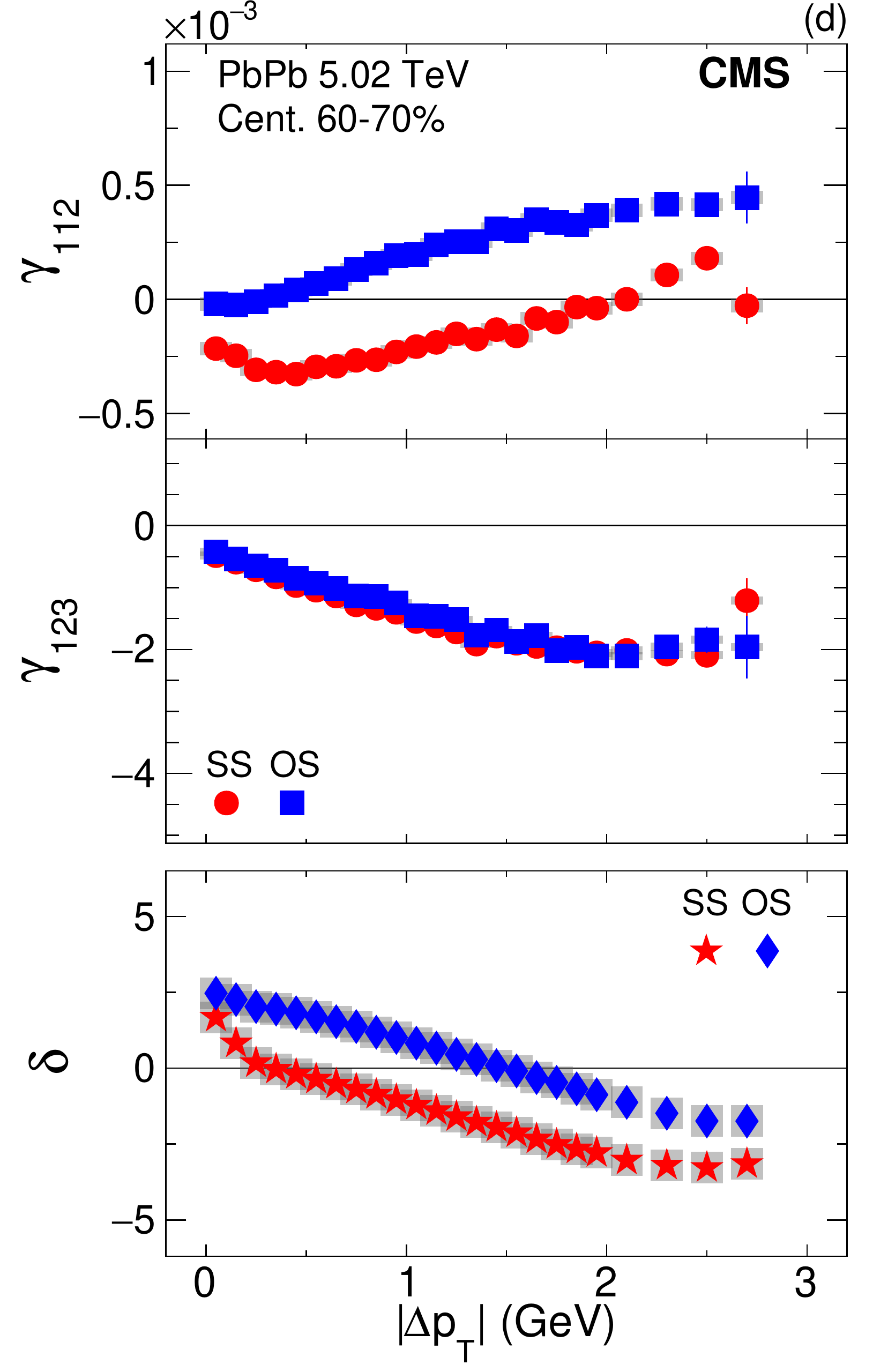}
  \includegraphics[width=0.32\textwidth]{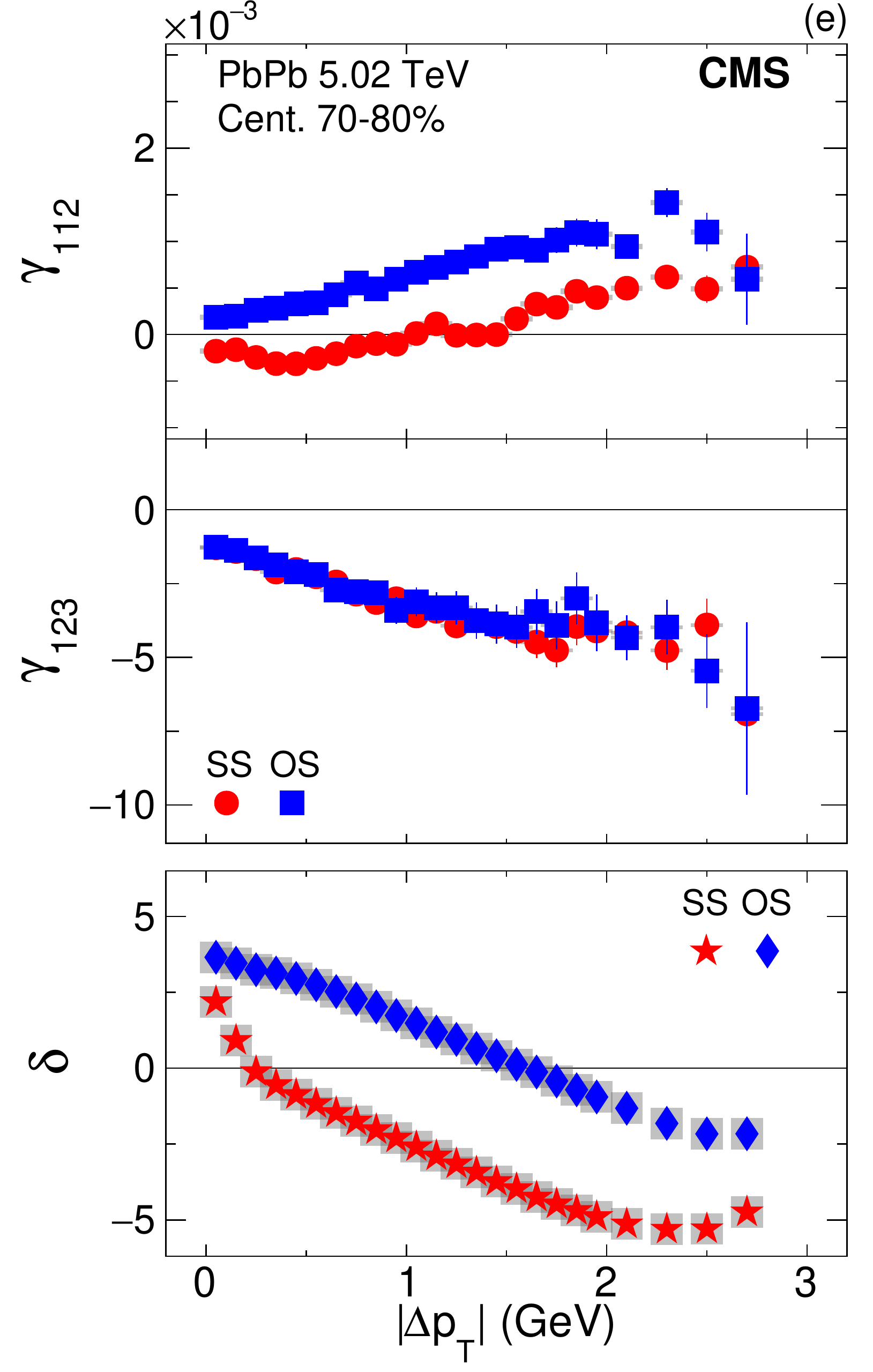}
  \caption{ \label{fig1_Cent_appendix_b}
The SS and OS three-particle correlators,
$\gamma_{112}$ (upper)  and $\gamma_{123}$ (middle), and two-particle
correlator, $\delta$ (lower), as a function of $\abs{\Delta\pt}$
for five centrality classes in \PbPb collisions at $\rootsNN = 5.02$\TeV. The SS and OS two-particle correlators are denoted by different markers. Statistical and systematic uncertainties
are indicated by the error bars and shaded regions, respectively.
   }
\end{figure*}

\begin{figure*}[thb]
\centering
  \includegraphics[width=0.32\textwidth]{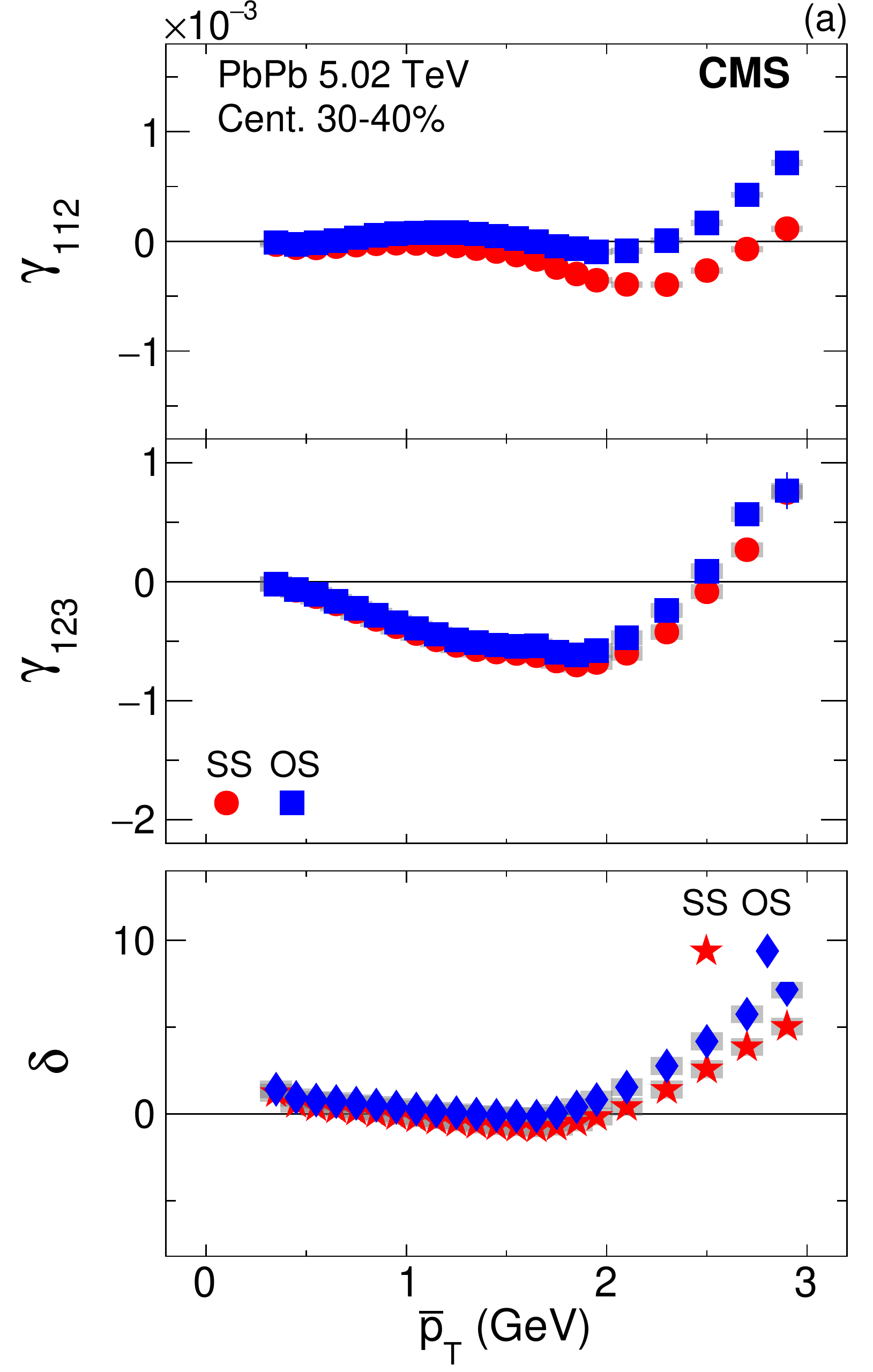}
  \includegraphics[width=0.32\textwidth]{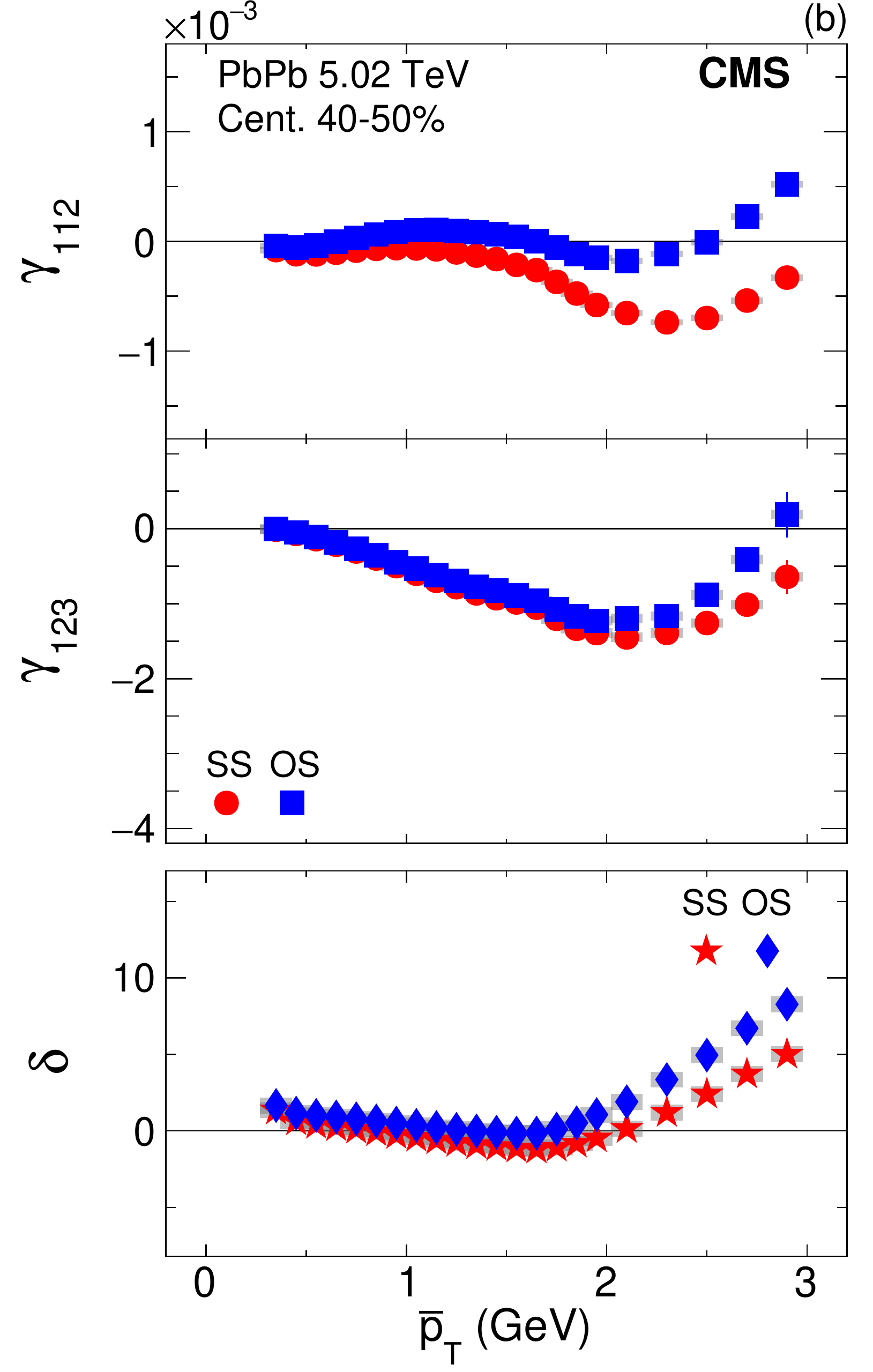}
  \includegraphics[width=0.32\textwidth]{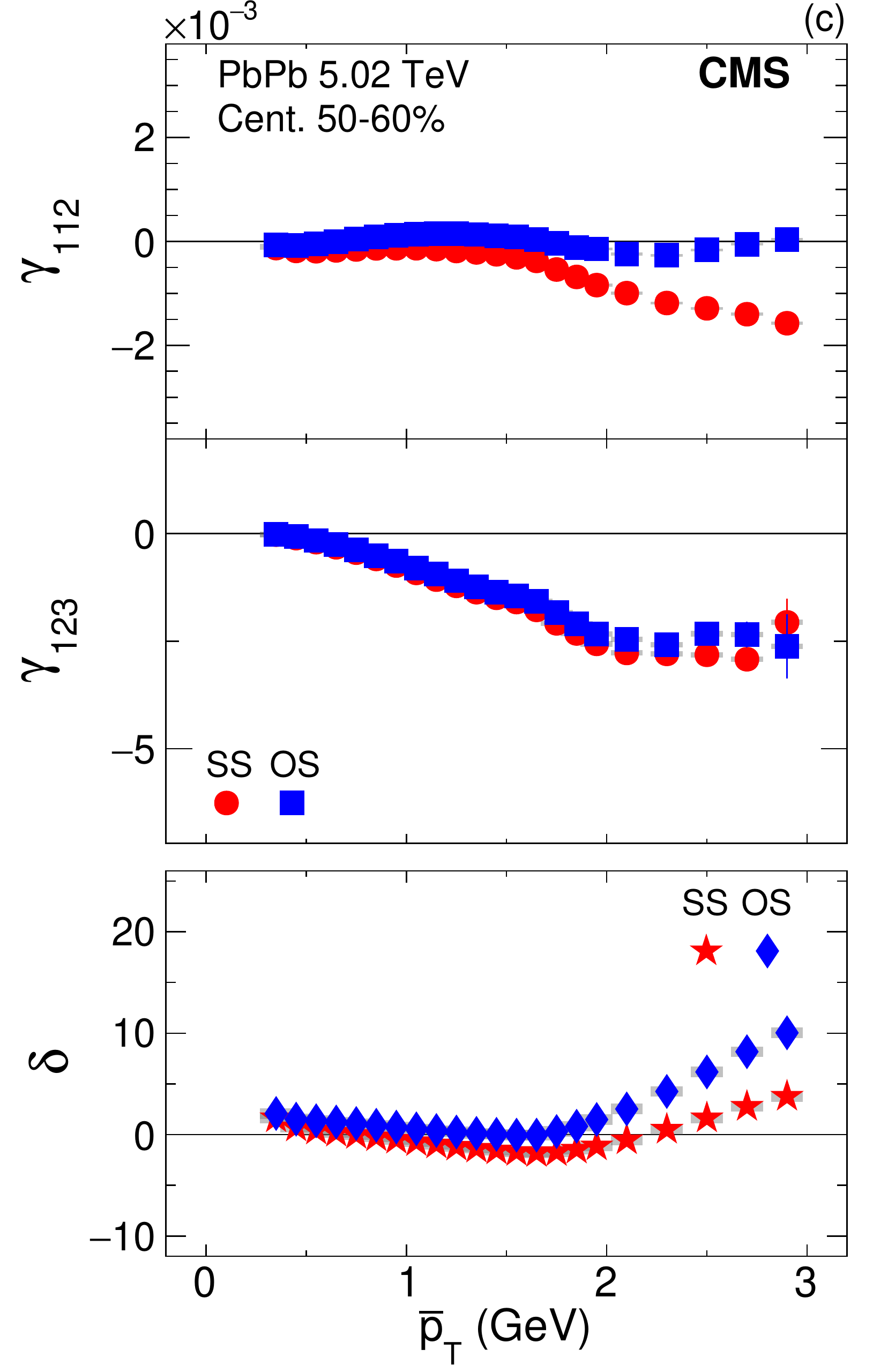}
  \includegraphics[width=0.32\textwidth]{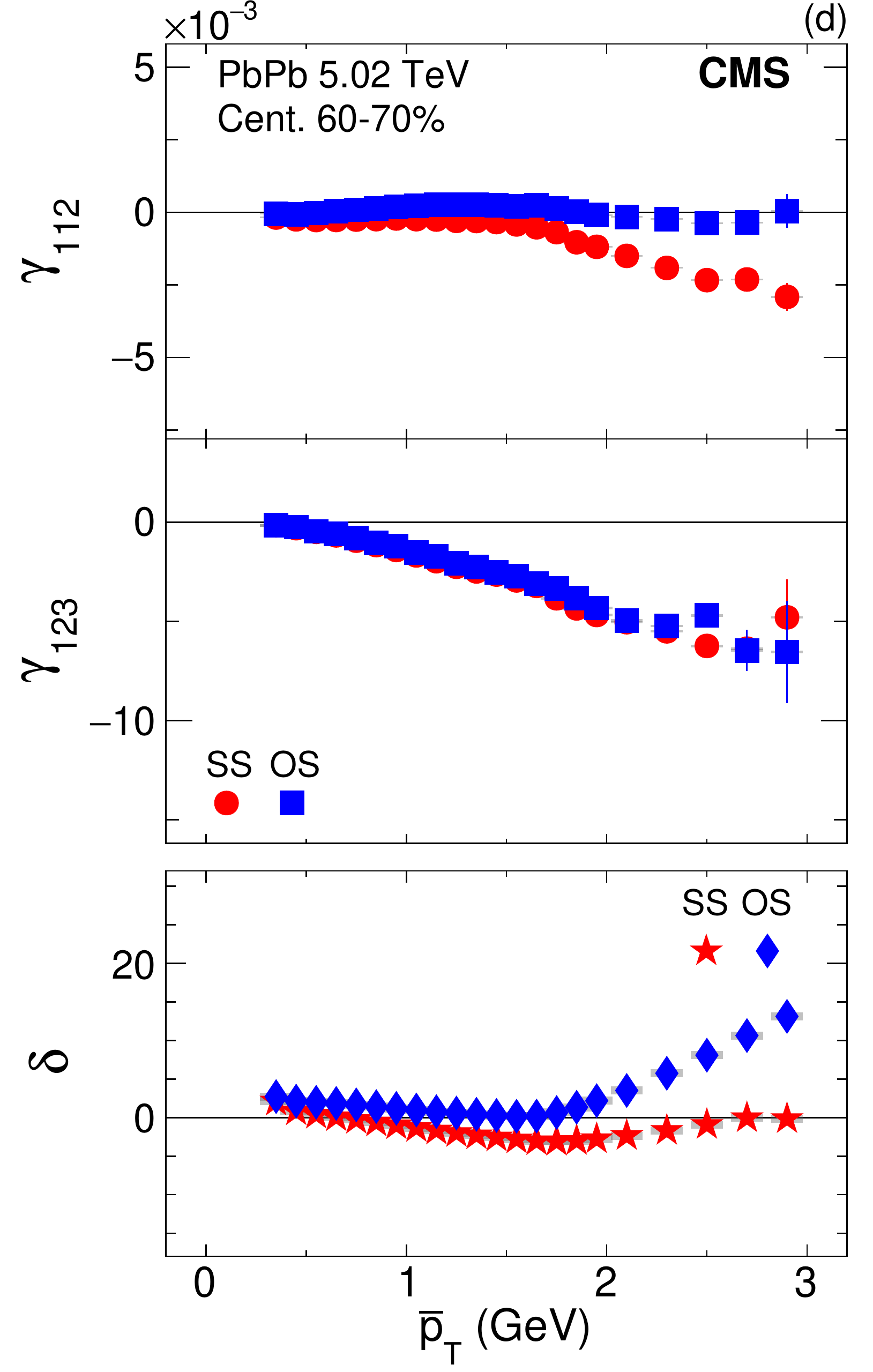}
  \includegraphics[width=0.32\textwidth]{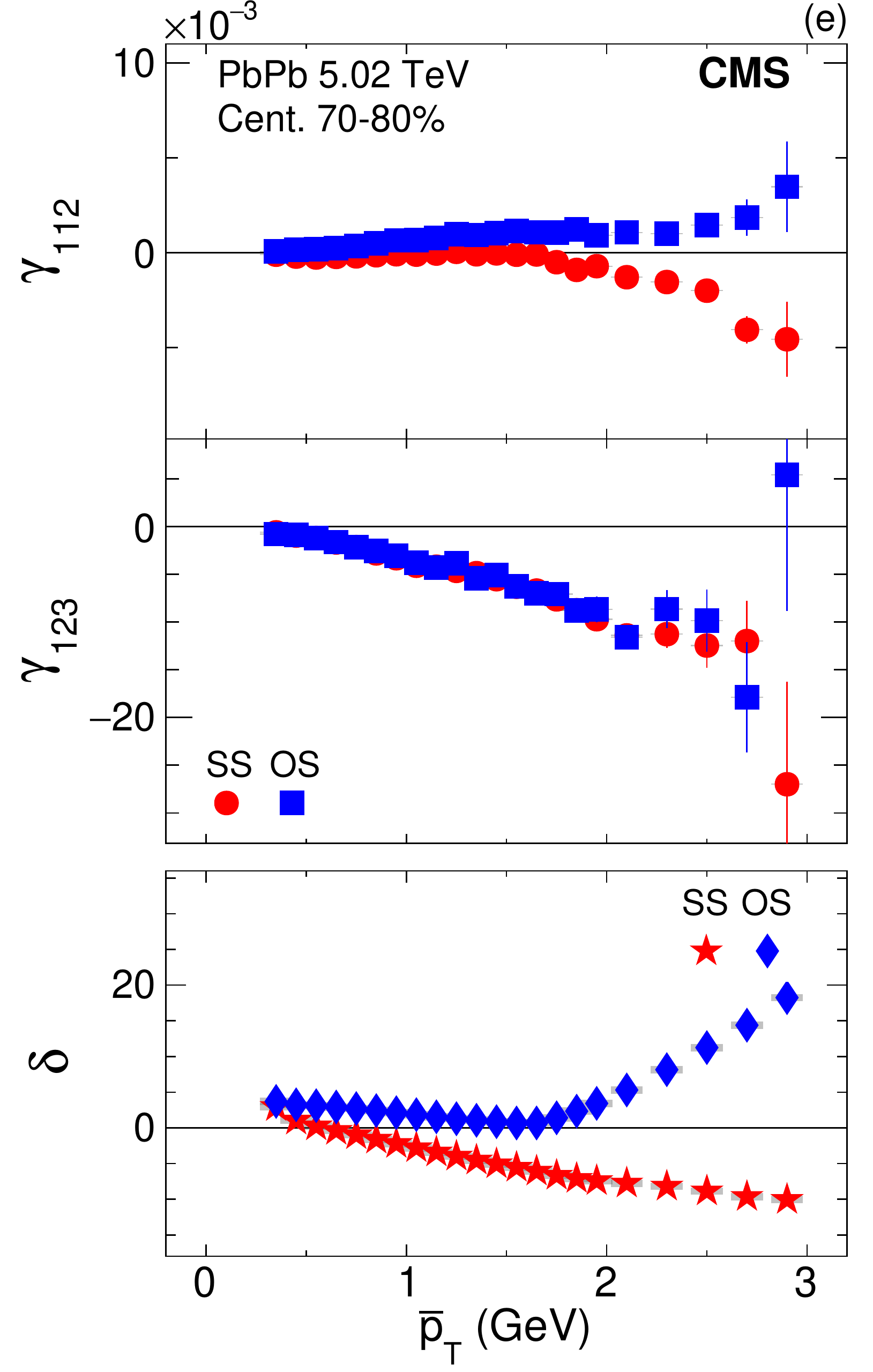}
  \caption{ \label{fig1_Cent_appendix_c}
The SS and OS three-particle correlators,
$\gamma_{112}$ (upper)  and $\gamma_{123}$ (middle), and two-particle
correlator, $\delta$ (lower), as a function of $\ptbar$
for five centrality classes in \PbPb collisions at $\rootsNN = 5.02$\TeV. The SS and OS two-particle correlators are denoted by different markers. Statistical and systematic uncertainties
are indicated by the error bars and shaded regions, respectively.
   }
\end{figure*}

The figures in Appendix~\ref{app:c} show the $\gamma_{112}$, $\gamma_{123}$, and the $\delta$ correlators as a function of $\abs{\deta}$, $|\Delta\pt|$, and \ptbar in \pPb collisions at $\rootsNN = 8.16$\TeV and \PbPb collisions at 5.02\TeV. In \pPb and \PbPb collisions, the results are shown for multiplicity ranges \noff = [120,150), [150,185), [185,250), and [250,300) in
Figs.~\ref{fig1_appendix_a} to~\ref{fig1_appendix_c}. In \PbPb collisions, the results are also shown for five centrality classes from 30--80\% in Figs.~\ref{fig1_Cent_appendix_a} to~\ref{fig1_Cent_appendix_c}.

\clearpage
\bibliography{auto_generated}

\cleardoublepage \section{The CMS Collaboration \label{app:collab}}\begin{sloppypar}\hyphenpenalty=5000\widowpenalty=500\clubpenalty=5000\textbf{Yerevan Physics Institute,  Yerevan,  Armenia}\\*[0pt]
A.M.~Sirunyan, A.~Tumasyan
\vskip\cmsinstskip
\textbf{Institut f\"{u}r Hochenergiephysik,  Wien,  Austria}\\*[0pt]
W.~Adam, F.~Ambrogi, E.~Asilar, T.~Bergauer, J.~Brandstetter, E.~Brondolin, M.~Dragicevic, J.~Er\"{o}, M.~Flechl, M.~Friedl, R.~Fr\"{u}hwirth\cmsAuthorMark{1}, V.M.~Ghete, J.~Grossmann, J.~Hrubec, M.~Jeitler\cmsAuthorMark{1}, A.~K\"{o}nig, N.~Krammer, I.~Kr\"{a}tschmer, D.~Liko, T.~Madlener, I.~Mikulec, E.~Pree, N.~Rad, H.~Rohringer, J.~Schieck\cmsAuthorMark{1}, R.~Sch\"{o}fbeck, M.~Spanring, D.~Spitzbart, W.~Waltenberger, J.~Wittmann, C.-E.~Wulz\cmsAuthorMark{1}, M.~Zarucki
% ...
\vskip\cmsinstskip
% ...
1:~~Also at Vienna University of Technology, Vienna, Austria\\
% ...

\end{sloppypar}
\end{document}